\documentclass[aps,onecolumn,nofootinbib,groupedaddress,superscriptaddress,11pt]{article}
\pdfoutput=1 
\usepackage{jheppub} 
\usepackage{float}

\usepackage{hyperref}
\usepackage{booktabs}
\usepackage{cleveref}
\usepackage[T1]{fontenc} 
\usepackage{siunitx} 
\usepackage{xcolor,graphicx}
\usepackage{dcolumn}
\usepackage{bm}
\usepackage[normalem]{ulem}
\usepackage{braket} 
\usepackage{amsmath}
\usepackage{cancel}
\usepackage{slashed}
\usepackage{multicol}
\usepackage{bbold}
\usepackage[splitrule]{footmisc}
\usepackage{multirow}

\counterwithin*{equation}{section}
\allowdisplaybreaks

\newcommand{\nnl}{\nonumber \\}

\newcommand{\beq}{\begin{equation}}
\newcommand{\eeq}{\end{equation}}
\newcommand{\ba}{\begin{array}}
\newcommand{\ea}{\end{array}}
\newcommand{\bea}{\begin{eqnarray}}
\newcommand{\eea}{\end{eqnarray} }
\newcommand{\be}{\begin{eqnarray}}
\newcommand{\ee}{\end{eqnarray} }
\newcommand{\bal}{\begin{align}}
\newcommand{\eal}{\end{align}}
\newcommand{\bi}{\begin{itemize}}
\newcommand{\ei}{\end{itemize}}
\newcommand{\ben}{\begin{enumerate}}
\newcommand{\een}{\end{enumerate}}
\newcommand{\bc}{\begin{center}}
\newcommand{\ec}{\end{center}}
\newcommand{\bt}{\begin{table}}
\newcommand{\et}{\end{table}}
\newcommand{\btb}{\begin{tabular}}
\newcommand{\etb}{\end{tabular}}

\newcommand{\cM}{{\mathcal M}}

\def\hc{{\rm h.c.}}

\usepackage{etoolbox,siunitx}
\robustify\bfseries

\usepackage{tikz}

\begin{document}

\title{Effective Field Theory in Long-Baseline Neutrino Oscillation Experiments} 

\author[a,b]{Joachim Kopp,}
\author[a,c]{Zahra Tabrizi,}
\author[e]{Salvador Urrea}

\affiliation[a]{Theoretical Physics Department, CERN, Geneva, Switzerland}
\affiliation[b]{PRISMA Cluster of Excellence \& Mainz Institute for Theoretical Physics, Johannes Gutenberg University, Staudingerweg 7, 55099 Mainz, Germany}
\affiliation[c]{PITT PACC, Department of Physics and Astronomy, University of Pittsburgh, 3941 O’Hara St., Pittsburgh, PA 15260, USA}
\affiliation[e]{IJCLab, Pôle Théorie (Bat. 210), CNRS/IN2P3, 91405 Orsay, France}

\emailAdd{jkopp@cern.ch}
\emailAdd{z$\_$tabrizi@pitt.edu}
\emailAdd{salvador.urrea@ijclab.in2p3.fr}

\abstract{We study the phenomenology of physics beyond the Standard Model in long-baseline neutrino oscillation experiments using the most general parametrisation of heavy new physics in the framework of Standard Model Effective Theory (SMEFT), as well as its counterpart below the electroweak scale, Weak Effective Field Theory (WEFT). We compute neutrino production, oscillation, and detection rates in these frameworks, consistently accounting for renormalisation group running as well as SMEFT/WEFT matching. We moreover use appropriately modified neutrino--nucleus cross sections, focusing specifically on the regime of quasi-elastic scattering. Compared to the traditional formalism of non-standard neutrino interactions (NSI), our approach is theoretically more consistent, and it allows for straightforward joint analyses of data taken at different energy scales and by different experiments including not only neutrino oscillation experiments, but also searches for charged lepton flavour violation, low-energy precision measurements, and the LHC. As a specific example, we carry out a sensitivity study for the DUNE experiment and compute projected limits on the WEFT and SMEFT Wilson coefficients. Together with this paper, we also release a public simulation package called ``GLoBES-EFT'' for consistently simulating long-baseline neutrino oscillation experiments in the presence of new physics parameterized either in WEFT or in SMEFT. GLoBES-EFT is available from \href{https://github.com/SalvaUrrea2/GLoBES-EFT}{GitHub}.
}

\maketitle

\section{Introduction}
\label{sec:introduction}

Upcoming long-baseline neutrino oscillation experiments like DUNE \cite{DUNE:2020txw,DUNE:2020ypp,DUNE:2020jqi}, HyperKamiokande \cite{Hyper-KamiokandeProto-:2015xww,Hyper-Kamiokande:2018ofw}, and JUNO \cite{JUNO:2021vlw} are poised to play a transformative role in neutrino physics. They will measure the remaining unknowns of neutrino oscillations, in particular the leptonic CP-violating phase $\delta_\text{CP}$, which is one of the two parameters of the Standard Model that have never been measured.\footnote{Here, we take the point of view that massive neutrinos are part of the Standard Model. This is motivated by the fact that the Glashow--Salam--Weinberg model with massless neutrinos could be considered ``standard'' for 15 years between 1983 (discovery of the $W$ and $Z$ bosons) and 1998 (discovery of neutrino oscillations), while the model with massive neutrinos has been the state of the art for 27~years already, thereby making it more standard than its predecessor without neutrino masses.} (The other one is the mass of the lightest neutrino mass eigenstate, which will likely be determined from cosmology in the coming years.) Long-baseline neutrino experiments, together with cosmological probes, will thereby complete the Standard Model -- and probe possible deviations from it. These deviations could come either in the form of new, very weakly coupled, particles that can be directly produced in neutrino oscillation experiments (sterile neutrinos with masses up to $\mathcal{O}(\si{GeV})$~\cite{Mohapatra:1986bd, GonzalezGarcia:1988rw, Wyler:1982dd, Akhmedov:1995ip, Akhmedov:1995vm, Barry:2011wb, Zhang:2011vh}, light dark matter~\cite{Antel:2023hkf}, axion-like particles~\cite{Peccei:1977hh, Peccei:1977ur, Weinberg:1977ma, Wilczek:1977pj}, \ldots), or they can originate from heavy new physics which subtly modifies neutrino interactions and oscillations.

Here, we focus on this second class of physics beyond the Standard Model. A robust framework to describe the manifestations of heavy new physics at low-energy scales is provided by Effective Field Theory (EFT), which extends the SM Lagrangian with non-renormalisable higher-dimensional operators that are suppressed by the scale of new physics, $\Lambda$. For experiments operating at energies below the QCD scale, the appropriate EFT is one that includes leptons, nucleons, and photons as degrees of freedom. Above the QCD scale, the model of choice is Weak Effective Field Theory (WEFT), which includes dynamical quarks (except for the top quark), but no electroweak bosons. Finally, above the electroweak scale, WEFT is embedded into the Standard Model Effective Field Theory (SMEFT), the most general extension of the SM Lagrangian with higher-dimensional operators. (Of course, if new particles, such as sterile neutrinos or light bosons, should exist at intermediate energy scales, they will need to be included as well.) In the context of neutrino physics, EFT operators of dimension six and higher can affect neutrino production, propagation, and detection processes, leading to measurable deviations in experimental observables. Such deviations, if observed, can offer indirect insights into the nature and scale of BSM physics.

To exemplify this, we will study specifically the constraints that DUNE can impose on EFT operators relevant to neutrino production, detection, and oscillations, using both the near and far detectors.\footnote{The DUNE experiment is particularly well suited for new physics searches thanks to the large energy range covered by its wide-band beam, its long baseline, and the superb event reconstruction capabilities offered by its liquid argon detectors. Moreover, beam fluxes and detector response functions for DUNE are public, allowing us to model the experiment reliably. On the downside, DUNE is smaller than HyperKamiokande, limiting its reach in searches that are purely statistics dominated. \label{fn:why-dune}} We use comprehensive phenomenological simulations of the DUNE experiment based on realistic neutrino fluxes and cross sections as well as a versatile oscillation engine, including at every step the appropriate modifications in the presence of new physics. We moreover use detailed detector response functions and a sophisticated model of systematic uncertainties. We present our results as projected constraints on WEFT and SMEFT Wilson coefficients, carefully taking into account renormalisation group running and WEFT/SMEFT matching. For the numerical analysis, we have developed a simulation package called GLoBES-EFT which extends the widely used GLoBES package \cite{Huber:2004ka, Huber:2007ji} by the capability to consistently simulate and analyse neutrino oscillation measurements across a wide range of energy scales in the presence of arbitrary dimension-6 operators. GLoBES-EFT is available from \href{https://github.com/SalvaUrrea2/GLoBES-EFT/}{GitHub} \cite{github}.

The plan of the paper is as follows: we begin in \cref{sec:eft-framework} by setting out the EFT frameworks we are going to use, thereby also fixing our notation, and we show how dimension-6 operators affect neutrino production, propagation, and detection. In \cref{sec:experimental-setup}, we describe our simulations of the DUNE experiment, and in \cref{sec:results} we present our main results in the form of sensitivity projections for DUNE. We conclude in \cref{sec:conclusion}. In the appendices, we collect the relevant renormalisation group equations and matching conditions, present the complete set of results for all WEFT and SMEFT constraints obtained in this work, and we document the GLoBES-EFT package.

\section{Theory Framework: EFT Ladder}
\label{sec:eft-framework}

\subsection{Standard Model Effective Field Theory}

SMEFT is the most general EFT framework for exploring new physics beyond the electroweak scale \cite{Buchmuller:1985jz}. It is formulated as a series of higher-dimensional operators constructed from the Standard Model's particle content while preserving its symmetries:
\begin{align}
    \mathcal{L}_{\rm SMEFT} = \mathcal{L}_{\rm SM}
                            + \mathcal{L}_{d=5} + \mathcal{L}_{d=6} + \dots
    \label{eq:SMEFT1}
\end{align}
Here, the operators of dimension $d$ are suppressed by factors of $1 / \Lambda^{d-4}$, where $\Lambda$ represents the scale of new physics. Besides the Weinberg operator~\cite{Weinberg:1979sa} at $d=5$, which generates Majorana neutrino masses, the most important contributions arise from the $d=6$ operators. These are conveniently expressed as:
\begin{align}
    \mathcal{L}_{d=6} = \sum_i \frac{c_i}{v^2} \mathcal{O}_i
\end{align}
where $v$ is the Higgs vacuum expectation value (vev) and $c_i$ are the dimensionless Wilson coefficients.\footnote{Here we follow the same conventions as, e.g., ref.~\cite{Breso-Pla:2023tnz}, of treating $v$ as the EFT power-counting parameter. This has the advantage that the magnitudes of the SMEFT Wilson coefficients and the WEFT Wilson coefficient they map onto are numerically comparable. The scale of the new particles that mediate the effective interactions needs to be $\gg v$ of course for SMEFT to be a consistent EFT.}

SMEFT has been studied extensively in the literature. The establishment of the Warsaw basis \cite{Grzadkowski:2010es} and the Silh basis \cite{Contino:2013kra} was a significant milestone, as it provided a systematic way to classify all $d=6$ operators without redundancies. There have been significant theoretical developments focused on the renormalization group evolution of these operators \cite{Jenkins:2013zja, Jenkins:2013wua, Alonso:2013hga}, studying the matching conditions between the SMEFT and low-energy effective field theories \cite{Jenkins:2017jig}, and preparing automated tools that generate Feynman rules \cite{Dedes:2017zog}. 

\begin{table}
    \centering
    \begin{tabular}{r@{\;}c@{\;}l}
        \toprule
        \multicolumn{3}{c}{4-leptons} \\
        \midrule
        $[O_{ll}]_{\alpha\beta\gamma\delta}$ &=& ${1 \over 2} 
        (\bar l_\alpha \gamma_\mu P_L l_\beta) (\bar l_\gamma \gamma^\mu P_L l_\delta)$
        \\[0.2cm]

        $[O_{l e}]_{\alpha\beta\gamma\delta}$ &=& $(\bar l_\alpha \gamma_\mu P_L l_\beta) 
        (\bar e_\gamma \gamma^\mu P_R e_\delta)$
        \\
        \bottomrule
    \end{tabular}
    \caption{Flavour-conserving 4-lepton operators in the SMEFT Lagrangian.}
    \label{tab:4l}
\end{table}

\begin{table}
    \centering
    \begin{tabular}{r@{\;}c@{\;}l|r@{\;}c@{\;}l}
        \toprule
        \multicolumn{6}{c}{2-lepton--2-quark} \\
        \midrule
        \multicolumn{3}{c}{chirality-conserving} & \multicolumn{3}{c}{chirality-violating} \\
        \midrule
        $[O_{l q}]_{\alpha\beta jk}$ &=& $(\bar l_\alpha \gamma_\mu P_L l_\beta) 
        (\bar q_j \gamma^\mu P_L q_k)$ &
        $[O^{(1)}_{l e q u}]_{\alpha\beta jk}$ &=& $(\bar l_{ \alpha}^m P_R e_\beta) \epsilon_{mn}
        (\bar q^n_j  P_R u_k)$
        \\[0.2cm]

        $[O^{(3)}_{l q}]_{\alpha\beta jk}$ &=& $(\bar l_\alpha \gamma_\mu \sigma^m P_L l_\beta)
        (\bar q_j \gamma^\mu \sigma^m P_L q_k)$ &
        $[O^{(3)}_{l e q u}]_{\alpha\beta jk}$ &=& $(\bar l_{ \alpha}^m \sigma_{\mu\nu} P_R e_\beta) \epsilon_{mn} (\bar q^n_j \sigma^{\mu\nu} P_R u_k)$
        \\[0.2cm]

        $[O_{l u}]_{\alpha\beta jk}$ &=& $(\bar l_\alpha \gamma_\mu P_L l_\beta)
        (\bar u_j \gamma^\mu P_R u_k)$ &
        $[O_{l e d q}]_{\alpha\beta jk}$ &=& $(\bar l_\alpha P_R e_\beta) (\bar d_j P_L q_k)$
        \\[0.2cm]

        $[O_{l d}]_{\alpha\beta jk}$ &=& $(\bar l_\alpha \gamma_\mu P_L l_\beta)
        (\bar d_j \gamma^\mu P_R d_k)$ \\
        \bottomrule
    \end{tabular}
    \caption{Flavour-conserving 2-lepton--2-quark operators in the SMEFT Lagrangian.}
    \label{tab:2l2q}
\end{table}

For connecting SMEFT operators to the observables at particle physics experiments, it is useful to rewrite \cref{eq:SMEFT1} in terms of the gauge boson mass eigenstates after electroweak symmetry breaking and to separate the Higgs field into its vev and the physical Higgs boson. The effects of the higher-dimensional operators will then appear as either new interaction terms not present in the SM Lagrangian or as corrections to the SM couplings. For the parametrization of the $d=6$ operators, we use the so-called Higgs basis~\cite{LHCHiggsCrossSectionWorkingGroup:2016ypw} (see also \cite{Azatov:2022kbs} for a recent review). The new $d=6$ operators relevant for neutrino interactions with charged leptons and quarks (which form a subset of all 4-lepton and 2-lepton--2-quark operators) are shown in \cref{tab:4l,tab:2l2q}. In these tables, $l_\alpha \equiv (\nu_\alpha, e_\alpha)$ are the $SU(2)$ doublets containing the left-handed lepton fields $q_j \equiv \big(\sum_k V^\dagger_{jk} u_k,d_j\big)$ are the corresponding quark doublets. $V$ denotes the Cabibbo--Kobayashi--Maskawa (CKM) matrix~\cite{Cabibbo:1963yz,Kobayashi:1973fv}. The vertex corrections to the fermions' electroweak gauge interactions that are relevant for this work are given by~\cite{Breso-Pla:2023tnz}: 
\begin{align}
    {\cal L}_{\rm SMEFT} \supset & -\sqrt{g_L^2 + g_Y^2} Z_\mu \bigg\{
        \Big[ \tfrac{1}{2} \mathbb{1} + \delta g_L^{We} + \delta g_L^{Z e} \Big]_{\alpha\beta}
        (\bar \nu_\alpha \gamma^\mu P_L \nu_\beta)
                                        \notag\\[0.05cm]
    &+ \Big[ \big(-\tfrac{1}{2} + \sin^2 \theta_W \big) \mathbb{1}
            + \delta g_L^{Z e} \Big]_{\alpha\beta} (\bar e_\alpha \gamma^\mu P_L e_\beta)
     + \Big[ \sin^2 \theta_W \mathbb{1} + \delta g_R^{Z e} \Big]_{\alpha\beta}
            (\bar e_\alpha \gamma^\mu P_R e_\beta)
                                        \notag\\[0.2cm]
    &+ \Big[ \big( \tfrac{1}{2} - \tfrac{2}{3} \sin^2 \theta_W \big) \mathbb{1}
                + \delta g_L^{Zu}  \Big]_{jk}  (\bar u_j \gamma^\mu P_L u_k)
     + \Big[-\tfrac{2}{3} \sin^2 \theta_W \mathbb{1}
                + \delta g_R^{Zu} \Big]_{jk} (\bar u_j \gamma^\mu P_R u_k)
                                        \notag\\[0.2cm]
    &+ \Big[ \big(-\tfrac{1}{2} + \tfrac{1}{3} \sin^2\theta_W \Big) \mathbb{1}
                + \delta g_L^{Zd} \Big]_{jk} (\bar d_j \gamma^\mu  P_L d_k)
     + \Big[ \tfrac{1}{3} \sin^2\theta_W \mathbb{1}
                + \delta g_R^{Zd} \Big]_{jk} (\bar d_j \gamma^\mu P_R d_k)
    \Big\}
                                        \notag\\[0.1cm]
    &+ {g_L \over \sqrt 2} \bigg\{
        W^{\mu+} \big[ \mathbb{1} + \delta g^{We}_L \big]_{\alpha\beta}
            (\bar \nu_\alpha \gamma_\mu P_L e_\beta)
      + W^{\mu+} \big[ V + \delta g^{Z u}_L - \delta g^{Z d}_L \big]_{jk}
            (\bar u_j \gamma_\mu P_L d_k)
                                        \notag\\
     &+ W^{\mu+} \big[ \delta g^{W q}_R \big]_{jk}
            (\bar u_j \gamma_\mu P_R d_k) +\hc
    \bigg\}
                                        \notag\\
     &+ {(g_L^2 + g_Y^2) v^2 \over 8} Z_\mu Z^\mu
      + {g_L^2 v^2 \over 4} \, (1 + 2 \, \delta m) W_\mu^+ W^{-\,\mu} \,.
    \label{eq:SMEFT2_vertex}
\end{align}
Here, $v \equiv (\sqrt{2} G_F)^{-1/2} \approx \SI{246}{GeV}$ is the vacuum expectation value of the SM Higgs field, $P_{L,R} = \frac{1}{2} (1 \mp \gamma^5)$ are the chirality projection operators, and $\theta_W$ is the weak mixing angle.  We denote the mass eigenstates of down-type quarks, up-type quarks, and charged leptons by $d_k$, $u_j$, and $e_\alpha$, respectively. The neutrino fields $\nu_\alpha$ are expressed in the flavour basis, and are related to the corresponding mass eigenstates via the Pontecorvo--Maki--Nakagawa–Sakata (PMNS) mixing matrix~\cite{Pontecorvo:1967fh, Maki:1962mu}: $\nu_\alpha = \sum_{i=1}^3 U_{\alpha i} \nu_i$. We sum over the generation indices denoted by Greek letters, $\alpha,\beta,\gamma,\delta = e,\mu,\tau$ for the leptons, and by Latin letters $j,k=1,2,3$ for the quarks. The parameters $\delta g_{L,R}$ are vertex corrections, and $\delta m = \frac{1}{2} \big( [\delta g_L^{We}]_{ee} + [\delta g_L^{We}]_{\mu\mu} \big) - \frac{1}{4} v^2[C_{ll}]_{e\mu\mu e}$ is a correction to the $W$ mass. The relation between the $\delta g_{L,R}$ and the Wilson coefficients is given in \cref{sec:matching}. It is important to note that not all the corresponding Wilson coefficients in SMEFT are independent. For instance $[C_{ll}]_{\alpha\beta\gamma\delta} = [C_{ll}]_{\gamma\delta\alpha\beta}$ and 
$[C_X]_{\alpha\beta\gamma\delta}  = [C_X]_{\beta\alpha\delta\gamma}^*$ (for $X=ll,le$).

\subsection{Weak Effective Field Theory}

SMEFT provides a convenient framework for probing high-energy physics above the weak scale. However, accelerator neutrino experiments operate at much lower energies $\mu \lesssim m_W$, where the most general EFT Lagrangian is the one Weak Effective Field Theory (WEFT). In WEFT, the electroweak gauge bosons, the Higgs boson, and the top quark are integrated out, and the electroweak symmetry is explicitly broken. Focusing on those WEFT operators which can be explored in long-baseline neutrino oscillation experiments, we are specifically interested in charged current (CC) and neutral current (NC) four-fermion interactions between neutrinos and regular matter. The CC WEFT Lagrangian is given by:
\begin{align} \label{eq:WEFT_CC}
    \mathcal{L}_{\rm WEFT,CC} \supset & -\frac{2 V_{jk}}{v^2} \bigg\{
        \Big[\mathbb{1} + \epsilon_L^{jk} \Big]_{\alpha\beta}
            \Big(\bar{u}^j \gamma^\mu P_L d^k\Big)
            \Big(\bar{e}_\alpha \gamma_\mu P_L \nu_\beta\Big)
      + \Big[\epsilon_R^{jk}\Big]_{\alpha\beta}
            \Big(\bar{u}^j \gamma^\mu P_R d^k\Big)
            \Big(\bar{e}_\alpha \gamma_\mu P_L \nu_\beta\Big) \notag\\
    & + \frac{1}{2} \Big[\epsilon_S^{jk} \Big]_{\alpha\beta}
            \Big(\bar{u}^j d^k\Big) \Big(\bar{e}_\alpha P_L \nu_\beta\Big)
      - \frac{1}{2} \Big[\epsilon_P^{jk} \Big]_{\alpha\beta}
            \Big(\bar{u}^j \gamma_5 d^k\Big) \Big(\bar{e}_\alpha P_L \nu_\beta\Big) \notag\\
    & + \frac{1}{4} \Big[\epsilon_T^{jk}\Big]_{\alpha\beta}
            \Big(\bar{u}^j \sigma^{\mu \nu} P_L d^k\Big)
            \Big(\bar{e}_\alpha \sigma_{\mu \nu} P_L \nu_\beta\Big)
      + \text{h.c.} \bigg\}\,.
\end{align}
Here $V_{jk}$ are again the entries of the CKM matrix and the tensor $\sigma^{\mu\nu}$ is defined as $\tfrac{i}{2} [\gamma^\mu,\gamma^\nu]$. The new CC interactions introduced in \cref{eq:WEFT_CC} are conventionally parametrised through dimensionless Wilson coefficients $[\epsilon_X^{jk}]_{\alpha\beta}$, where the indices $j$, $k$ label quark generations and $\alpha$, $\beta$ correspond to the charged lepton and neutrino flavours, respectively. Finally, the subscript $X$ identifies the Lorentz structure of the operator, where $X = L, R, S, P, T$ corresponds to left-handed, right-handed, scalar, pseudo-scalar, and tensor interactions, respectively. 

The relevant WEFT Lagrangian for the NC interactions between neutrinos and other fields is given by
\begin{align}
    \mathcal{L}_{\rm WEFT,NC} \supset  -\frac{2}{v^2} &\bigg\{
        \Big[g^f_L \mathbb{1}+\epsilon_L^{m,f}\Big]_{\alpha\beta}\Big(\bar{\nu}_\alpha \gamma_\mu P_L \nu_\beta\Big)
            \Big(\bar{f} \gamma^\mu P_L f\Big)
     \nonumber\\ &+ \Big[g^f_R \mathbb{1}+\epsilon_R^{m,f}\Big]_{\alpha\beta}
            \Big(\bar{\nu}_\alpha \gamma_\mu P_L \nu_\beta\Big)
            \Big(\bar{f} \gamma^\mu P_R f\Big) \bigg\}\,,
    \label{eq:WEFT_NC}
\end{align}
where $f$ can be an electron, an up-quark or a down-quark ($e$, $u$, $d$) in the medium. NC couplings to heavier quarks and charged leptons are irrelevant for neutrinos propagating through ordinary matter. The SM part of \cref{eq:WEFT_NC} depends on the couplings $g_L^f = T_3^f - Q_f \sin ^2 \theta_W$ and $g_R^f = -Q_f \sin^2 \theta_W$, where $T_3^f$ and $Q_f$ are the hypercharge and electric charge of fermion $f$, respectively. Note, however, that the SM terms are flavour-universal, therefore they contribute only an unphysical overall phase to the neutrino oscillation amplitudes. The Wilson coefficients for NC interactions are denoted $\epsilon_L^{m,f}$. (The superscript `m' stands for `matter effects'.)

In \cref{sec:matching} we have summarised the matching relations between the WEFT Wilson coefficients $[\epsilon_X^{jk}]_{\alpha\beta}$, $[\epsilon_X^{m,f}]_{\alpha\beta}$ and vertex corrections $\delta g$ on the one side, and the Wilson coefficients of SMEFT in the Warsaw basis on the other. In this work, we perform the matching between SMEFT and WEFT at a renormalization scale $\mu \sim m_Z$, and we then run the WEFT coefficients in \cref{eq:WEFT_CC,eq:WEFT_NC} down to $\mu = \SI{2}{GeV}$, which is the relevant scale for accelerator neutrino experiments. For running and operator mixing we follow ref.~\cite{Gonzalez-Alonso:2017iyc} and we use the $\overline{\rm MS}$ regularization scheme.

\section{Event Rate}
\label{sec:eft-neutrino-exp}

The new CC interactions in the Lagrangian in \cref{eq:WEFT_CC} can affect the event spectra observed in a neutrino oscillation experiment by modifying the neutrino production and detection processes.  Let us assume that the neutrino is produced together with a charged lepton $e_\alpha$ in the decay of a ``source'' particle, $S \to \nu_i e_\alpha S'$,\footnote{Note that we work with neutrino mass eigenstates here to be consistent with the notation in refs.~\cite{Falkowski:2019xoe, Falkowski:2019kfn, Falkowski:2021bkq}.} where $S'$ denotes any additional decay products. At the parton level, this production process involves a decay of the form $u_j \to \nu_i e_\alpha d_k$ (e.g.\ in kaon decay) or an interaction $u_j \bar{d}_k \to \nu_i e_\alpha$ (e.g.\ in pion decay). We denote the production amplitude $\mathcal{M}_{\alpha i}^{S,jk}$. For an accelerator-based neutrino beam, potentially relevant source particles are $\pi^\pm$, kaons  ($K^\pm$ and $K_L$) and $\mu^\pm$ for $\smash{\overset{\scriptscriptstyle{(-)}}{\nu}_{\!e}}$  and $\smash{\overset{\scriptscriptstyle{(-)}}{\nu}_{\!\mu}}$ production; and charm meson ($D_s^\pm$, $D^\pm$) decays for $\smash{\overset{\scriptscriptstyle{(-)}}{\nu}_{\!\tau}}$. The J-PARC neutrino beam (\SI{30}{GeV} primary proton energy)~\cite{T2K:2011qtm} is dominated by pions, while the DUNE beam also contains a substantial number of kaons thanks to the larger primary proton energy (\num{60}--\SI{120}{GeV})~\cite{DUNE:2020ypp}. Charm mesons only become important at even larger primary proton energies, for instance in SHiP (\SI{400}{GeV} protons form the SPS at CERN)~\cite{SHiP:2015vad} or in collider neutrino experiments (FASER~\cite{FASER:2018eoc}, SND@LHC~\cite{SNDLHC:2022ihg}). Muon decays are always present but subdominant as most muons are stopped before they can decay.

Neutrinos are detected through their CC scattering into a charged lepton $e_\beta$ on a target nucleus $T$: $\nu_i T \to e_\beta T'$, where $T'$ collectively denotes the additional interaction products. The corresponding parton level interaction is of the form $\nu_i d_k \to u_j e_\beta$, and we denote the detection amplitude as $\mathcal{M}_{\beta i}^{D,jk}$. Given that neutrino oscillation experiments are not sensitive to the absolute neutrino mass scale, following~\cref{eq:WEFT_CC} the PMNS matrix elements can be factored out and the production and detection amplitudes can be written as
\begin{align}
    \begin{split}
    \cM_{\alpha i}^{S,jk} &= U_{\alpha i}^{*} A^{S,jk}_{L,\alpha}
        + \sum_{X=L,R,S,P,T} [\epsilon_X^{jk} U]_{\alpha i}^* A^{S,jk}_{X,\alpha} \,,
                                                        \\[0.2cm]
    \cM_{\beta i}^{D,jk}  &= U_{\beta i} A^{D,jk}_{L,\beta}
        + \sum_{X=L,R,S,P,T} [\epsilon_X^{jk} U]_{\beta i} A^{D,jk}_{X,\beta} \,.
    \end{split}
    \label{eq:Mdecomposition}
\end{align}
In these equations, $A^{S,jk}_{X,\alpha}$ and $A^{D,jk}_{L,\beta}$ are \emph{reduced matrix elements} from which the PMNS matrix elements and the WEFT Wilson coefficients have been factored out. They are independent of the neutrino mass and flavour, but depend on kinematic variables, the quark flavours participating in the interactions, and on its Lorentz structure.

NC interactions are not used as detection processes in oscillation analyses, but modifications to these interactions as in \cref{eq:WEFT_NC} do affect neutrino propagation in matter through the Mikheyev--Smirnov--Wolfenstein (MSW) effect \cite{Wolfenstein:1977ue, Mikheyev:1985zog}. The neutrino oscillation Hamiltonian in the flavour basis including new NC interactions is
\begin{align}
    H &= \frac{1}{2 E_\nu}\,  U \, {\rm diag} \begin{pmatrix}
                                         0 \\
                                           & \Delta m^2_{21} \\
                                           &                 & \Delta m^2_{31}
                                    \end{pmatrix} \, U^\dagger
       + V_{\rm PMNS} \,,
    \label{eq:NCHamiltonian}
\end{align}
with 
\begin{align}
   (V_{\rm PMNS})_{\alpha\beta} = \frac{1}{v^2} n_e
       \Big( \delta_{\alpha e}
           + \frac{n_f}{n_e} \big[ \epsilon_V^{m,f} \big]_{\alpha\beta} \Big) \,,
\end{align}
where $n_f$ denotes the number density of particle species $f$ in the background matter. Recall that neutrino matter effects only probe vector interactions, i.e., the combination $\epsilon_V^{m,f} = \epsilon_L^{m,f} + \epsilon_R^{m,f}$. The terms describing couplings to electrons, up quarks, and down quarks (the only constituents of ordinary matter) combine into an effective coupling
\begin{align}
    \epsilon_{\alpha\beta}^{\oplus} \equiv
      [\epsilon_V^{m,e}]_{\alpha\beta}
    + \left(2 + Y_n^\oplus\right) [\epsilon_V^{m,u}]_{\alpha\beta}
    + \left(1 + 2 Y_n^\oplus\right) [\epsilon_V^{m,d}]_{\alpha\beta} ,
    \label{eq:epsilon_earth}
\end{align}
where $Y_n^\oplus = n_n / n_e = 1.051$ is the average ratio of neutron and electron number densities along a neutrino trajectory through the Earth's outer layers according to the PREM model~\cite{Dziewonski:1981xy}. Since oscillation experiments are sensitive to the matter term only up to a global phase, among the three diagonal operators, only two independent combinations (for instance $\epsilon_{ee}^{\oplus} - \epsilon_{\mu\mu}^{\oplus}$ and $\epsilon_{\tau\tau}^{\oplus} - \epsilon_{\mu\mu}^{\oplus}$) can be directly probed~\cite{Gonzalez-Garcia:2011vlg}.

\subsection{Neutrino Oscillations in Presence of New Physics}
\label{sec:eft-neutrino-exp}

The new physics parameters we wish to probe at neutrino experiments are encoded in the production amplitude $\cM_{\alpha n}^{S,jk}$, the detection amplitude $\cM_{\alpha n}^{S,jk}$, and the oscillation Hamiltonian $H$ in \cref{eq:NCHamiltonian}. These in turn determine the experimental observables, in particular the differential event rates. For neutrinos emitted with flavour $\alpha$, propagating through matter, and detected with flavour $\beta$, the differential event rate at a distance $L$ from the source is~\cite{Falkowski:2019kfn, Falkowski:2021bkq}\footnote{Refs.~\cite{Falkowski:2019kfn, Falkowski:2021bkq} only consider neutrino oscillations in vacuum and study the effects of CC interactions on neutrino fluxes and interaction cross sections. Here, we generalize their expressions to consider the effect of new NC interactions in neutrino propagation.}
\begin{align}
  \frac{dR_{\beta}}{dE_\nu} &= N_T \sigma_{\beta}^{\rm SM}(E_\nu)
      \sum_{\alpha,S} \phi_{\alpha}^{S,{\rm SM}}(E_\nu) \,
                      \tilde{P}_{\alpha\beta}^{S}(E_\nu, L)\,,
  \label{eq:rate}
\end{align}
where $E_\nu$ is the neutrino energy, $N_T$ is the number of target particles, $\sigma_{\beta}^{\rm SM}(E_\nu)$ is the SM cross section for $\nu_\beta$ scattering on the target material,\footnote{Here and in the following, we assume that only a single detection process, namely quasi-elastic scattering, is relevant. (See \cref{sec:Detection} for a discussion of this point.) If several detection processes were considered, $\sigma_{\beta}^{\rm SM}$ and $\tilde{P}_{\alpha\beta}^{S}(E_\nu, L)$ should get another superscript `D' labeling the detection process, and the sum should run over $D$ as well.} $\smash{\phi_{\alpha}^{S,{\rm SM}}(E_\nu)}$ is the flux of $\nu_\alpha$ that the detector would see if there were no oscillations, and the sum runs over initial neutrino flavours $\alpha$ and over classes of parent particles, $S = \pi, K, D$. There is some arbitrariness in how we group together different production processes here. We choose to define classes of processes based on the underlying quark-level interaction as processes that at the quark level are identical depend on the same combination of WEFT Wilson coefficients. We absorb geometric acceptance effects into the flux. Expressions for the SM cross sections and fluxes relevant for DUNE are given in \cref{sec:nu-flux,sec:Detection}.

The most important ingredient in \cref{eq:rate} and the main topic of this study is the pseudo-probability $\tilde{P}^{S}_{\alpha\beta}$, which is the oscillation probability at a distance $L$ from the source, in the presence of the SM and new matter effects, multiplied by correction factors that describe the modifications of the flux and cross section as a result of any new interactions beyond the SM. Due to these corrections, $\tilde{P}^{S}_{\alpha\beta}$ does not necessarily satisfy the conditions $\smash{\sum_\alpha \tilde{P}^{S}_{\alpha\beta} = 1}$ and $\smash{\sum_\beta \tilde{P}^{S}_{\alpha\beta} = 1}$, hence the term pseudo-probability. It is given by
{\small
\begin{align}
  \tilde P^S_{\alpha\beta} (E_\nu,L)
    &= \sum_{n,m} e^{-i ({\lambda_n - \lambda_m}) L} \notag\\[-0.1cm]
    &\hspace{-1.5cm}
     \times \bigg[ \tilde U_{\alpha n}^* \tilde U_{\alpha m}
          +\! \sum_{X,j,k} p_{XL,\alpha}^{S,jk} [\epsilon_X^{jk} \tilde U]_{\alpha n}^* \tilde U_{\alpha m}
          +\!\!  \sum_{X,j,k} p_{XL,\alpha}^{S,jk*} \tilde U_{\alpha n}^* [\epsilon_X^{jk} \tilde U]_{\alpha m}
          +\!\!\! \sum_{X,Y,j,k} p_{XY,\alpha}^{S,jk} [\epsilon_X^{jk} \tilde U]_{\alpha n}^*
                                                [\epsilon_Y^{jk} \tilde U]_{\alpha m} \bigg] \nnl
    &\hspace{-1.5cm}
     \times \bigg[ \tilde U_{\beta n} \tilde U_{\beta m}^*
          +\! \sum_{X,r,s} d_{XL,\beta}^{rs}  [\epsilon_X^{rs} \tilde U]_{\beta n} \tilde U_{\beta m}^*
          +\! \sum_{X,r,s} d_{XL,\beta}^{rs*} \tilde U_{\beta n} [\epsilon_X^{rs} \tilde U]^{*}_{\beta m}
          +\! \sum_{X,Y,r,s} d_{XY,\beta}^{rs} [\epsilon_X^{rs} \tilde U]_{\beta n}
                                             [\epsilon_Y^{rs} \tilde U]^{*}_{\beta m} \bigg],
  \label{eq:tildeP}
\end{align}}%
where $\tilde{U}$ is the $3 \times 3$ neutrino mixing matrix in the matter basis, that is, the matrix which diagonalizes the Hamiltonian in \cref{eq:NCHamiltonian}. The matrix $\tilde{U}$ differs from the vacuum mixing matrix $U$, both due to the SM matter effects and due to the new neutral-current interactions ($\epsilon_V^{m,f}$). The oscillation phase in the first line depends on the eigenvalues $\lambda_i$ of the oscillation Hamiltonian, \cref{eq:NCHamiltonian}. In vacuum, $\lambda_i = m_i^2 /  (2 E_\nu)$, with $m_i$ the neutrino mass eigenvalues. The ``production coefficient'' $p_{XY,\alpha}^{S, jk}$ gives the correction of the neutrino flux of flavour $\alpha$ coming from the source $S$ as a result of the $XY$ interaction for quark flavours $jk$; similarly, the ``detection coefficient'' $d_{XY,\beta}^{rs}$ parametrizes the new physics-induced correction to the CC interaction cross section of neutrinos with flavour $\beta$. The terms with $Y=L$ in \cref{eq:tildeP} describe interference between a new interaction and the SM amplitude, proportional to one power of $\epsilon$, while the terms with $X,Y \neq L$ correspond to pure new physics contributions, proportional to two powers of $\epsilon$. In the following sections, we will discuss the production and detection coefficients in more detail.

\subsection{Neutrino Production}
\label{sec:Productions}

The production coefficients give the ratio between the differential neutrino fluxes with and without the presence of new physics. For an experiment with multiple source particles $S$ that decay into neutrinos, they are defined as:
\begin{align}
    p_{XY,\alpha}^{S,jk} &\equiv
      \frac{\int dE_S \sum_i \frac{\phi^{S_i}(E_S)}{E_S}
             \beta^S_i(E_S) \int\! d\Pi_{P'_i} A_{X,\alpha}^{S_i,jk} 
                                      A_{Y,\alpha}^{S_i,jk*}}
           {\int dE_S \sum_{i'}\frac{\phi^{S_{i'}}(E_S)}{E_S}
            \beta^S_{i'}(E_S) \int\! d\Pi_{P'_{i'}} |A_{L,\alpha}^{S_i,jk}|^2} \,.
    \label{eq:pxy}
\end{align}
Here, $S_i$ denotes any decay modes of $S$ that contribute to the transition $u^j \bar d^k \to e_\alpha \nu$, and $\phi^{S_i}(E_S)$ is the normalized energy distribution of parent particles $S_i$ with lab-frame energy $E_S$. For example, for kaon decays, the sum runs over the two-body fully leptonic and three-body semi-leptonic decays of $K^\pm$ as well as the semi-leptonic three-body decay of $K_L$. The factor $\beta^S_i(E_S)$ is the experimental acceptance for each production channel $i$, defined as the ratio of the relative flux fraction of channel $i$ in each experiment, over the true value of the branching ratio. The reduced amplitudes $A_{X,\alpha}^{S_i,jk}$ have been defined in~\cref{eq:Mdecomposition}. Finally, $d\Pi_{P'_i}$ is the phase space measure for channel $i$, excluding the integration over the neutrino energy. (The full phase space measure would be $d\Pi_{P_i}\equiv d\Pi_{P'_i} dE_\nu$.)

\subsubsection{Charged Pion Decays}
\label{sec:PionDecay}

Charged pion decays make the most important contribution to the neutrino flux at DUNE and HyperKamiokande. At the parton level, $\pi^+$ decay corresponds to the process $u \bar d \to e^+_\alpha \nu$, which is sensitive to new physics proportional to the $[\epsilon_X^{ud}]_{\alpha\beta}$ Wilson coefficients. As the leptonic two-body decay is the only relevant $\pi^\pm$ decay channel here, \cref{eq:pxy} simplifies tremendously to \cite{Falkowski:2021bkq}:
\begin{align}
    p_{XY,\alpha}^{\pi,ud} =
        \frac{A_{X,\alpha}^{\pi,ud} A_{Y,\alpha}^{\pi,ud*}}
             {|A_{L,\alpha}^{\pi,ud}|^2} \,.
    \label{eq:pxy-pi-decay}
\end{align}
The relevant amplitudes are
\begin{align}
    A^{\pi,ud}_{L,\alpha} &= -A^{\pi, ud}_{R,\alpha}
        = \frac{V_{ud}}{v^2} (\bar{u}_{\nu} \gamma^\mu P_L v_{e_\alpha})
          \bra{0} \bar{d} \gamma_\mu \gamma_5 u \ket{\pi^+(p_\pi)} \,,  \nnl
    A^{\pi,ud}_{P,\alpha} &=
         -\frac{V_{ud}}{v^2} (\bar{u}_{\nu} P_R v_{e_\alpha})
          \bra{0} \bar{d} \gamma_5 u \ket{\pi^+(p_\pi)} \,,  \nnl 
    A^{\pi,ud}_{S,\alpha} &= A^{\pi,ud}_{T,\alpha} = 0 , 
    \label{eq:pi-decay-amplitudes}
\end{align}
where $p_\pi$ is the pion 4-momentum and $v_{e_\alpha}$, $\bar{u}_{\nu}$ are the Dirac spinor wave functions of the charged lepton and the neutrino, respectively. The hadronic matrix elements can be parametrized as
\begin{align}
    \bra{0} \bar{d} \gamma^\mu \gamma_5 u \ket{\pi^+(p_\pi)}
        = i p_\pi^\mu f_{\pi} \,,
    \qquad 
    \bra{0} \bar{d} \gamma_5 u \ket{\pi^+(p_\pi)}
        = -i \frac{m_{\pi}^2}{m_u + m_d}  f_{\pi} \,,
  \label{eq:pi-decay-matrix-elements}
\end{align}
where $f_\pi = 130.2(0.8)$~MeV~\cite{Aoki:2019cca} is the pion decay constant and $m_\pi$, $m_u$, $m_d$ are the charged pion, up quark, and down quark masses, respectively. Plugging \cref{eq:pi-decay-amplitudes,eq:pi-decay-matrix-elements} into \cref{eq:pxy-pi-decay}, we obtain for the production coefficients~\cite{Falkowski:2019kfn,Falkowski:2021bkq}
\begin{align}
    p_{LL,\alpha}^{\pi,ud} &=  p_{RR,\alpha}^{\pi,ud} = -p_{LR,\alpha}^{\pi,ud} = 1
                                                            \,,\notag\\[0.2cm]
    p_{PL,\alpha}^{\pi,ud} &= -p_{PR,\alpha}^{\pi,ud}
                       = -\frac{m_\pi^2}{m_{e_\alpha} (m_u+m_d)} 
                       \simeq -27\ (-5600)
                              &\quad\text{for $\alpha=\mu\ (e)$} \,,
                                                         \label{eq:pioncoefficients} \\
    p_{PP,\alpha}^{\pi,ud} &= \frac{m_\pi^4}{m_{e_\alpha}^2 (m_u+m_d)^2}
                       \simeq 730 \ (3.1 \times 10^7)
                              &\quad\text{for $\alpha=\mu\ (e)$} \,. \notag
\end{align}
For the numerical evaluation, we have used $m_u+m_d \simeq \SI{6.82(9)}{MeV}$~\cite{Aoki:2019cca}. Note that $p_{PP,\alpha}^{\pi,ud} =\big(p_{PL,\alpha}^{\pi,ud}\big)^2$. We see that for both $PL$ and $PP$ interactions, the production coefficients enjoy substantial chiral enhancement, which can be understood as the absence of the chiral suppression present in the SM. This enhancement implies that long-baseline neutrino experiments employing neutrino beams from pion decay are particularly sensitive to $[\epsilon_P^{ud}]_{\alpha\beta}$ for $\alpha=e,\mu$. The dependence of the pseudoscalar production coefficients on $m_\pi/m_{e_\alpha}$ also implies that the enhancement is much larger for $\nu_e$ than for $\nu_\mu$. Therefore, even though in the SM the fluxes of electron (anti)neutrinos are at least two orders of magnitude smaller than those of muon (anti)neutrinos at DUNE and HyperKamiokande (see e.g. \cref{fig:FluxND} below), they are still of great importance to the search for pseudoscalar new physics.

\begin{figure}
  \centering
  \includegraphics[width=1\textwidth]{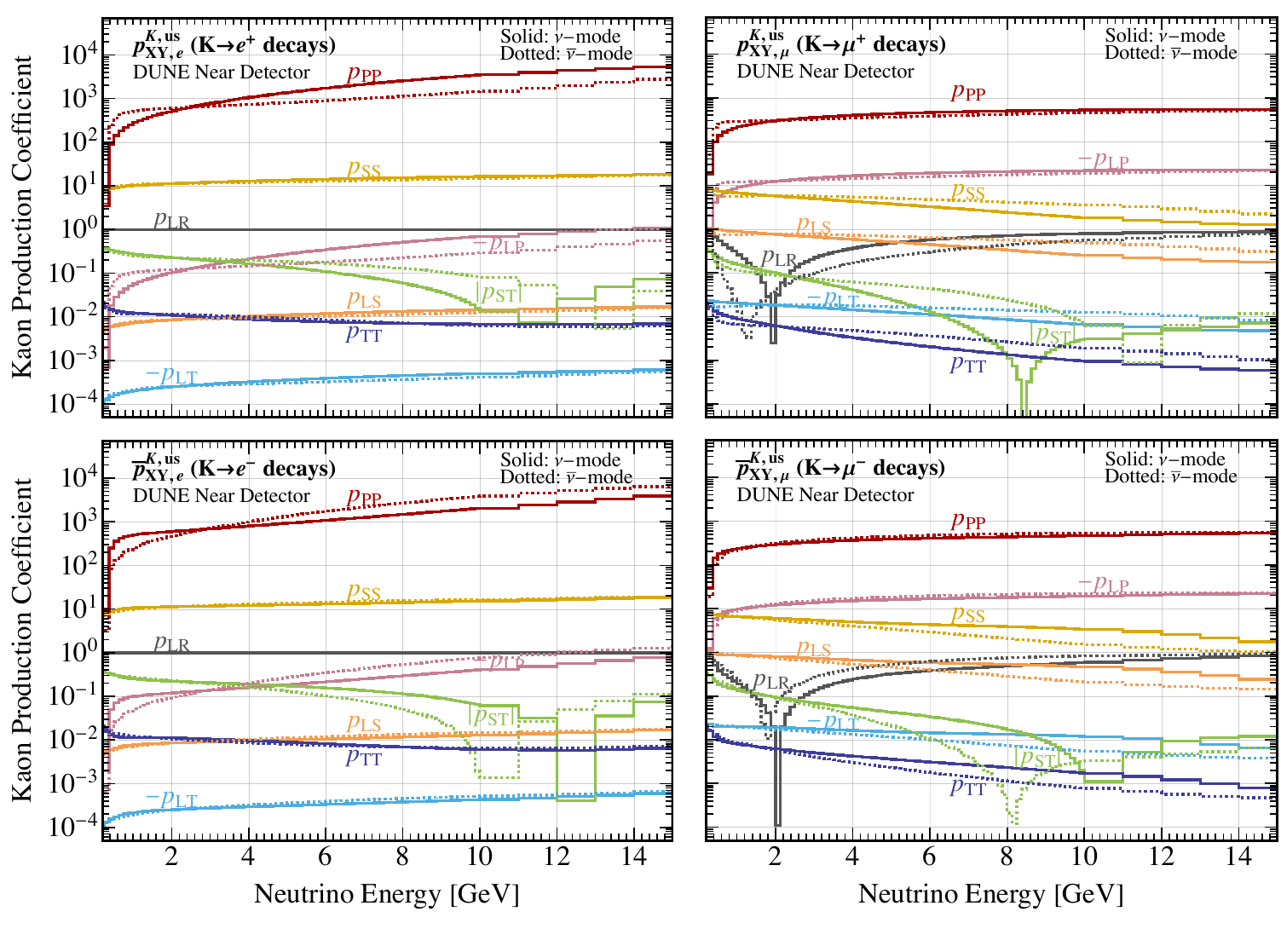}
  \caption{Production Coefficients for kaon decay. Left: decay to electrons. Right: decay to muons. Bottom panels show the decays to anti-neutrinos.}
  \label{fig:KaonProdND}
\end{figure}

\subsubsection{Kaon Decays}
\label{sec:KaonDecay}

Kaon decays are the second most important neutrino source at DUNE. Mediated at the parton level by $u  \bar s \to e^+_\alpha \nu$, they are sensitive to the $[\epsilon_X^{us}]_{\alpha\beta}$ Wilson coefficients. Several decay channels contribute in the sum over $i$ in \cref{eq:pxy}, namely charged kaons through their two-body ($K^+\to e_\alpha^+ \nu$) and three-body ($K^+\to \pi^0 e_\alpha^+ \nu$) decays, as well as $K_L$ through its semi-leptonic decay mode $K_{L} \to \pi^- e_\alpha^+ \nu$. This makes the calculation of the production coefficients for kaon decays more involved than for pion decay. The amplitudes $A_{X,\alpha}^{S_i,us}$ for two-body $K^\pm$ decay are similar to the ones for pion decay given in \cref{eq:pi-decay-amplitudes}, while the expressions for the three-body decays are more complex; they can be found for instance in Appendix~A of ref.~\cite{Falkowski:2021bkq}. 

We plot the production coefficients $p_{XY,\alpha}^{K,us}$ for kaon decays as a function of the neutrino energy in \cref{fig:KaonProdND}, where we have followed Section~2.1.2 of ref.~\cite{Falkowski:2021bkq} for the numerical evaluation. We observe a spread of seven orders of magnitude depending on the Lorentz structure of the new interaction. The strongest enhancement ($> \num{1000}$ compared to the SM) is found for pseudoscalar interactions, which are chirally enhanced. Note that the coefficients $p_{XY,\alpha}^{K,us}$ are different for the near and far detectors because the fluxes  $\phi_{S_i}(E_S)$ and the efficiency/acceptance factors $\beta^S_i(E_S)$ in \cref{eq:pxy} are different for the two detectors. For the same reason, the production coefficients also depend on the beam configuration (forward or reverse horn current), the target and horn geometry, the primary proton energy, etc.

\subsubsection{Charm Decays}
\label{sec:CharmDecay}

Neutrinos from charm decay are very much subdominant in accelerator-based long-baseline neutrino experiments due to the relatively low primary proton energies. In HyperKamiokande, this neutrino flux is completely negligible, whlie in DUNE, a very small flux of tau neutrinos is expected, the main source of which are fully leptonic $D^\pm_s$ decays. At the parton level these decays are mediated by $c  \bar s \to e^+_\alpha \nu$, and hence they are sensitive to the $[\epsilon_X^{cs}]_{\alpha \beta}$ Wilson coefficients. The corresponding production amplitudes can be computed in the same way as for pion decay, see \cref{eq:pi-decay-amplitudes}. We find the following non-zero production coefficients:
\begin{align}
    p_{LL,\alpha}^{{D_s},cs} &=  p_{RR,\alpha}^{{D_s},cs} = -p_{LR,\alpha}^{{D_s},cs} = 1
                                                            \,,\notag\\[0.2cm]
    p_{PL,\alpha}^{{D_s},cs} &= -p_{PR,\alpha}^{{D_s},cs}
                       = -\frac{m_{D_s}^2}{m_{e_\alpha} (m_c+m_s)} 
                       = -1.59,\,-26.71,\,-5.52 \times 10^3 
                              &\quad\text{for $\alpha=\tau,\,\mu,\,e$} \,,
                                                         \label{eq:Dscoefficients} \\
    p_{PP,\alpha}^{{D_s},cs} &= \frac{m_{D_s}^4}{m_{e_\alpha}^2 (m_c+m_s)^2}
                       =  2.52,\,713.37,\,3.05 \times 10^7 
                              &\quad\text{for $\alpha=\tau,\,\mu,\,e$} \,. \notag
\end{align}
For the masses we have used $m_{D_s} = \SI{1.968(34)}{GeV}$, $m_c = \SI{1.280(13)}{GeV}$ and $m_s = \SI{92.9(7)}{MeV}$ here~\cite{ParticleDataGroup:2024cfk}.

An additional, though even smaller, contribution to the neutrino flux comes from leptonic $D^\pm$ decays, giving DUNE some sensitivity to $[\epsilon_X^{cd}]_{\alpha\beta}$. The corresponding production coefficients are: 
\begin{align}
    p_{LL,\alpha}^{{D^+},cd} &=  p_{RR,\alpha}^{{D^+},cd} = -p_{LR,\alpha}^{{D^+},cd} = 1
                                                            \,,\notag\\[0.2cm]
    p_{PL,\alpha}^{{D^+},cd} &= -p_{PR,\alpha}^{{D^+},cd}
                       = -\frac{m_{D^+}^2}{m_{e_\alpha} (m_c+m_d)} 
                       = -1.53,\,-25.75,\,-5.32 \times 10^3 
                              &\quad\text{for $\alpha=\tau,\,\mu,\,e$} \,,
                                                         \label{eq:Dscoefficients} \\
    p_{PP,\alpha}^{{D^+},cd} &= \frac{m_{D^+}^4}{m_{e_\alpha}^2 (m_c+m_d)^2}
                       =  2.34,\,663.17,\,2.84 \times 10^7 
                              &\quad\text{for $\alpha=\tau,\,\mu,\,e$} \,, \notag
\end{align}
where we have used $m_d = \SI{4.71}{MeV}$, and $m_{D^+} = \SI{1.869(65)}{GeV}$~\cite{ParticleDataGroup:2024cfk}.

\subsection{Neutrino Detection}
\label{sec:Detection}

In terms of neutrino detection processes, long-baseline neutrino oscillation experiments operate in the most interesting, but also most challenging, energy range around $E_\nu \sim \SI{1}{GeV}$. In this energy range quasi-elastic scattering (QES), resonance production (RES), and deep-inelastic scattering (DIS) are all relevant. Moreover, multi-nucleon effects and the structure of the target nucleus are relevant. Here, we focus on QES, which is the most important detection process especially in HyperKamiokande, but also in DUNE. We have previously computed quasi-elastic neutrino scattering cross sections in the presence of new physics in ref.~\cite{Kopp:2024yvh}. (A corresponding study for the DIS regime can be found in \cite{Falkowski:2021bkq}.) Here, we summarize the results from \cite{Kopp:2024yvh} and use them to compute the detection coefficients $d_{XY,\beta}^{rs}$.

The reduced matrix elements for charged current quasi-elastic (CCQE) interactions of the form $\nu_\beta + n \to e_\alpha^- + p^+$ are
\begin{align}
    \begin{split}
        A_{L,\alpha}
            &= -\frac{2V_{ud}}{v^2} \Big[
                    \bar{u}_{e_\alpha}(p_{e_\alpha}) \gamma^\mu P_L u_\nu(p_\nu)
                    \Big]
                    \braket{p(p_p)|\bar{q}_u \gamma_\mu P_L q_d|n(p_n)} \,, \\
        A_{R,\alpha}
            &= -\frac{2V_{ud}}{v^2} \Big[
                    \bar{u}_{e_\alpha}(p_{e_\alpha}) \gamma^\mu P_L u_\nu(p_\nu)
                    \Big]
                    \braket{p(p_p)|{\bar q}_u \gamma_\mu P_R q_d|n(p_n)} \,, \\
        A_{S,\alpha}
            &= -\frac{V_{ud}}{v^2} \Big[
                    \bar{u}_{e_\alpha}(p_{e_\alpha}) P_L u_\nu(p_\nu) \Big]
                    \braket{p(p_p)|{\bar q}_u  q_d|n(p_n)} \,, \\ 
        A_{P,\alpha}
            &= \frac{V_{ud}}{v^2} \Big[
                   \bar{u}_{e_\alpha}(p_{e_\alpha}) P_L u_{\nu}(p_\nu) \Big]
                   \braket{p(p_p)|{\bar q}_u \gamma_5 q_d|n(p_n)}
        \,,  \\
        A_{T,\alpha}
            &= -\frac{V_{ud}}{2v^2} \Big[
                    \bar{u}_{e_\alpha}(p_{e_\alpha}) \sigma^{\mu\nu} P_L u_\nu(p_\nu)
                    \Big]
                    \braket{p(p_p)|{\bar q}_u \sigma_{\mu\nu} q_d|n(p_n)} \,.
    \end{split}
    \label{eq:A-detection}
\end{align}
Here $u_{e_\alpha}$ and $u_\nu$ denote the spinor wave functions for charged leptons and neutrinos, respectively, $q_{u,d}$ are the quark field operators, and $\ket{p(p_p)}$, $\ket{n(p_n)}$ are nucleon states. The reduced matrix elements for anti-neutrino interactions, $\bar\nu_\beta + p \to e_\alpha^+ + n$, can be obtained by replacing $u_{e_\alpha} \to v_\nu$ and $u_\nu \to v_{e_\alpha}$, as well as taking the complex conjugate of the hadronic currents and the CKM element $V_{ud}$ in \cref{eq:A-detection}.

The hadronic currents appearing in \cref{eq:A-detection} are discussed in detail in sec.~2.2 of ref.~\cite{Kopp:2024yvh}. There, they are first expressed in terms of Lorentz-invariant form factors (see eqs.~(2.10)--(2.14) of ref.~\cite{Kopp:2024yvh}), and subsequently the form factors are written as functions of the momentum transfer $Q^2$. Much of the uncertainty in neutrino--nucleon cross sections comes from the axial form factor of the nucleon. Here, we use a lattice QCD prediction for this form prediction \cite{RQCD:2019jai}.

The total CCQE neutrino--nucleon scattering cross section for neutrinos of flavour $\beta$ can be written as
\begin{align}
    \sigma_\beta = \sigma_\beta^\text{SM}
                 + \sum_X \hat\sigma^\text{int}_{LX,\beta}
                 + \sum_{X,Y} \hat\sigma^\text{NP}_{XY,\beta} \,,
\end{align}
where $\sigma_\beta^\text{SM}$ is the SM cross section, $\hat\sigma^\text{int}_{LX,\beta}$ contains interference terms between SM and new interactions, and $\hat\sigma^\text{NP}_{XY,\beta}$ are the pure new physics contributions. The hat on the latter two terms indicates that these terms can be negative.

The SM cross section is
\begin{align}
    \sigma^{\rm{SM}}_\beta= \hat\sigma_{LL,\beta} \equiv \frac{1}{2 \kappa} \sum_\text{spins}
        \int\!d\Pi_D \, |A_{L,\beta}|^2 \,. 
    \label{eq:sigma-SM}
\end{align}
where $\kappa \equiv 4 E_\nu m_{n,p}$, with $m_{n,p}$ the nucleon masses, and the phase space measure is $d\Pi_D \equiv \big[ \prod_j d^3p_j/(2\pi)^3 \big] \times (2\pi)^4 \delta^{(4)}(p_i - p_f)$. In analogy, the SM--new physics interference term is
\begin{align}
    \hat\sigma^\text{int}_{LX,\beta} &\equiv \frac{1}{2 \kappa} \sum_\text{spins}
        \int\!d\Pi_D (A_{L,\beta} A^*_{X,\beta}) \,
    &\text{($X = L,R,S,P,T$)}
    \label{eq:IntDet}
\end{align}
and the pure new physics term is
\begin{align}
    \hat\sigma^\text{NP}_{XY,\beta} &\equiv \frac{1}{2 \kappa} \sum_\text{spins}
        \int\!d\Pi_D (A_{X,\beta} A^*_{Y,\beta}) \,. 
    &\qquad\qquad \text{($X, Y \neq L$)}
    \label{eq:QuadDet}
\end{align}
Analogous formulas hold for the anti-neutrino cross sections $\bar\sigma^{\text{SM}}_\beta$, $\hat{\bar\sigma}^\text{int}_{LX,\beta}$, and $\hat{\bar\sigma}^\text{NP}_{XY,\beta}$.

The discussion so far has revolved around neutrino--nucleon cross sections, whereas what we actually need are neutrino--\emph{nucleus} cross sections. To obtain these, we need to fold the matrix element products $A_{X,\beta} A^*_{Y,\beta}$ in \cref{eq:sigma-SM,eq:IntDet,eq:QuadDet} with a nuclear spectral function, as explained in more detail in ref.~\cite{Kopp:2024yvh}. Here, we use the spectral function of oxygen from ref.~\cite{Benhar:1994hw, Benhar:1989aw} as a proxy for the spectral function of argon, and we then normalize the final cross sections to the atomic number of argon. In the following, we will implicitly assume the folding has been done, and we will take $\sigma_\beta^\text{SM}$ and $\hat\sigma_{XY,\beta}$ to mean the neutrino--\emph{nucleus} cross sections.

We can now define the detection coefficients for CCQE neutrino--nucleus interactions, $d_{XY,\beta}^{jk}$, which give the ratios of the (partial) cross sections that include new physics to the SM cross section. Since for CCQE process only scattering on up and down quarks matters, we can drop the quark flavour indices and simply write
\begin{align}
    d_{XY,\beta} \equiv d_{XY,\beta}^{\rm{CCQE},ud} &= \frac{\hat\sigma_{XY,\beta}}{\sigma_\beta^{\rm{SM}}}\,,
    &\qquad\text{($X, Y = L, R, S, P, T$)}
\end{align}
where $\hat\sigma_{XY,\beta}$ is either the interference term given in \cref{eq:IntDet}, or the quadratic term in \cref{eq:QuadDet}. The corresponding detection coefficients for anti-neutrinos will be denoted by ${\bar{d}}_{XY,\beta}^{\,\rm{CCQE},ud}$.

\begin{figure}
    \centering
    \includegraphics[width=0.5\textwidth]{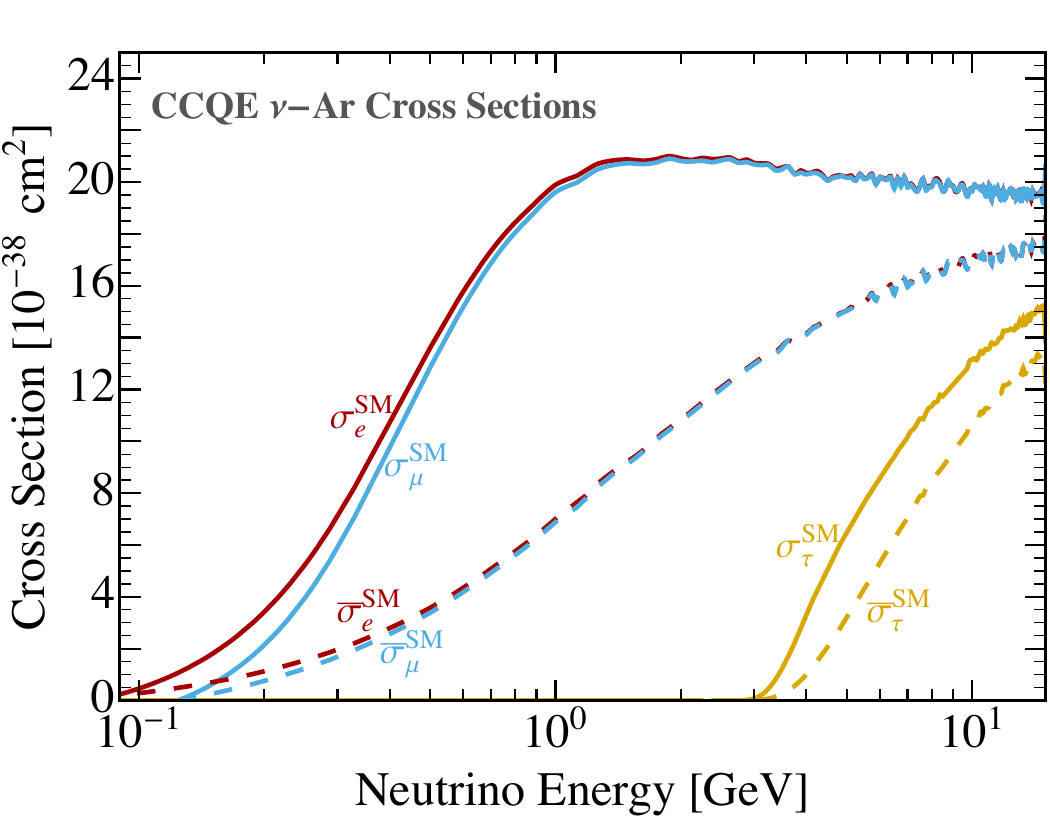}
    \caption{SM cross sections for neutrino (solid) and anti-neutrino (dashed) charged-current quasi-elastic (CCQE) scattering on argon. Red, blue, and orange curves correspond to the $\nu_e$, $\nu_\mu$ and $\nu_\tau$ flavours, respectively. Figure based on ref.~\cite{Kopp:2024yvh}.}
    \label{fig:xsections-SM}
\end{figure}

\begin{figure}
  \centering
  \includegraphics[width=1\textwidth]{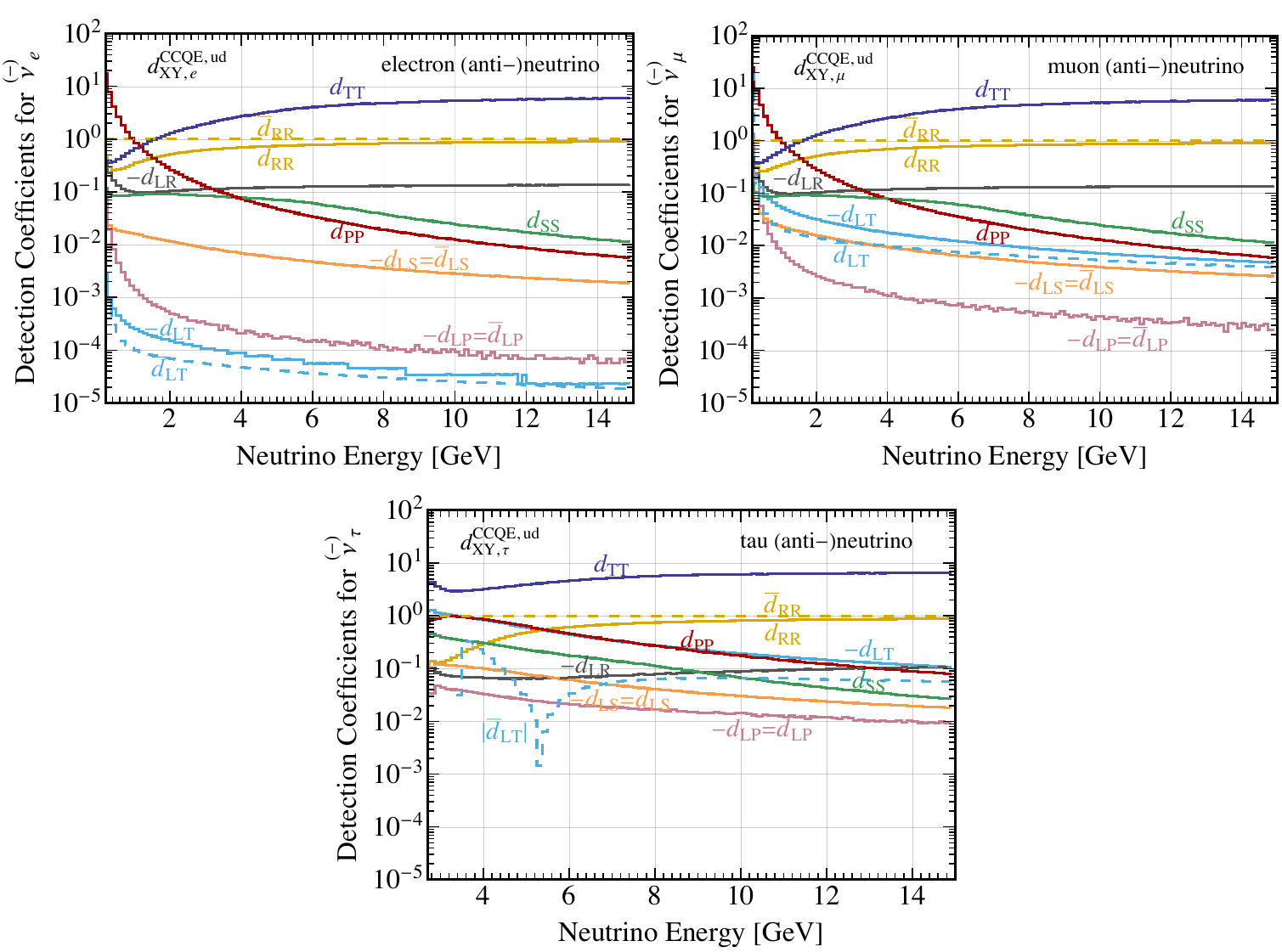}
  \caption{Detection coefficients for CCQE scattering. Upper left: electron neutrinos; upper right: muon neutrinos; bottom: tau neutrinos. Solid curves are for neutrinos, dashed curves for anti-neutrinos. Figure based on ref.~\cite{Kopp:2024yvh}}
  \label{fig:detCoeff}
\end{figure}

We collect numerical results in \cref{fig:xsections-SM,fig:detCoeff}. In \cref{fig:xsections-SM}, we plot the SM CCQE cross sections for all neutrino and anti-neutrino flavours as a function of neutrino energy.  The differences between the cross sections for $\nu_e$ (red), $\nu_\mu$ (blue) and $\nu_\tau$ are entirely due to kinematics (in particular the different charged lepton masses). Differences between neutrinos and anti-neutrinos reflect the fact that neutrinos scatter on neutrons, while anti-neutrinos scatter on protons.  The CCQE detection coefficients are shown in \cref{fig:detCoeff}. Our main findings are:
\begin{itemize}
  \item The interference term $\hat\sigma^\text{int}_{LX,\beta}$ in \cref{eq:IntDet}, and therefore the detection coefficient $d_{LX,\beta}$, can be negative. In fact, for ${\bar{d}}_{LT,\tau}$ there is a sign change at about $5$~GeV. 
  
  \item For $LR$ interaction we have $d_{LR,\beta}={\bar{d}}_{LR,\beta}$, while for $LS$ and $LP$ interactions the relation $d_{LS/LP,\beta} = -{\bar{d}}_{LS/LP,\beta}$ holds. Moreover, $d_{RR,\beta} = {\bar{d}}_{LL,\beta}$ and $d_{SS/PP/TT,\beta} = {\bar{d}}_{SS/PP/TT,\beta}$.
  
  \item The largest detection coefficient is the one for pure tensor (TT) interactions. The enhancement with respect to the SM is due to the fact that the tensor form factors entering in $\hat\sigma^\text{NP}_{TT,\beta}$ are larger than the SM form factors at large $Q^2$, as is evident from fig.~1 in ref.~\cite{Kopp:2024yvh}. This translates to better sensitivity to $\epsilon_T^{ud}$ on the detection side. 
  
  \item We also note an enhancement of $\mathcal{O}(10)$ for the PP interaction at very low energies. This is due to the large pseudoscalaor form factor of the nucleon, which is $\sim 350$ at $Q^2=0$. As this form factor scales with $(Q^2)^{-2}$, it drops quickly at larger momentum transfer though. This means that if there is new physics that mediates pseudoscalar interactions, we should expect an excess of events at very low energies. 
  
  \item For the RR interaction the detection coefficient is $\mathcal{O}(1)$, and while there is no enhancement, we can still expect good sensitivity to $\epsilon_R^{ud}$. 
  
\end{itemize}
A comment is in order regarding the detection coefficients involving a scalar current. These coefficients depend on the induced scalar form factor $\tilde G_S(Q^2)$ of the nucleon, which is fraught with large uncertainties \cite{Kopp:2024yvh}. While $\tilde G_S(Q^2)$ is expected to be small, it here appears with a large prefactor, making it potentially relevant. In the following, we estimate $\tilde G_S(Q^2)$ using the constituent quark model as discussed in ref.~\cite{Kopp:2024yvh}.

\subsection{Indirect New Physics Effects}
\label{eq:indirect-np}

New physics can affect neutrino experiments not only ``directly'' by impacting neutrino productions, oscillation probabilities, and detection cross sections, but also ``indirectly'' by biasing external measurements that are used to determine fundamental constants whose values in turn enter the calculation of neutrino fluxes, oscillation probabilities, and detection cross sections. In \cref{eq:rate} we have separated the expression for the differential event rate into the pseudo-probability $\tilde{P}_{\alpha\beta}^S$ (which encapsulates all direct new physics effects coming from either WEFT or SMEFT Wilson coefficients) as well as the SM flux and cross section. In doing so, we implicitly assumed that the SM predictions for $\phi_{\alpha}^{S, \mathrm{SM}}$ and $\sigma_\beta^{\mathrm{SM}}$ can be reliably computed within the SM without any contamination from new physics. However, this is generally not the case because fluxes and cross sections depend on fundamental parameters such as the Fermi constant $G_F$ and the CKM element $V_{ud}$, which cannot be determined from first principles but must be extracted from experiments. However, the corresponding experiments (muon decays in the case of $G_F$, superallowed nuclear beta decays for $V_{ud}$) may be affected by new physics, leading to biased results.\footnote{$G_F$ and $V_{ud}$ enter also the calculation of $\tilde{P}_{\alpha\beta}^S$ through the production and detection coefficients. There, however, any new physics-induced corrections to their values are higher-order in the Wilson coefficients.}

Starting with $\sigma_\beta^{\mathrm{SM}}$, we note that direct new physics effects modify the cross section in an energy-dependent manner unless the Lorentz structure matches exactly that of the SM (see \cref{fig:detCoeff} above). Indirect new physics, on the other hand, only leads to an overall rescaling which can be absorbed into the overall normalization uncertainty and is therefore largely undetectable. Hence, we can safely neglect indirect new physics in detection.

Turning to $\phi_\alpha^{S, \mathrm{SM}}$, we first note that analyses using far detector data will base their flux predictions on measurements at the near detector. As new physics in neutrino production affects the near and far detectors in the same way (modulo effects of the different acceptances and efficiencies of the two detectors), they largely cancel out. Hence, the following discussion is relevant only to analyses focused on near detector data alone. This is justified as we find that, whenever new physics affects neutrino production, the corresponding effects are most easily discernible in the near detector thanks to its much larger statistical sample. 

Theoretical neutrino flux predictions are typically based on measured decay widths of the relevant parent particles. The auxiliary measurements that determine these decay widths may be affected by new physics, but often they are affected in the same way as neutrino production in long-baseline experiments. In this case, the new physics effects cancel, leading to reduced sensitivity to the new physics. This can be avoided by computing the decay width in a more ab initio way, relying on experimental data only for the determination of the CKM matrix elements while taking hadronic form factors from lattice QCD calculations. In this work, we only need the CKM elements $V_{ud}$, $V_{us}$, $V_{cd}$, and $V_{cs}$, which mediate the decays of $\pi$, $K$, $D$, and $D_s$ mesons, respectively. To determine them, we closely follow the procedure of ref.~\cite{Descotes-Genon:2018foz}.

Using the Wolfenstein parametrization of the CKM matrix \cite{Wolfenstein:1983yz} up to $\mathcal{O}(\lambda^4)$, all CKM elements of interest to us are determined by the single parameter $\lambda$:
\begin{align}
    \begin{split}
        V_{ud} &=  V_{cs} = 1 - \frac{1}{2} \lambda^2 -\frac{1}{8} \lambda^4+ \mathcal{O}(\lambda^6)\,, \\
        V_{us} &= -V_{cd} = \lambda + \mathcal{O}(\lambda^5)\,.
    \end{split}
    \label{eq:ckm-elements-approx}
\end{align}
We need to carefully distinguish between quantities extracted from measurements that are possibly affected by new physics (e.g.\ $V_{jk}^\text{exp}$, $\lambda^\text{exp}$) and the true quantities appearing in the model Lagrangian (e.g.\ $V_{jk}$, $\lambda$). We write
\begin{align}
    \begin{split}
        V_{jk}  &\equiv V_{jk}^\text{exp} - \delta V_{jk} \,, \\
        \lambda &\equiv \lambda^\text{exp} - \delta \lambda \,,
    \end{split}
    \label{eq:separation}
\end{align}
where $\delta V_{jk}$ and $\delta\lambda$ are the correction terms that include the new physics effects. Combining \cref{eq:ckm-elements-approx,eq:separation}, we can write
\begin{align}
    \begin{split}
        \delta V_{ud} &= \delta V_{cs} = 
            - \delta\lambda \, \big[ 
                  \lambda^\text{exp} + \tfrac{1}{2}(\lambda^\text{exp})^3 \big] \,,\\
        \delta V_{us} &= -\delta V_{cd} = \delta \lambda \,.
    \end{split}
\end{align}
Experimentally, $\lambda^\text{exp}$ can be extracted by first using the ratio $\Gamma(K^- \to \mu^- \bar\nu_\mu) / \Gamma(\pi^- \to \mu^- \bar\nu_\mu)$ to determine $|V^{\text{exp}}_{us}/V^{\text{exp}}_{ud}| = 0.23131 \pm 0.00050$. Using \cref{eq:ckm-elements-approx}, this translates into  $\lambda^\text{exp} = 0.22537 \pm 0.00046$~\cite{Descotes-Genon:2018foz}. The indirect new physics corrections to $\lambda$ can be written as \cite{Descotes-Genon:2018foz}
\begin{align}
    \delta\lambda &= \frac{1}{2} \big[\lambda^\text{exp}
        - (\lambda^\text{exp})^3 \big] \Delta_{K/\pi} \\
     &= 0.1070(2) \times \Delta_{K/\pi} \nonumber\,,
\end{align}
where
\begin{multline}
    \Delta_{K/\pi} = 2 \, \Re \Big(
        \big[\epsilon_L^{us} - \epsilon_R^{us} \big]_{\mu\mu}-\big[\epsilon_L^{ud} - \epsilon_R^{ud} \big]_{\mu \mu} 
    \Big) \\
  - \frac{2}{m_\mu} \bigg(
        \frac{m_{K^\pm}^2 \Re\big([\epsilon_P^{u s}]_{\mu \mu}\big)}{m_u+m_s}
      - \frac{m_{\pi^\pm}^2 \Re\big([\epsilon_P^{ud}]_{\mu \mu}\big)}{m_u+m_d}
    \bigg)
  + \mathcal{O}(\Lambda^{-4})\,.
\end{multline}

Returning to the neutrino flux prediction, as the fluxes are computed using the experimentally measured $V_{jk}^{\text{exp}}$, they need to be corrected by replacing
\begin{align}
    V_{jk}^\text{exp} \to V_{jk} = V_{jk}^\text{exp} - \delta V_{jk} ,
\end{align}
which is equivalent to multiplying the flux component coming from couplings to quark flavours $(jk)$ by a factor $(1 - 2\delta V_{jk} / V_{jk}^\text{exp})$. (The factor is due to the quadratic dependence of the fluxes on the CKM elements.)

The possible presence of indirect new physics effects implies that, when we refer to a particular bound as a DUNE or HyperKamiokande constraint, it should really be interpreted as a combined bound from DUNE/HyperKamiokande and measurements of $\Gamma(K^- \to \mu^- \bar\nu_\mu) / \Gamma(\pi^- \to \mu^- \bar\nu_\mu)$.

\section{Experimental Setup and Simulation Details}
\label{sec:experimental-setup}

We are now ready to compute the sensitivity of DUNE to new physics in the context of WEFT and SMEFT. (The reasons for focusing specifically on DUNE here have been outlined in footnote~\ref{fn:why-dune} on \cpageref{fn:why-dune} above.) Here, we outline how we simulate the DUNE experiment.

\subsection{Neutrino Flux}
\label{sec:nu-flux}

DUNE will feature one of the most intense neutrino beams in the world, produced at the Long-Baseline Neutrino Facility (LBNF), where a high-intensity beam of \SI{120}{GeV} protons will be dumped onto a target to generate pions and kaons, which are then focused in the forward direction by a set of magnetic horns. The decays of these mesons yield a neutrino beam with a peak energy between \SI{2}{GeV} and \SI{3}{GeV}.

In our analysis, we assume the nominal beamline configuration detailed in the DUNE Technical Design Report (TDR)~\cite{DUNE:2020ypp, DUNE:2020jqi, DUNE:2021cuw}, and we assume 3.5~years of neutrino mode (forward horn current or FHC) and 3.5~years of antineutrino mode (reverse horn current or RHC) operation. As the new physics-induced modifications to the neutrino flux depend on the neutrino's parent meson, we separate the fluxes from DUNE's Geant4 \cite{Agostinelli:2002hh, Allison:2006ve, Allison:2016lfl} based simulation of the LBNF beamline (G4LBNF) \cite{DUNE:2020lwj} according to the parent meson species. To increase statistics and improve the smoothness of the flux prediction, we re-decay each parent meson 1000 times. We retain only neutrinos whose trajectories cross the ND or the FD.

The publicly available DUNE flux predictions do not include neutrinos from charm decay, given that their contribution to conventional oscillation analyses is negligible. For new physics searches, however, neutrinos from $D$ meson decays can be quite important. First, because new physics might couple predominantly to second and third generation quarks, and second because $D_s$ mesons are the only source of $\nu_\tau$ in LBNF, potentially making them a unique tool to search for new effects involving $\tau$ neutrinos. Therefore, we have generated charm meson fluxes ourselves using Pythia (v8.3.07)~\cite{Bierlich:2022pfr} with the SoftQCD flag on to simulate the interactions of \SI{120}{GeV} protons impinging on a carbon target. This approach is valid as charm mesons decay promptly, meaning the resulting neutrino flux is not influenced by the magnetic horns (which are designed to focus pions and kaons). Our simulations are conservative in the sense that they include only the primary proton--proton interactions, while neglecting secondary interactions in the target and the surrounding material.  While such secondary interactions can significantly enhance the production of light mesons such as $\pi^0$ and $\pi^\pm$, their relevance for heavy flavour production is much more limited. We track the trajectories of the neutrinos and keep only those that intersect the ND or the FD.

\begin{figure}
  \centering
  \includegraphics[width=1\textwidth]{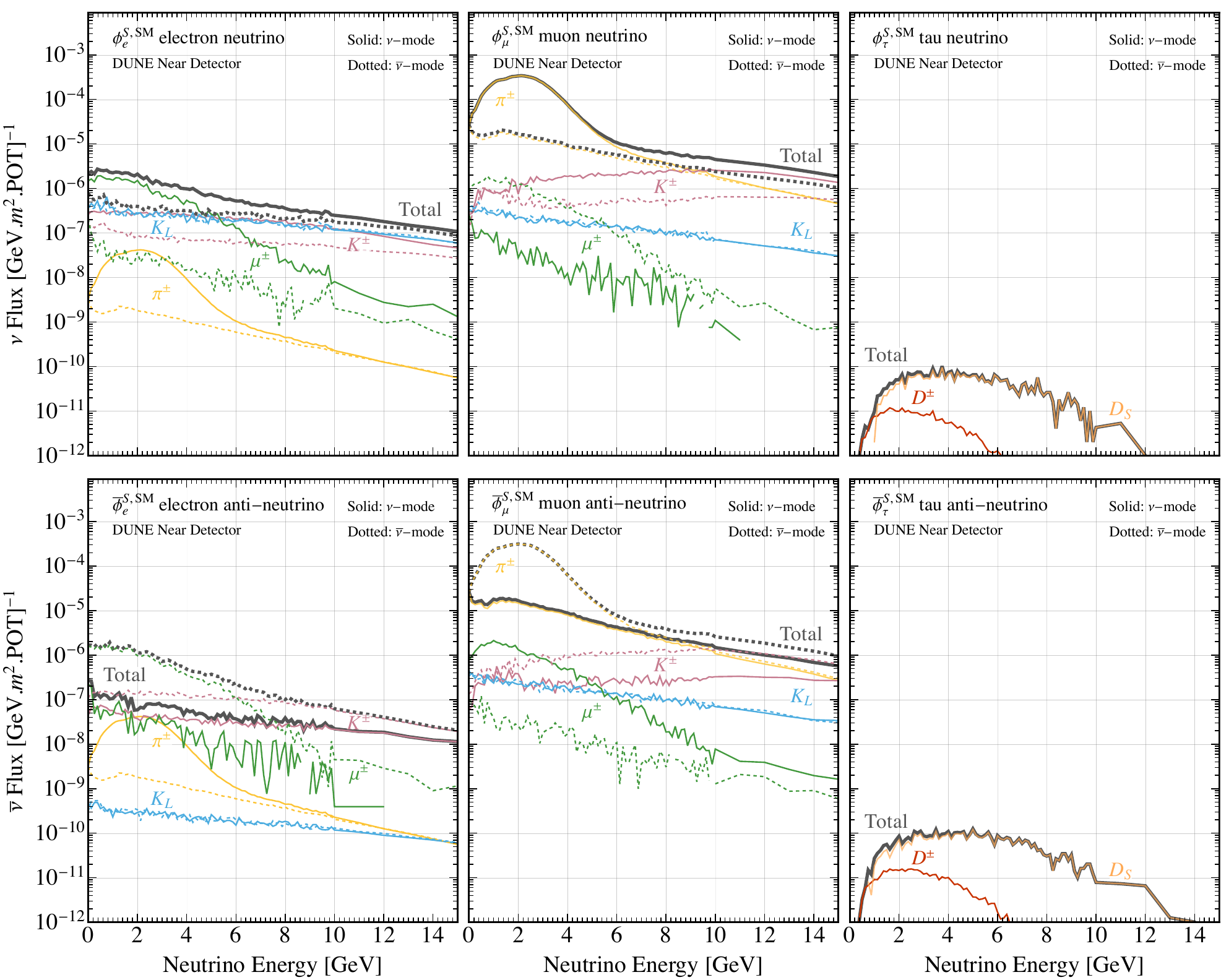}
  \caption{Electron, muon, and $\tau$ neutrino (upper panels) and antineutrino (lower panels) fluxes in the SM at the DUNE ND-LAr near detector. We show both the total fluxes (black) and the individual contributions from each parent species ($\pi^{\pm}$, $K^{\pm}$, $\mu^{\pm}$, $D^{\pm}$, and $D_s$). Solid lines correspond to forward horn current (the neutrino-dominated beam, "$\nu$-mode"), while dotted lines are for reverse horn current (anti-neutrino-dominated beam, "$\bar\nu$-mode").}
  \label{fig:FluxND}
\end{figure}

In \cref{fig:FluxND} we show the expected on-axis neutrino and antineutrino fluxes at the DUNE near detector, separated by parent particle and shown for both forward horn current (FHC, neutrino-dominated beam) and reverse horn current (RHC, anti-neutrino-dominated beam) modes. The fluxes at the far detector are very similar after accounting for the geometric rescaling with $1/L^2$.

\subsection{Signal and Background Modelling}
\label{sec:sig-and-bg}

The neutrino fluxes from \cref{fig:FluxND}, separated by parent particle, form one of the central inputs to end-to-end simulations of DUNE, which we carry out using the GLoBES library~\cite{Huber:2004ka, Huber:2007ji}, supplemented with ancillary files provided by the DUNE collaboration~\cite{DUNE:2021cuw}. These ancillary files notably provide a sophisticated model of the detector response including energy smearing, energy-dependent efficiencies, etc.

The files DUNE provides describe only the far detector, while we also need a simulation of the near detector to determine the experiment's sensitivity to new charged-current interactions, which is dominated by the near detector. We assume here that the response of the near detector is identical to that of the far detector; we only change the baseline from \SI{1284.9}{km} to \SI{574}{m} and the fiducial detector mass from \SI{40}{kt} to \SI{67.2}{tons}. We moreover increase systematic uncertainties in the near detector to 25\% to conservatively account for our imperfect understanding of the beam, which in a near detector-dominated analysis remains unmitigated (see \cref{sec:statistics} below).

To incorporate new physics in the SMEFT and WEFT frameworks, we have extended GLoBES and developed the ``GLoBES-EFT'' simulation package, which simulates long-baseline neutrino oscillation experiments in the presence of new dimension-6 operators. Specifically, GLoBES-EFT implements the pseudo-probabilities $\tilde P^S_{\alpha\beta} (E_\nu,L)$ from \cref{eq:tildeP}, along with SMEFT/WEFT matching and renormalization group evolution of the Wilson coefficients. The details of GLoBES-EFT and how to use it can be found in~\cref{sec:GLoBES-EFT}, and the package is available for download from \href{https://github.com/SalvaUrrea2/GLoBES-EFT/}{GitHub} \cite{github}. The production and detection coefficients are computed outside GLoBES and then imported in tabulated forms.

On the detection side, we focus on quasi-elastic neutrino interactions and use the corresponding cross sections in the presence of new physics from ref.~\cite{Kopp:2024yvh} to calculate the detection coefficients $d_{XY,\beta}^{rs}$ (see in \cref{sec:Detection}). We do not include processes involving resonance production (e.g.\ $\nu + p \to e + (\Delta(1232) \to p + \pi)$) as computing the corresponding cross-sections in the presence of general new physics would be more challenging. In particular, it would require knowledge of hadronic transition matrix elements with non-standard Lorentz structures. These matrix elements are already a major source of uncertainty for quasi-elastic scattering \cite{Kopp:2024yvh}; in the regime of QCD resonance production, they would pose an even greater problem. We also do not consider neutrino interactions in the DIS regime, even though they are in principle more straightforward to predict (see e.g.\ ref.~\cite{Falkowski:2021bkq}). But as the DIS regime falls within the high-energy tail of the DUNE flux, where statistics is lower and oscillations are suppressed, it is less important to our study. An additional reason for not considering DIS here is that it is often impossible to distinguish between lower-energy DIS and RES, implying that an analysis considering one of these processes, but not the other, is impossible.

In our analysis, we distinguish three types of QE events samples: $\nu_e$-like, $\nu_\mu$-like, and $\nu_{\tau}$-like events:
\begin{itemize}
    \item \textbf{$\nu_e$-like events.} These events correspond to CC $\nu_e$ interactions and are characterised by a nucleon and an electron emerging from the interaction vertex, with the electron producing an electromagnetic shower in the detector. $\nu_e$-like events include electron neutrinos produced via standard oscillations, electron neutrinos produced via new physics contributions to the appearance (pseudo-)probabilities $\tilde{P}_{\alpha e}^{S}$, and intrinsic $\nu_e$ already present in the beam, which are sensitive to the disappearance probabilities $\tilde{P}^{S}_{ee}$. In addition, this sample may contain a small number of misidentified events from CC $\nu_\mu$, CC $\nu_\tau$, and neutral-current (NC) interactions. Misidentification can occur, for example, when a CC $\nu_\mu$ interaction produces a muon with a very short track, or when a NC interaction generates a $\pi^0$ that decays into two photons, with only one photon converting visibly in the TPC, producing a signal similar to that of a CC $\nu_e$ event. We account for detection efficiencies, energy smearing functions, and misidentification rates using the auxiliary files provided by the DUNE collaboration \cite{DUNE:2021cuw}. 

    \item \textbf{$\nu_\mu$-like events.} These events are characterised by a long, straight muon track emerging from the interaction vertex. In the SM, they originate almost exclusively from genuine $\nu_\mu$ and $\bar\nu_\mu$ in the beam, whereas in presence of new physics, they can also come from $\nu_e$, $\bar\nu_e$, $\nu_\tau$, or $\bar\nu_\tau$ producing a muon through flavour-changing interactions. $\nu_\mu$-like events probe the disappearance pseudo-probabilities $\tilde P^{S}_{\alpha\mu}$. Misidentified events from CC $\nu_{\tau}$ and NC interactions may also contribute to this sample. Contamination from $\nu_e$ misidentification is negligible. As for the $\nu_e$ sample, we incorporate the relevant misidentification probabilities, detection efficiencies, and energy smearing effects based on the auxiliary files from ref.~\cite{DUNE:2021cuw}. 

    \item \textbf{$\nu_\tau$-like events.} Ref.~\cite{DUNE:2021cuw} does not consider $\nu_\tau$, given that they are irrelevant to standard oscillation analyses. In presence of new physics, however, $\nu_\tau$ interactions can be important. In modeling them, we follow the approach outlined in refs.~\cite{Coloma:2021uhq, DeGouvea:2019kea}. At energies of a few GeV, the $\tau$ lepton from a CC $\nu_\tau$ interaction travels only a few millimetres before decaying, rendering direct track reconstruction in liquid argon unfeasible. $\tau$ detection instead has to rely on the $\tau$ decay products, which are either one or three hadrons, or a charged lepton accompanied by missing energy. Given that hadronic decays constitute approximately 65\% of all $\tau$ decays, and that leptonic decays (each with a branching ratio around 17\%~\cite{Zyla:2020zbs}) are more easily confused with standard CC $\nu_e$ and $\nu_\mu$ events, we focus exclusively on the hadronic decay modes. For these, we assume a conservative signal efficiency of 30\%. We model the reconstructed energy in $\nu_\tau$ events with a Gaussian distribution centred at $0.45 E_\nu^{\text{true}}$ and a standard deviation of $0.25 E_\nu^{\text{true}}$, where $E_\nu^{\text{true}}$ is the true neutrino energy \cite{DeGouvea:2019kea}. The dominant background arises from neutral-current interactions, which can produce hadronic signatures resembling $\tau$ decays. To estimate this contamination, we use the NC migration matrices from ref.~\cite{DUNE:2021cuw} and apply a constant misidentification rate of 0.5\%, following ref.~\cite{DeGouvea:2019kea}. Additionally, a 20\% normalization uncertainty is assigned to all event samples involving $\tau$ detection. Lastly, we take the SM CC $\nu_\tau$ cross sections from the auxiliary files of ref.~\cite{Alion:2016uaj}, which were generated using GENIE v2.8.4~\cite{Andreopoulos:2015wxa, Tena-Vidal:2021rpu}.
\end{itemize}
\Cref{tab:samples} summarises the contributions to each event sample, outlining the associated neutrino oscillation probabilities probed in our analysis. Each event sample can be impacted by new physics, as discussed in \cref{sec:eft-neutrino-exp}.

\begin{table}
    \centering
    \scalebox{0.96}{
    \begin{tabular}{c|cccc cc}
    \toprule
    Sample & Contribution & Parent meson & Probability probed & \multicolumn{2}{c}{Systematic} \\
    \cmidrule(lr){5-6}
    & & & & ND & FD \\
    \midrule
    \multirow{15}{*}{$\nu_e$-like} 
      & \multirow{3}{*}{$\nu_e$ appearance} 
        & $K$ & $\tilde{P}_{\mu e}^{K}$ & \multirow{3}{*}{0.25} & \multirow{3}{*}{0.02} \\
        & & $\pi$ & $\tilde{P}_{\mu e}^{\pi}$ & & \\
        & & $D,D_s$ & $\tilde{P}_{\mu e}^{S}, \tilde{P}_{\tau e}^{S}$ & & \\ \cline{2-6}
      & \multirow{3}{*}{$\nu_e$ intrinsic} 
        & $K$ & $\tilde{P}_{ee}^{K}$ & \multirow{3}{*}{0.25} & \multirow{3}{*}{0.02} \\
        & & $\pi$ & $\tilde{P}_{ee}^{\pi}$ & & \\
        & & $D,D_s$ & $\tilde{P}_{ee}^{S}$ & & \\ \cline{2-6}
      & \multirow{3}{*}{$\nu_\mu$ mis-ID} 
        & $K$ & $\tilde{P}_{\mu\mu}^{K}, \tilde{P}_{e\mu}^{K}$ & \multirow{3}{*}{0.25} & \multirow{3}{*}{0.05} \\
        & & $\pi$ & $\tilde{P}_{\mu\mu}^{\pi}, \tilde{P}_{e\mu}^{\pi}$ & & \\
        & & $D,D_s$& $\tilde{P}_{\mu\mu}^{S}, \tilde{P}_{e\mu}^{S}, \tilde{P}_{\tau\mu}^{S}$ & & \\ \cline{2-6}
      & \multirow{3}{*}{$\nu_\tau$ mis-ID} 
        & $K$ & $\tilde{P}_{e\tau}^{K}, \tilde{P}_{\mu\tau}^{K}$ & \multirow{3}{*}{0.25} & \multirow{3}{*}{0.20} \\
        & & $\pi$ & $\tilde{P}_{e\tau}^{\pi}, \tilde{P}_{\mu\tau}^{\pi}$ & & \\
        & & $D,D_s$ & $\tilde{P}_{\tau\tau}^{S}, \tilde{P}_{e\tau}^{S}, \tilde{P}_{\mu\tau}^{S}$ & & \\ \cline{2-6}
      & \multirow{3}{*}{NC} 
        & $K$ & $\tilde{P}_{e\alpha}^{K}, \tilde{P}_{\mu\alpha}^{K}$ & \multirow{3}{*}{0.25} & \multirow{3}{*}{0.10} \\
        & & $\pi$ & $\tilde{P}_{e\alpha}^{\pi}, \tilde{P}_{\mu\alpha}^{\pi}$ & & \\
        & & $D,D_s$ & $\tilde{P}_{e\alpha}^{S}, \tilde{P}_{\mu\alpha}^{S}, \tilde{P}_{\tau\alpha}^{S}$ & & \\ \hline
    \multirow{9}{*}{$\nu_\mu$-like} 
      & \multirow{3}{*}{$\nu_\mu$ disappearance} 
        & $K$ & $\tilde{P}_{\mu\mu}^{K}$,$\tilde{P}_{e\mu}^{K}$ & \multirow{3}{*}{0.25} & \multirow{3}{*}{0.05} \\
        & & $\pi$ & $\tilde{P}_{\mu\mu}^{\pi}$,$\tilde{P}_{e\mu}^{\pi}$ & & \\
        & & $D,D_s$ & $\tilde{P}_{\mu\mu}^{S}$,$\tilde{P}_{e\mu}^{S}$,$\tilde{P}_{\tau\mu}^{S}$ & & \\ \cline{2-6}
      & \multirow{3}{*}{$\nu_\tau$ mis-ID} 
        & $K$ & $\tilde{P}_{e\tau}^{K}, \tilde{P}_{\mu\tau}^{K}$ & \multirow{3}{*}{0.25} & \multirow{3}{*}{0.20} \\
        & & $\pi$ & $\tilde{P}_{e\tau}^{\pi}, \tilde{P}_{\mu\tau}^{\pi}$ & & \\
        & & $D,D_s$ & $\tilde{P}_{\tau\tau}^{S}, \tilde{P}_{e\tau}^{S}, \tilde{P}_{\mu\tau}^{S}$ & & \\ \cline{2-6}
      & \multirow{3}{*}{NC} 
        & $K$ & $\tilde{P}_{e\alpha}^{K}, \tilde{P}_{\mu\alpha}^{K}$ & \multirow{3}{*}{0.25} & \multirow{3}{*}{0.10} \\
        & & $\pi$ & $\tilde{P}_{e\alpha}^{\pi}, \tilde{P}_{\mu\alpha}^{\pi}$ & & \\
        & & $D,D_s$ & $\tilde{P}_{e\alpha}^{S}, \tilde{P}_{\mu\alpha}^{S}, \tilde{P}_{\tau\alpha}^{S}$ & & \\ \hline
    \multirow{6}{*}{$\nu_\tau$-like} 
      & \multirow{3}{*}{$\nu_\tau$ appearance} 
        & $K$ & $\tilde{P}_{e\tau}^{K}, \tilde{P}_{\mu\tau}^{K}$ & \multirow{3}{*}{0.25} & \multirow{3}{*}{0.20} \\
        & & $\pi$ & $\tilde{P}_{e\tau}^{\pi}, \tilde{P}_{\mu\tau}^{\pi}$ & & \\
        & & $D,D_s$ & $\tilde{P}_{e\tau}^{S}, \tilde{P}_{\mu\tau}^{S}$ & & \\ \cline{2-6}
      & \multirow{3}{*}{NC} 
        & $K$ & $\tilde{P}_{e\alpha}^{K}, \tilde{P}_{\mu\alpha}^{K}$ & \multirow{3}{*}{0.25} & \multirow{3}{*}{0.10} \\
        & & $\pi$ & $\tilde{P}_{e\alpha}^{\pi}, \tilde{P}_{\mu\alpha}^{\pi}$ & & \\
        & & $D,D_s$ & $\tilde{P}_{e\alpha}^{S}, \tilde{P}_{\mu\alpha}^{S}, \tilde{P}_{\tau\alpha}^{S}$ & & \\
        \bottomrule
    \end{tabular}
    }
    \caption{Contributions to the different experimental event samples in DUNE, separated by parent meson ($K$, $\pi$, $D$, $D_s$). We do not explicitly list neutrinos from muon decay here as we do not consider new physics operators modifying muon decay in this work; neutrinos from muon decay are, however, included in our simulations. The fourth column lists the oscillation (pseudo-)probabilities probed by each contribution. For $K$ and $\pi$, probabilities with an initial $\tau$ flavour ($\tilde{P}_{\tau\alpha}$) are omitted because decays of these mesons are below the $\tau$ threshold. The systematic uncertainties for the Near Detector (ND) and Far Detector (FD) are listed in the final two columns. Note that some of these systematic errors are correlated between different channels, as detailed in \cref{tab:sys} below.}
    \label{tab:samples}
\end{table}

\begin{figure}
  \centering
  \includegraphics[width=1\textwidth]{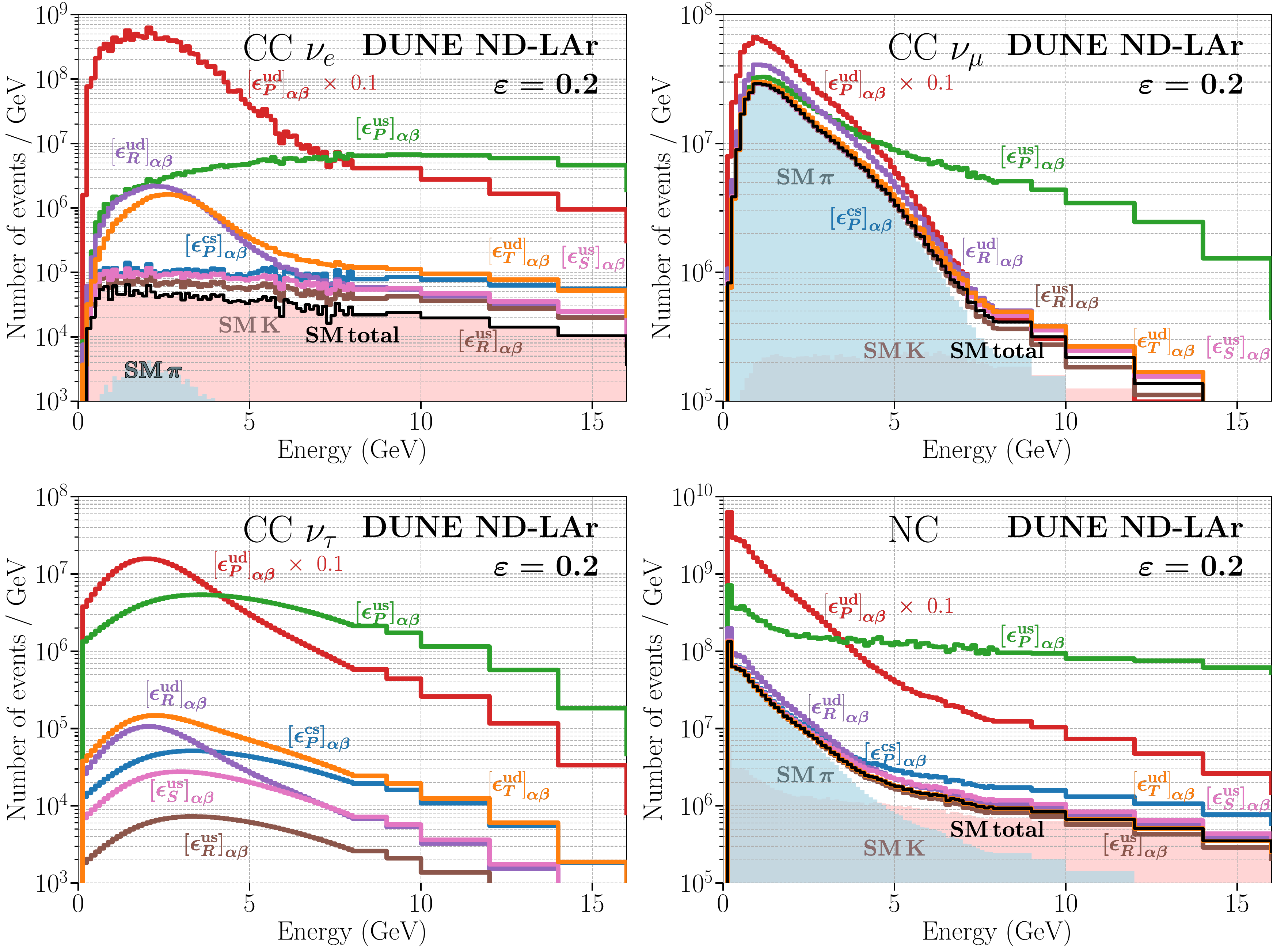}
  \caption{Predicted QE event spectra at the DUNE ND-LAr near detector. Red and blue shaded histograms correspond to the SM rates from the pion-decay and kaon-decay fluxes, with the black line indicates the sum of these contributions. Coloured lines illustrate the rates in presence of new physics, one operator at a time and with the corresponding Wilson coefficient set to $\epsilon = 0.2$. For some operators, the number of events has been rescales as indicated in the plot to improve readability. In this figure, we have turned off the indirect new physics effects.}
  \label{fig:hist_WEFT_ND}
\end{figure}

\begin{figure}
  \centering
  \includegraphics[width=1\textwidth]{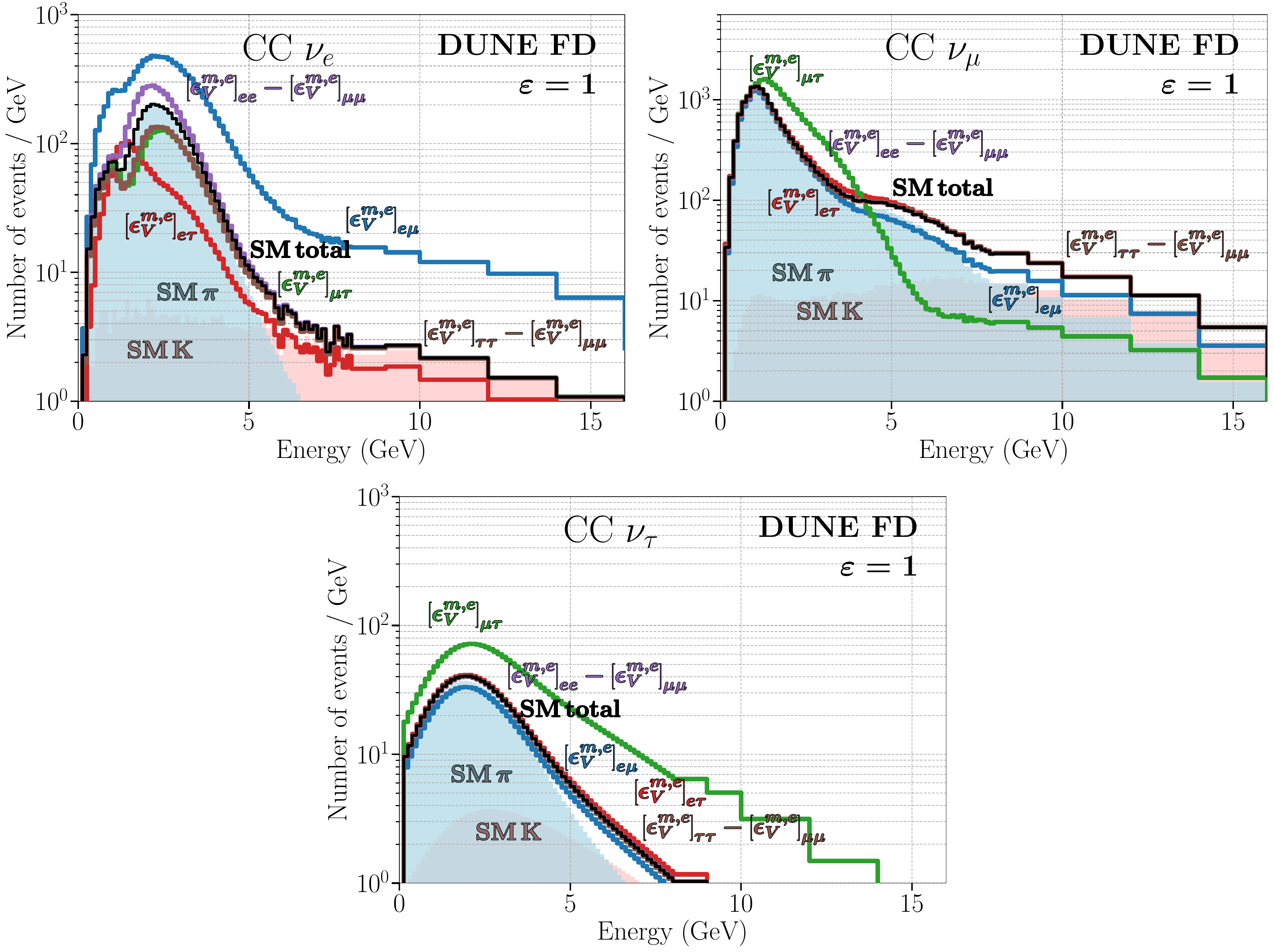}
  \caption{Predicted QE event spectra at the DUNE far detector. Red and blue shaded histograms correspond to the SM rates from the pion-decay and kaon-decay fluxes, with the black line indicates the sum of these contributions. Coloured lines illustrate the rates in presence of new physics, one operator at a time and with the corresponding Wilson coefficient set to $\epsilon = 1$.}
  \label{fig:hist_WEFT_FD}
\end{figure}

\Cref{fig:hist_WEFT_ND} illustrates the possible impact of new CC dim-6 WEFT operators on the event spectra in the different categories at the DUNE near detector. Similarly, \cref{fig:hist_WEFT_FD} displays the event rates at the DUNE far detector with and without new physics, where we focus on NC WEFT operators to which only the FD is sensitive as they affect neutrino propagation through matter, but not the CC neutrino production and detection processes. Besides the three event categories included in our analysis ($\nu_e$-like, $\nu_\mu$-like, $\nu_\tau$-like), we show in the bottom right panels of \cref{fig:hist_WEFT_ND} also the rates of neutral current events, which are included in our analysis only as a background to the CC samples. The effect of CC new physics on NC events is exclusively coming from modified neutrino production. We do not model the impact of NC NSI on the NC neutrino--nucleus scattering cross sections (and hence do not show NC events in \cref{fig:hist_WEFT_FD}) -- a small new physics correction to a small background would be negligible in our analysis. In case of a discovery of new physics, it would be highly interesting to analyse neutral current events separately to better disentangle new physics in the source from new physics in the detector and to narrow down the operator responsible for the observed deviation. The experimental signature of NC events is a recoil nucleon, possible additional hadrons, and no charged lepton.

\Cref{fig:hist_WEFT_ND,fig:hist_WEFT_FD} reflect our expectations from \cref{fig:KaonProdND,fig:detCoeff}: the largest enhancement is seen for new pseudoscalar interactions due to the absence of chiral suppression in pion decay. New interactions involving strange quarks affect kaon decays, but not pion decays; hence, the event excess in this case extends over a broad energy range and does not exhibit the peak around \SIrange{1}{2}{GeV} characteristic for neutrinos from pion decay.

\subsection{Statistical Analysis}
\label{sec:statistics}

\begin{table}
  \centering
  \begin{tabular}{lc}
    \toprule
    Source of uncertainty                                      & $1\sigma$ error \\
                                                               & (FD after ND corrections) \\
    \midrule                                                  
    appearance signal ($\nu$ mode)                             & 0.02 \\
    appearance signal ($\bar\nu$ mode)                         & 0.02 \\
    disappearance signal ($\nu$ mode)                          & 0.05 \\
    disappearance signal ($\bar\nu$ mode)                      & 0.05 \\
    intrinsic $\nu_e$, $\bar\nu_e$ background ($\nu$ mode)     & 0.05 \\
    intrinsic $\nu_e$, $\bar\nu_e$ background ($\bar\nu$ mode) & 0.05 \\
    misidentified $\nu_\mu$, $\bar\nu_\mu$ + NC background     & 0.05 \\
    misidentified $\nu_\tau$, $\bar\nu_\tau$ background        & 0.20 \\
    NC background to disappearance channels                    & 0.10 \\
    \bottomrule
  \end{tabular}
  \caption{Systematic normalization uncertainties in the far detector, following refs.~\cite{DUNE:2020ypp, DUNE:2021cuw}. Error estimates are based on the assumption that the near detector is able to effectively control systematics, see text for details.}
  \label{tab:sys}
\end{table}

To estimate DUNE's sensitivity to new interactions beyond the Standard Model, we carry out a frequentist maximum-likelihood fit to an Asimov data set, as implemented in GLoBES \cite{Huber:2004ka,Huber:2007ji}.

Systematic uncertainties are implemented following ref.~\cite{DUNE:2021cuw} through the nine nuisance parameters listed in \cref{tab:sys}. They bias the normalization, but not the spectral shape, of the corresponding event categories. The systematic errors given in \cref{tab:sys} are for the far detector and are based on the assumption that a highly capable suite of near detectors is available to precisely characterize the beam and measure neutrino cross sections. Simulating just the far detector, while taking into account the effect of the near detector through such an effective systematic error budget, is justified in searches for new NC interactions, which affect only the far detector. It is problematic, though, in the presence of new CC interactions as these affect both the ND and the FD. Therefore, in this case, we explicitly simulate the near detector as discussed in \cref{sec:sig-and-bg}. We use the same set of nuisance parameters as in the FD, but we set their uncertainties all to 0.25, a very conservative value. Even with such conservative assumptions, we find that constraints on new CC interactions remain entirely dominated by the ND thanks to its much larger statistical power. Therefore, when constraining exclusively CC interactions, we do not include FD data at all.

There are certain SMEFT operators that modify both NC and CC interactions, for instance $O_{lq}^{(3)}$. In this case, only the FD is sensitive to the modified oscillation probabilities, while both detectors are affected by modified neutrino production and detection rates. In fact, with a FD prediction based on ND data, the effect of modified CC interactions will even cancel out approximately (up to CC--NC interference effects and up to differences in acceptance between the two detectors). One might think that only a joint ND+FD fit can accurately constrain such combined CC+NC interactions. However, it turns out that the NC component of all SMEFT operators of this type maps onto the $\epsilon_L$ Wilson coefficient in WEFT. As such, their effect is degenerate with a bias in the matter density along the beam trajectory and hence largely unobservable.

\section{Results}
\label{sec:results}

Having set out our methodology, we are now ready to present sensitivity projections for DUNE. The goal is to identify those WEFT and SMEFT operators to which long-baseline experiments are particularly sensitive, and where long-baseline oscillations have an edge over other probes of physics beyond the SM.

We emphasize that -- especially when SMEFT is our starting point -- our constraints are based on simulations that consistently take into account new physics effects in neutrino production, oscillation, and detection.  Moreover, we present constraints in terms of Wilson coefficients evolved to the standard scales of $\mu = \SI{2}{GeV}$ for WEFT and $\mu = \SI{1}{TeV}$ for SMEFT, thereby allowing for straightforward comparisons with results from other, non-neutrino, experiments. This is often not the case for limits presented in the context of NSI formalism.

\subsection{Constraints on WEFT Operators}

\begin{figure}[H]
  \includegraphics[width=0.96\linewidth]{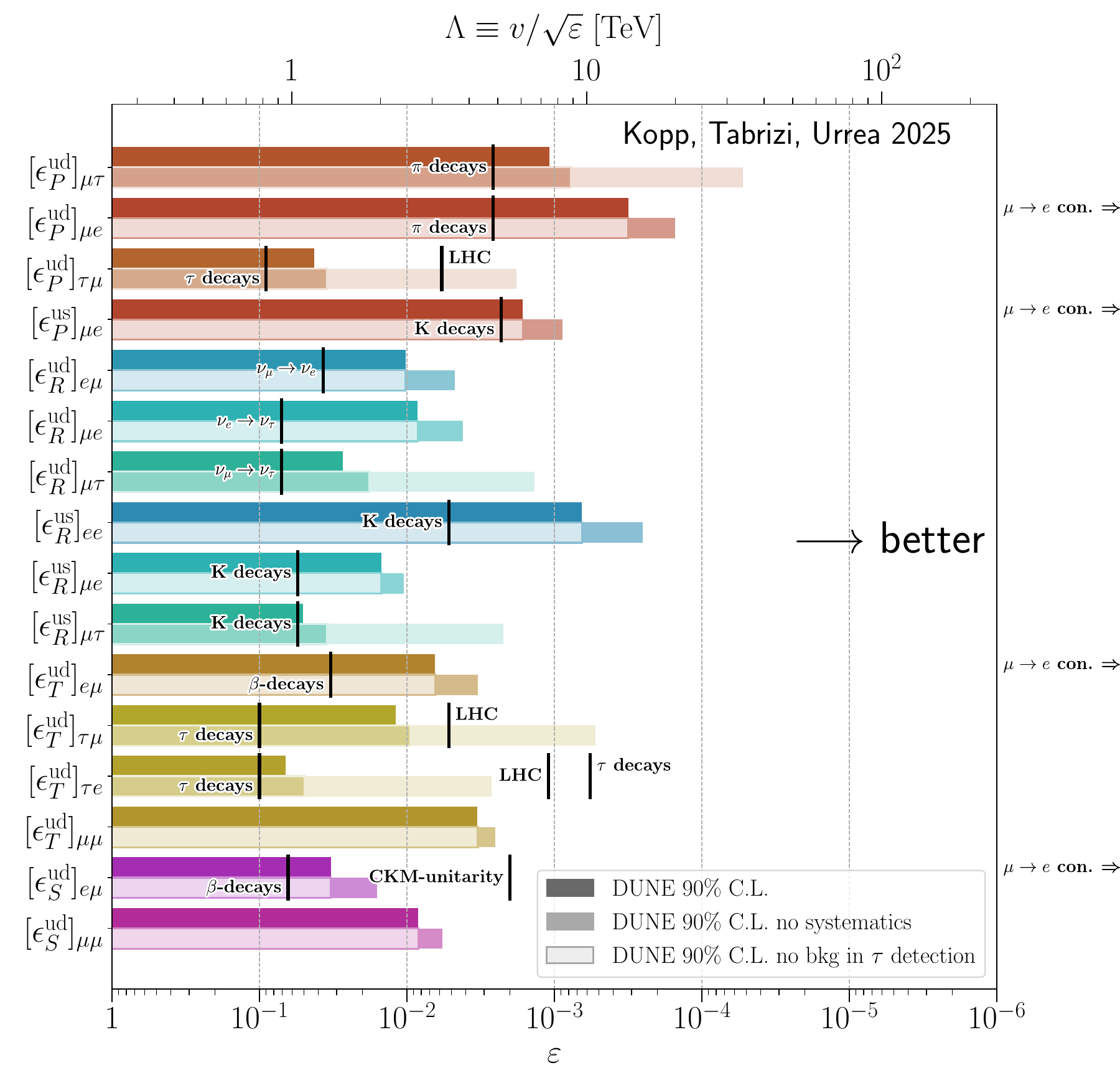}
  \caption{Expected DUNE sensitivity to new charged-current interactions within the Weak Effective Field Theory (WEFT) framework, assuming only one operator is relevant at a time. The plot focuses on those operators for which DUNE is most competitive with other probes. Our full set of projections is shown in \cref{fig:CC_P,fig:CC_R,fig:CC_S,fig:CC_T_ud} below.  For each operator, we show in dark the expected sensitivity under conservative assumptions on systematic errors (25\%) and backgrounds. The lighter bars have been computed under the assumption of \emph{no} systematic uncertainties, and the transparent bars show how results would improve under the highly optimistic assumption of background-free $\nu_\tau$ identification (while systematic uncertainties are kept at the conservative 25\% level). Vertical lines show current constraints, which are mostly taken from ref.~\cite{Falkowski:2021bkq}. The LHC constraints shown here and in the following plots have been obtained in the same way as those presented in ref.~\cite{Falkowski:2021bkq}, but are based on a more recent $\tau$+MET search by the ATLAS collaboration \cite{ATLAS:2024tzc}.}
  \label{fig:money}
\end{figure}

We begin with WEFT, which is the ``native'' EFT for accelerator-based long-baseline neutrino oscillation experiments, i.e., the EFT valid at the energy scales at which these experiments are operating. 
\Cref{fig:money} highlights our most important results in the context of WEFT. It shows the projected constraints on CC Wilson coefficients (see \cref{eq:WEFT_CC}) for those operators for which DUNE is expected to improve upon current experimental bounds. The full set of constraints on all CC operators is shown in \cref{fig:CC_P,fig:CC_R,fig:CC_S,fig:CC_T_ud} below. We report constraints both in terms of the dimensionless coefficients $\epsilon$ and in terms of the associated scale of new physics, defined as $\Lambda \equiv v / \sqrt{\epsilon}$. Our main conclusions are:
\begin{enumerate}
    \item DUNE will be able to probe new physics at the multi-TeV scale, in some cases significantly improving on previous constraints from oscillations or rare decays can be expected;

    \item Reduced systematic uncertainties would lead to improved constraint, though the improvement in most cases is not dramatic;

    \item Improved $\nu_\tau$ identification would be a game-changer for operators that modify the $\nu_\tau$ flux ($[\epsilon_X^{qq'}]_{\alpha\tau}$) or lead to $\tau$ production off the $\nu_\mu$ flux ($[\epsilon_X^{ud}]_{\tau\mu}$).
\end{enumerate}
In the following sections, and in \cref{fig:CC_P,fig:CC_R,fig:CC_S,fig:CC_T_ud}, we study DUNE's expected sensitivity to dim-6 WEFT operators in more detail. 

\subsubsection{Charged-Current Pseudoscalar Interactions}
\label{sec:results-P}

\begin{figure}
  \centering
  \includegraphics[width=0.99\textwidth]{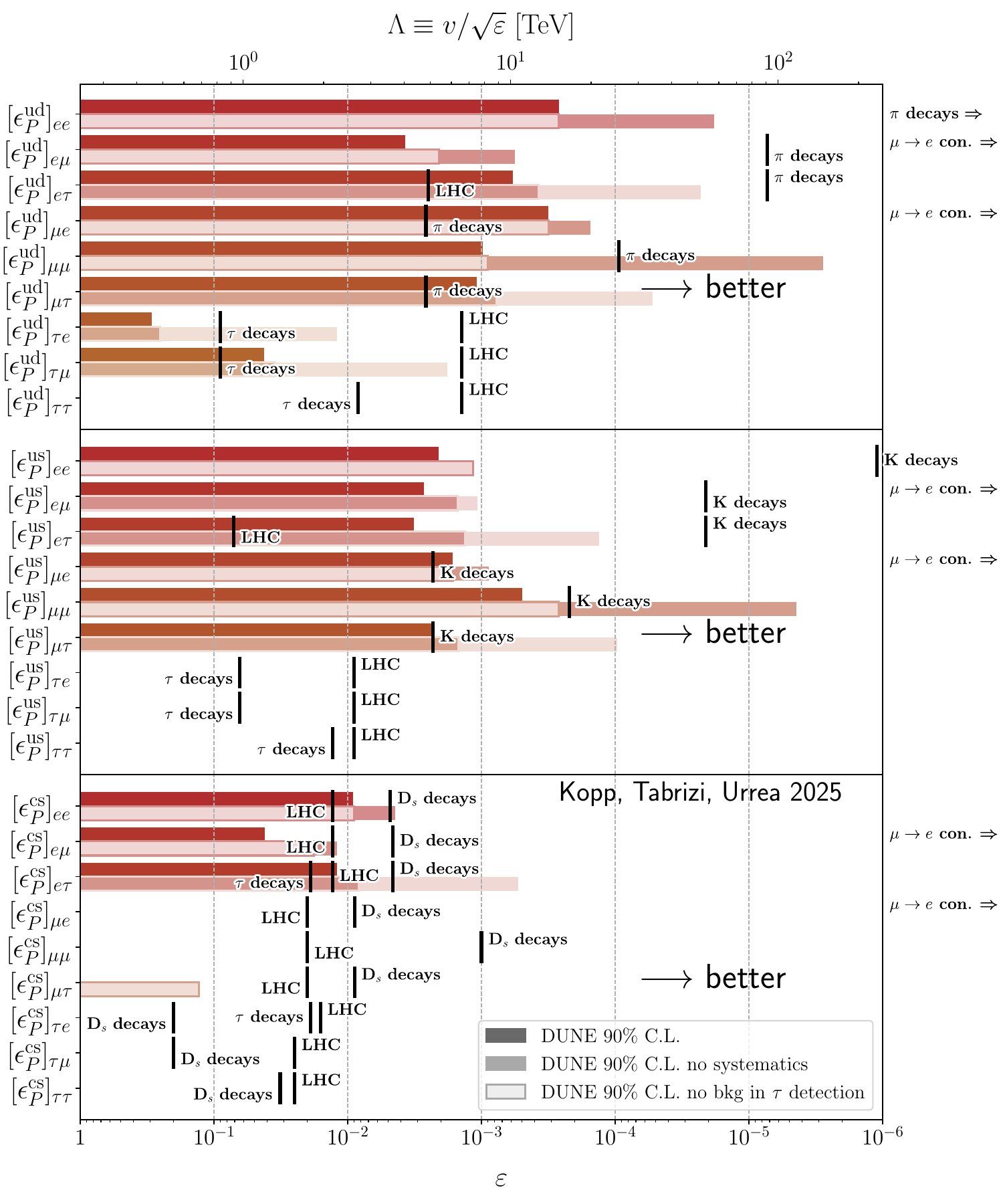}
  \caption{Expected DUNE sensitivity to new charged-current pseudoscalar interactions within the Weak Effective Field Theory (WEFT) framework, assuming only one operator is on at a time. The colour coding is the same as in \cref{fig:money}.}
  \label{fig:CC_P}
\end{figure}

We begin with pseudoscalar operators, projected constraints for which are shown in~\cref{fig:CC_P}, including couplings to first- and second-generation quarks ($ud$, $us$, and $cs$). Among all operator classes, pseudoscalar interactions receive some of the strongest projected bounds because they lift the chiral suppression in purely leptonic meson decays like $\pi \to e \nu_e$ and $\pi \to \mu \nu_\mu$. This leads to strong enhancement of the new physics effect compared to SM terms, reflected in large production coefficients (see~\cref{eq:pioncoefficients,eq:Dscoefficients,fig:KaonProdND}).

Considering first couplings to $u$ and $d$ quarks, we see that DUNE is expected to improve upon current bounds for several operators. Notably, the projected sensitivities to $[\epsilon_P^{ud}]_{\mu e}$ and $[\epsilon_P^{ud}]_{\mu\tau}$ surpass those from precision measurements of pion decay. However, $[\epsilon_P^{ud}]_{\mu e}$ remains more tightly constrained by $\mu \to e$ conversion in nuclei. For $[\epsilon_P^{ud}]_{\mu\mu}$, the reach is limited by large systematic uncertainties; only with a substantial reduction in systematics would DUNE become competitive with pion decay limits. For $[\epsilon_P^{ud}]_{\tau\mu}$ the sensitivity arises exclusively from the detection side via CCQE production of $\tau$ leptons off the large $\nu_\mu$ flux. For this operator, DUNE will be able to improve upon current limits from tau decay studies, but would be competitive with LHC constraints only if $\tau$ detection were background-free, which seems impossible at this point. For the remaining operators, current limits from pion decays and $\mu \to e$ conversion are considerably stronger than DUNE's projected sensitivity. 

The $us$ sector exhibits similar behaviour. DUNE is expected to be competitive with existing limits only for $[\epsilon_P^{us}]_{\mu\tau}$, while for all other entries, kaon decay, LHC limits, or $\mu\to e$ conversion far exceed DUNE's projected sensitivity. The lack of sensitivity of DUNE to $[\epsilon_P^{us}]_{\tau\alpha}$ arises from the fact that kaons do not decay into $\nu_\tau$, and, unlike in the $ud$ case, $us$ operators are irrelevant for CCQE scattering due to the lack of strange quarks in the nucleon.

The $cs$ sector is even more challenging due to the very small number of charm mesons produced in the LBNF target and due to the negligible charm and strangeness content of the nucleon precluding any effect of $[\epsilon_P^{cs}]_{\alpha\beta}$ on the detection side. As a result, DUNE’s projected bounds on $[\epsilon_P^{cs}]_{\alpha\beta}$ remain weaker than existing constraints from the LHC, tau decays, $\mu\to e$ conversion, and $D_s$ decays. Even weaker sensitivity is expected for $[\epsilon_P^{cd}]_{\alpha\beta}$ as $D^\pm$ fluxes are even lower than $D_s$ fluxes; hence, we do not show sensitivities to $[\epsilon_P^{cd}]_{\alpha\beta}$ here.

\subsubsection{Charged-Current Right-Handed Interactions}

\begin{figure}
  \centering
  \includegraphics[width=1\textwidth]{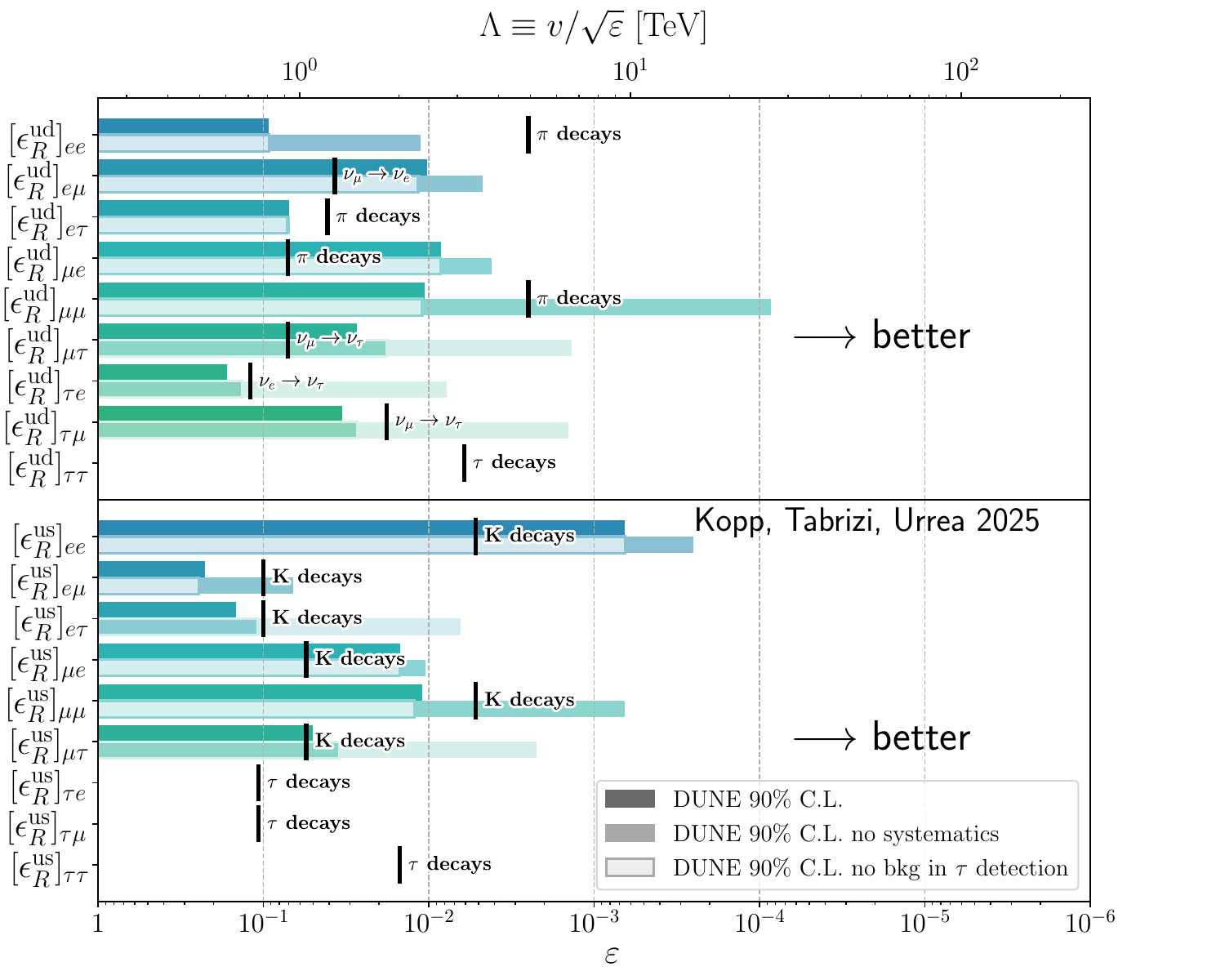}
  \caption{Expected DUNE sensitivity to new charged-current right-handed interactions within the Weak Effective Field Theory (WEFT) framework, assuming only one operator is on at a time. The colour coding is the same as in \cref{fig:money}.}
  \label{fig:CC_R}
\end{figure}

We next consider right–handed operators, whose projected sensitivities are displayed in~\cref{fig:CC_R}. Because these couplings do not lead to chiral enhancement in meson decays, limits are in general weaker than in the pseudoscalar case. In the $ud$ sector, DUNE nevertheless will be able to improve limits on several flavour-violating interactions. The bound on $[\epsilon_R^{ud}]_{\mu\tau}$, to which DUNE is sensitive mostly through the detection of $\nu_\tau$ produced via flavour-violating pion decays, is projected to surpass the current $\nu_\mu\!\to\nu_\tau$ appearance limit by a factor of two.  Similarly, constraints on $[\epsilon_R^{ud}]_{e\mu}$ and $[\epsilon_R^{ud}]_{\mu e}$ are expected to exceed current limits from $\nu_\mu\!\to\nu_e$ appearance and $\pi$ decays.  In contrast, for the remaining entries of $\epsilon_R^{ud}$, DUNE constraints will stay below existing bounds. This is true even for the diagonal couplings $[\epsilon_R^{ud}]_{ee}$ and $[\epsilon_R^{ud}]_{\mu\mu}$, even though the corresponding amplitudes interfere with the SM amplitude. 

In the $us$ sector, DUNE's sensitivity arises from modifications to neutrino production in kaon decays. For $[\epsilon_R^{us}]_{ee}$ and $[\epsilon_R^{us}]_{\mu e}$, DUNE is expected to outstrip the existing limits from kaon decays. In contrast, the projected limits on $[\epsilon_R^{us}]_{\tau\mu}$ and $[\epsilon_R^{us}]_{\tau e}$ are worse than the current $\tau$-decay bounds, where DUNE finds constraints $\mathcal{O}(1)$.

For right-handed coupling to $cs$ and $cd$ quark, DUNE’s projected limits cannot compete with current constraints from $D$ and $D_s$ decays because production rates of $D$ and $D_s$ mesons at LBNF are orders of magnitude smaller than those of kaons and pions. We therefore do not show results for $\epsilon_R^{cs}$ and $\epsilon_R^{cd}$ here.

\subsubsection{Charged-Current Scalar Interactions}
%
\begin{figure}
  \centering
  \includegraphics[width=1\textwidth]{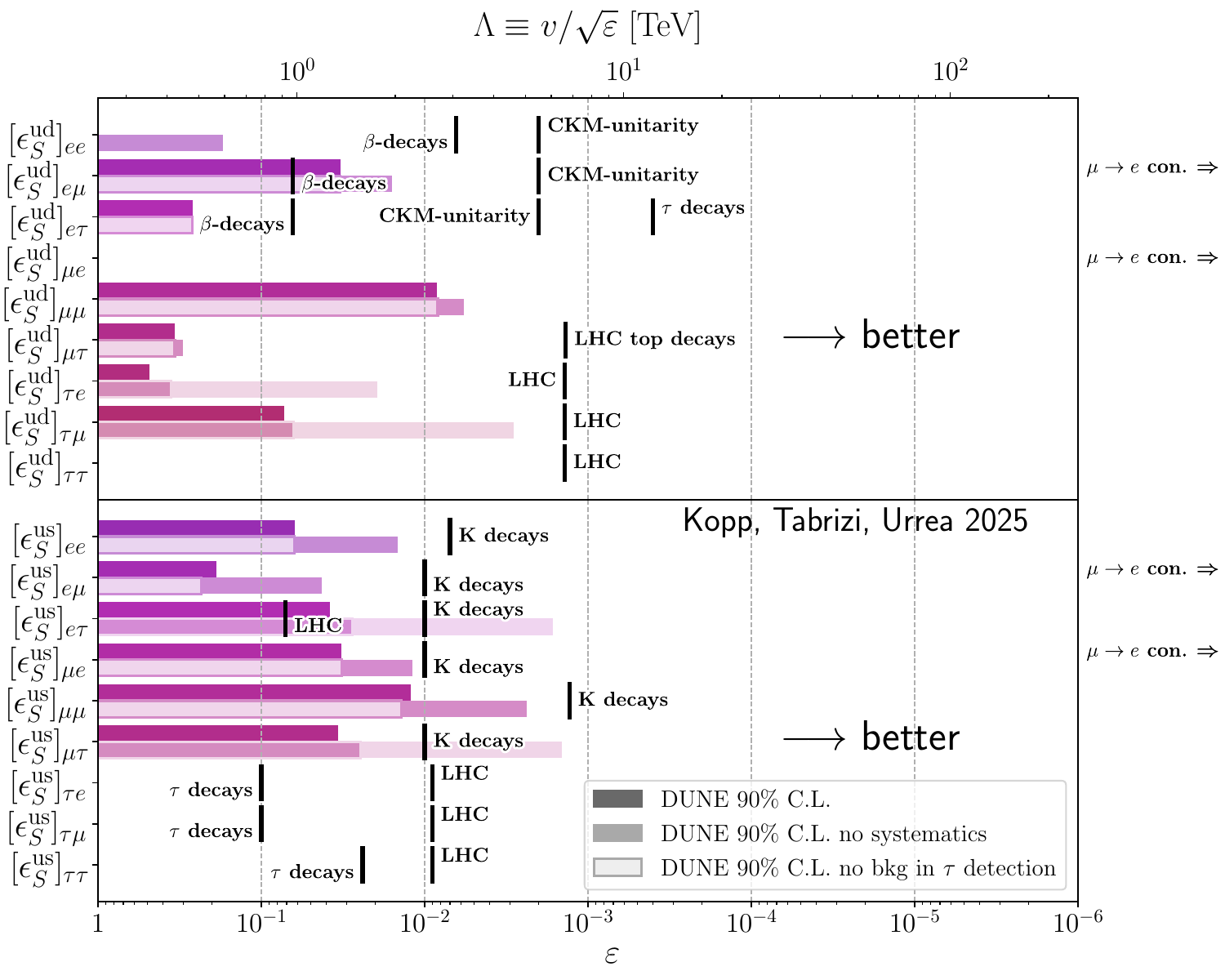}
  \caption{Expected DUNE sensitivity to new charged-current scalar interactions within the Weak Effective Field Theory (WEFT) framework, assuming only one operator is on at a time. The colour coding is the same as in \cref{fig:money}.}
   \label{fig:CC_S}
\end{figure}

We next analyse the projected limits on scalar interactions, shown in~\cref{fig:CC_S}.  For new scalar couplings to $u$ and $d$ quarks, the production coefficients for pion decay vanish due to the pseudoscalar nature of the pion, but the detection coefficients can be large if we include the contribution from the induced scalar term and estimate it as discussed at the end of \cref{sec:Detection} and in ref.~\cite{Kopp:2024yvh}. This enhancement allows DUNE to surpass the $\beta$–decay bound on $[\epsilon_S^{ud}]_{e\mu}$, although the limit remains weaker than the indirect constraint from CKM unitarity. For $[\epsilon_S^{ud}]_{\mu\mu}$, where no competitive bound exists at present, DUNE becomes the leading probe.

The situation in the $us$ sector is somewhat more favourable: non-vanishing scalar contributions appear in the three-body kaon decays $K \to e\nu\pi$, altering the $\nu_e$ and $\nu_\mu$ fluxes. DUNE therefore attains meaningful limits on several $\epsilon_S^{us}$ coefficients. Even so, these sensitivities are still less stringent than those derived from kaon decay.

We do not quote sensitivities for $\epsilon_S^{cs}$: the corresponding production coefficients vanish in $D_s$ decays (since the $D_s$ is a pseudoscalar), and the detection coefficients are numerically negligible, rendering DUNE insensitive.

\subsubsection{Charged-Current Tensor Interactions}
\label{sec:WEFT_tensor_results}

\begin{figure}
  \centering
  \includegraphics[width=1\textwidth]{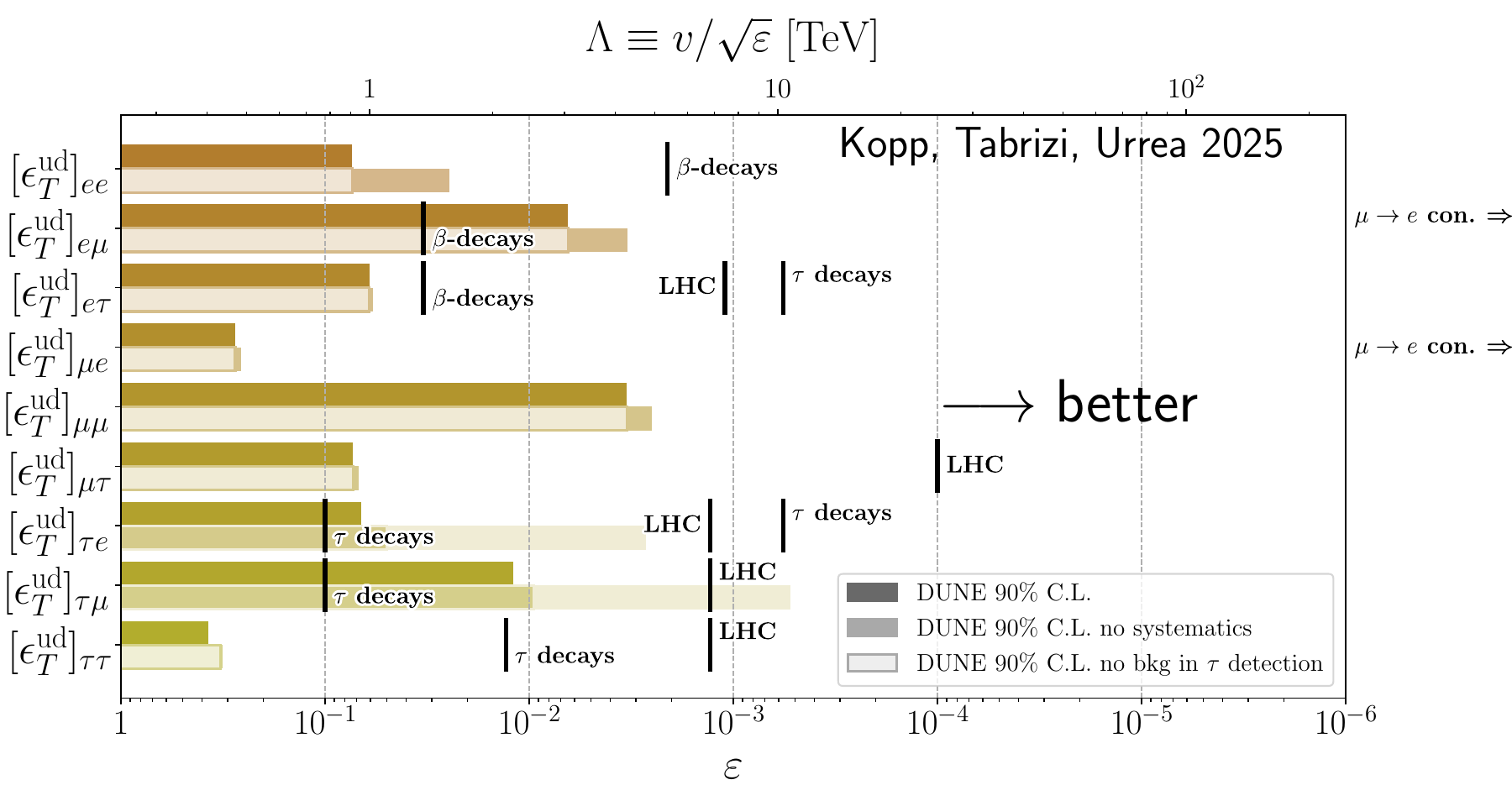}
  \caption{Expected DUNE sensitivity to new charged-current tensorial interactions within the Weak Effective Field Theory (WEFT) framework, assuming only one operator is on at a time. The colour coding is the same as in \cref{fig:money}.}
   \label{fig:CC_T_ud}
\end{figure}

DUNE's sensitivity to tensor interactions is displayed in~\cref{fig:CC_T_ud}. In the $ud$ sector, the production coefficients $p_{TX}^{ud}$ and $p_{XT}^{ud}$ vanish, so the sensitivity arises exclusively from modified detection processes, where the detection coefficients are large (see \cref{fig:detCoeff}). As a result, DUNE will provide limits that are meaningful but not competitive with existing constraints for most entries of $\epsilon_T^{ud}$. The only exception is the diagonal entry $[\epsilon_T^{ud}]_{\mu\mu}$, for which no direct bound exists in the literature. Therefore, DUNE will provide the leading constraint on this operator. DUNE will also be able to improve upon the $\beta$-decay bound on $[\epsilon_T^{ud}]_{e\mu}$, even though a much stronger constraint exists from $\mu \to e$ conversion in nuclei.

For couplings of the form $\epsilon_T^{us}$, the production coefficients are small, leading to sensitivities much weaker than those shown for $\epsilon_T^{ud}$; they are therefore not displayed.

For $\epsilon_T^{cs}$, the production coefficients vanish altogether and there is no contribution to detection via CCQE processes. Hence, DUNE is largely insensitive to these couplings as well and we do not show them here.

\subsubsection{Neutral-Current Interactions}
\label{sec:WEFT_NC_results}

\begin{figure}
  \centering
  \includegraphics[width=1\textwidth]{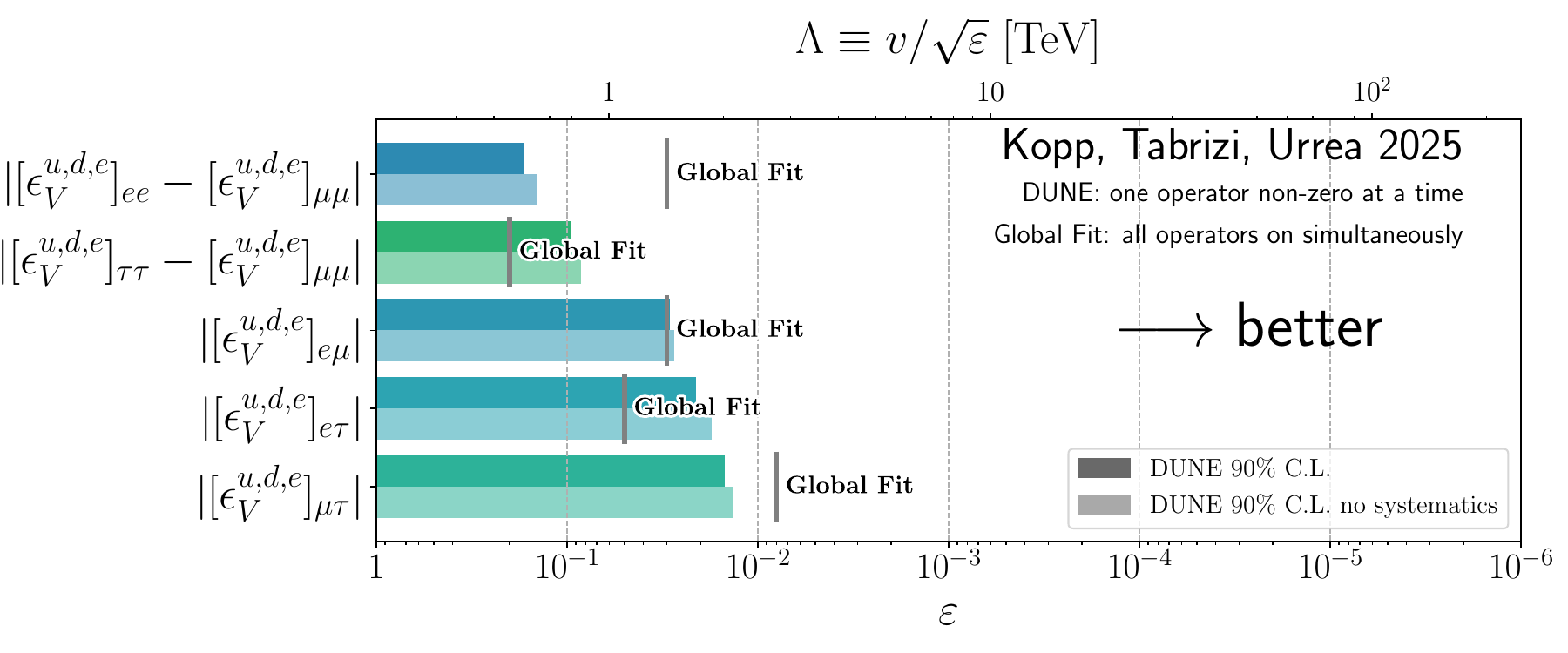}
  \caption{Expected DUNE sensitivity to the Wilson coefficients of new neutral-current vector couplings to electrons within the Weak Effective Field Theory (WEFT) framework. Bounds on NC interactions with $u$ and $d$ quarks can be directly obtained by dividing the constraints shown here by a factor of 3, see \cref{eq:epsilon_earth}. Each coloured bar represents the limit on a specific interaction, assuming all other new interactions are negligible. The shorter, darker bars correspond to conservative assumptions on the systematic uncertainties (we take the systematic uncertainties from ref.~\cite{DUNE:2021cuw} for the DUNE far detector), while the lighter-coloured band represents the hypothetical case with no systematic uncertainties. We also show in black the current constraint taken from the global fit in ref.~\cite{Coloma:2023ixt}. Note, however, that in this global fit, all NC Wilson coefficients are allowed to vary simultaneously, while in our analysis, we take only one of them non-zero at a time.}
  \label{fig:NC_R_L_e}
\end{figure}

Finally, \cref{fig:NC_R_L_e} shows the constraints on the NC Wilson coefficients from \cref{eq:WEFT_NC}. In the literature, the corresponding interactions are commonly referred to as non-standard neutrino interactions (NSI). We recall from \cref{sec:eft-neutrino-exp} that only vector interactions affect neutrino propagation via the MSW effect and that the only relevant combination of $\epsilon_{\alpha \beta}^{m,e}$, $\epsilon_{\alpha \beta}^{m,u}$, and $\epsilon_{\alpha \beta}^{m,d}$ in an experiment with a shallow neutrino trajectory is $\epsilon_{\alpha \beta}^{\oplus}$, as defined in \cref{eq:epsilon_earth}. Moreover, among the three lepton flavour diagonal entries of $\epsilon^{\oplus}$, only two effective combinations ($\epsilon_{ee}^{\oplus} - \epsilon_{\mu\mu}^{\oplus}$ and $\epsilon_{\tau\tau}^{\oplus} - \epsilon_{\mu\mu}^{\oplus}$), are relevant since an overall phase does not affect neutrino propagation.

We see that DUNE performs best for flavour-changing operators, with sensitivities at the $\epsilon \lesssim \text{few} \times 10^{-2}$ level, but also provides meaningful constraints on flavour-conserving interactions.

In \cref{fig:NC_R_L_e}, we also show the results of the global NSI fit from ref.~\cite{Coloma:2023ixt}. This global fit includes oscillation data as well as scattering data from neutrino--electron elastic scattering and coherent elastic neutrino--nucleus scattering (CE$\nu$NS). The comparison with DUNE needs to be taken with a large grain of salt, though, as in ref.~\cite{Coloma:2021uhq} all NC Wilson coefficients were allowed to vary simultaneously, while here, we assume only one of them is non-zero at a time. With this caveat in mind, we can nevertheless conclude that DUNE will play an important role in further constraining new physics in the neutrino sector. Its full power, however, will only be realized when combined with other, orthogonal, data sets in the context of global fits.


\subsection{Constraints on SMEFT Operators}

While studying new physics in the context of the low-energy Wilson coefficients in WEFT is common practice in neutrino physics, it is important to keep in mind that WEFT needs to be embedded into a higher-energy framework, in particular one that respects electroweak symmetry. Therefore, we now move on to analysing DUNE's sensitivity to new physics in the context of SMEFT. We will quote constraints on SMEFT operators defined at the scale of $\mu = \SI{1}{TeV}$, facilitating comparisons with constraints from elsewhere. We use the renormalization group equations from ref.~\cite{Gonzalez-Alonso:2017iyc} to run the theory down to the electroweak scale, where we perform the matching onto WEFT as described in \cref{sec:matching}. We then continue the running in WEFT down to \SI{2}{GeV}, where the observables are computed and compared to simulated DUNE measurements. A given SMEFT operator will typically generate several WEFT operators and can thereby contribute to neutrino production (via meson decays), propagation (via matter effects), and/or detection (via modified cross sections) simultaneously. The detailed mapping between WEFT and SMEFT operators can be found in \cref{sec:matching} and the summary of the RG running and mixing results is given in \cref{sec:running}.

\begin{figure}
  \includegraphics[width=1\textwidth]{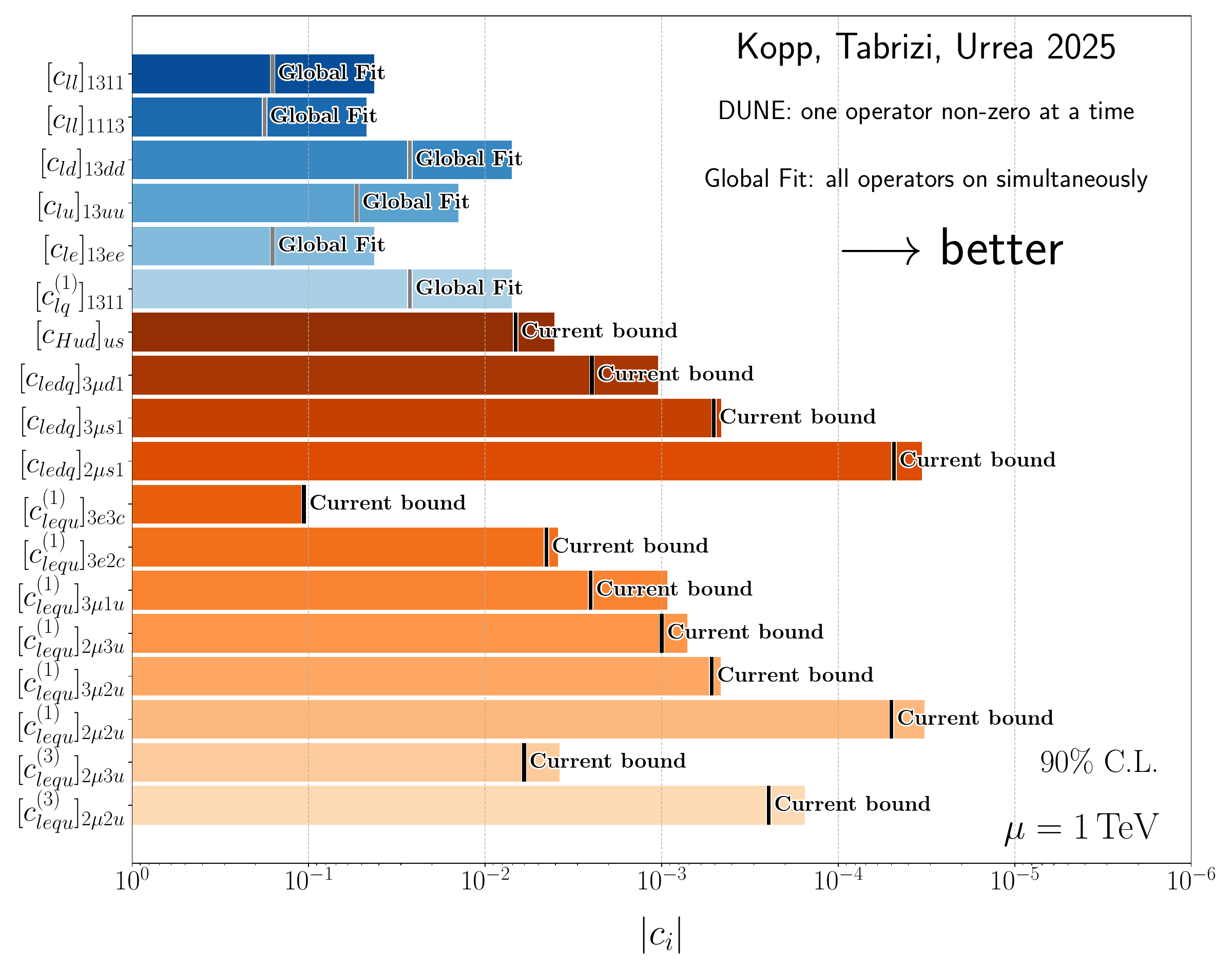}
  \caption{Expected DUNE sensitivity to new physics in the context of the Standard Model Effective Field Theory (SMEFT), assuming only one SMEFT operator is relevant at a time. The plot focuses on those operators for which DUNE is most competitive with other probes. Our full set of projections is shown in \cref{fig:smeft1,fig:smeft2,fig:smeft3,fig:smeft4,fig:smeft5}. Operators leading to NC WEFT are shown in cold colors, while those leading to CC WEFT are shown in warm colors. Vertical lines labeled ``Current bound'' indicate existing constraints, mostly from refs.~\cite{Falkowski:2021bkq,Coloma:2023ixt}. Vertical lines labeled Global Fit show the global fit constraint from ref.~\cite{Coloma:2023ixt}.}
  \label{fig:smeft-money1}
\end{figure}

We present the highlights of our analysis in \cref{fig:smeft-money1}, which shows the SMEFT operators for which DUNE is expected to improve upon current constraints. The complete set of results is summarised in \cref{fig:smeft1,fig:smeft2,fig:smeft3,fig:smeft4,fig:smeft5} in \cref{sec:full_set_constraints}. In these figures, each bar shows the 90$\%$~CL sensitivity of DUNE to one of the SMEFT Wilson coefficients $c_i$, assuming all other Wilson coefficients are set to zero. The coloured vertical markers indicate the best current bounds from either low-energy or high-energy experiments or the global fit to oscillation data and CE$\nu$NS~\cite{Coloma:2023ixt}. Note the index conventions for the Wilson coefficients: fermion singlets are denoted $e$, $\mu$, $\tau$, $u$, $d$, $s$, $c$, while for fermion doublets the generation is identified by an index $1$, $2$, or $3$. 

Several SMEFT operators give rise to new charged-current (CC) interactions after electroweak symmetry breaking. Among these, the operators to which DUNE is most sensitive are $O_{ledq}$ and $O_{lequ}^{(1)}$, which at low energies generate chirally enhanced pseudoscalar WEFT couplings (as well as scalar couplings which do not enjoy enhancement).

The SMEFT operator $O_{lequ}^{(3)}$ gives rise primarily to charged-current tensor interactions when matched onto WEFT. As discussed in \cref{sec:WEFT_tensor_results}, most tensor interactions in WEFT are already strongly constrained by present data, with the exception of $[\epsilon_T^{ud}]_{\mu\mu}$, which currently lacks a bound. Therefore, at the WEFT level, the DUNE constraint on $[\epsilon_T^{ud}]_{\mu\mu}$ is expected to be world-leading. The corresponding SMEFT operator, $[O_{lequ}^{(3)}]_{2\mu1u}$, however, is \emph{not} unconstrained due to the non-negligible operator mixing between $O_{lequ}^{(3)}$ and $O_{lequ}^{(1)}$ that occurs when running from \SI{1}{TeV} down to the electroweak scale, as given in \cref{eq:RGEsmeft}. Through this mixing, $O_{lequ}^{(3)}$ also induces pseudoscalar and scalar interactions, which are strongly constrained. Nevertheless, we find that, for some entries of $O_{lequ}^{(3)}$, the DUNE bound will still be slightly stronger than the current bounds obtained indirectly from operator mixing.

Another important CC operator is $O_{Hud}$, which generates flavour-universal right-handed interactions in WEFT. The constraint on $O_{Hud}$ is dominated by the limit on the most tightly constrained diagonal entry of $\epsilon_R^{\mathrm{ud}}$, which is $[\epsilon_R^{\mathrm{ud}}]_{\mu\mu}$ for couplings to $ud$ quarks and $[\epsilon_R^{\mathrm{us}}]_{ee}$ for couplings to $us$ quarks. For the $ud$ case, pion decay provides a yet stronger constraint, though, while for couplings to $us$, DUNE has a slight edge.

Lastly, for SMEFT operators that generate neutral-current operators in WEFT, DUNE's sensitivity arises from their impact on neutrino propagation through matter, which is probed at the far detector. Among these are four-lepton operators such as $c_{ll}, c_{ld}, c_{lu}, c_{le}$, and four-fermion operators such as $c^{(1)}_{lq}$. Their phenomenology is the same as that of the corresponding WEFT operators to which they match ($\left[\epsilon_V^{m}\right]_{\alpha \beta}$). As we saw in \cref{sec:WEFT_NC_results}, DUNE will be able to improve the existing limits on $\left[\epsilon_V^{m}\right]_{e \tau}$, and therefore the same is true for the SMEFT operators that give rise to it, as shown in \cref{fig:smeft-money1}.

\section{Summary and Conclusion}
\label{sec:conclusion}

In this paper, we have scrutinized the sensitivity of long-baseline neutrino oscillation experiments like DUNE and HyperKamiokande to new physics in the context of Effective Field Theory (EFT). Phenomenologically, we consider modified neutrino production, oscillation, and detection processes. In contrast to the traditional non-standard interaction (NSI) approach, our formulation is firmly grounded in a theoretically consistent EFT description. Notably, we embed the low-energy Weak Effective Field Theory (WEFT) into the $SU(2)$-invariant Standard Model Effective Field Theory (SMEFT), paying careful attention to renormalization-group effects and to WEFT/SMEFT matching. Our approach allows for consistent global analyses that combine the data from neutrino oscillation experiments, low-energy precision measurements, and high energy colliders.

As a concrete example, we have applied the EFT formalism to the DUNE experiment, quantifying its sensitivity to a comprehensive set of WEFT and SMEFT operators. Our most important results are summarized in \cref{fig:money,fig:smeft-money1}.

These projections are based on a detailed simulation of DUNE incorporating realistic fluxes, detector effects, and systematic uncertainties. To implement this analysis, we have developed a versatile simulation package called \href{https://github.com/SalvaUrrea2/GLoBES-EFT/}{\texttt{GLoBES-EFT}}, which evolves the SMEFT or WEFT Wilson coefficients to the $\mathcal{O}(\si{GeV})$ energy scale at which long-baseline experiments operate, and then carries out a detailed simulation and statistical analysis.

\vspace{1em}
\noindent
Our most important findings are:
\begin{enumerate}
    \item DUNE is able probe new physics at the multi-TeV scale, with the sensitivity even reaching up to \SI{10}{TeV} for some specific flavour-changing couplings.

    \item Constraints are particularly strong for new pseudoscalar interactions, which eliminate chiral suppression in pion and kaon decay.

    \item For some (but obviously not all) operators, DUNE constraints will be world-leading.
    
    \item When new charged-current (CC) interactions are present, the constraint is always dominated by the near detector thanks to its huge statistics.

    \item It is important not to neglect indirect new physics effects, that is, effects arising due to new physics contamination in measurements of $G_F$, $V_{ud}$, and other fundamental parameters that enter the DUNE simulation.
\end{enumerate}

We hope that this study, together with the release of \texttt{GLoBES-EFT}, will help advance new physics searches in long-baseline experiments to the next level, further cementing the important role that these experiments play in the global hunt for physics beyond the Standard Model.

\begin{acknowledgments}
   It is a pleasure to thank Noemi Rocco for enlightening discussions and for collaboration in the early stages of this work. We are moreover indebted to André de Gouvêa and Alexandre Sousa for providing invaluable advice regarding $\nu_\tau$ reconstruction in DUNE. Finally, we would like to thank the DUNE collaboration, and in particular Laura Fields, for making the results of the groundbreaking DUNE beamline simulations publicly available in the form of full simulated event records.
   The work of ZT is supported by Pitt PACC and CERN's Theoretical Physics department. 
\end{acknowledgments}

\appendix
\section{Matching Between SMEFT and WEFT}
\label{sec:matching}

In the following, we collect the matching conditions between SMEFT and WEFT operators, which we impose at $\mu \sim m_Z$. For the CC WEFT Lagrangian in \cref{eq:WEFT_CC}, the matching between $[\epsilon_X^{j k}]_{\alpha\beta}$ and the SMEFT Wilson coefficients of $d=6$ operators in the Warsaw basis is given by~\cite{Falkowski:2019xoe}
\begin{align}
    \left[\epsilon_L^{j k}\right]_{\alpha \beta} &\approx \frac{1}{V_{j k}}\left( V_{j k} \left[c_{H l}^{(3)}\right]_{\alpha \beta} + V_{i k} \left[c_{H q}^{(3)}\right]_{j i} \delta_{\alpha \beta} - V_{i k} \left[c_{l q}^{(3)}\right]_{\alpha \beta j i} \right), \\
    \left[\epsilon_R^{j k}\right]_{\alpha \beta} &\approx \frac{1}{2 V_{j k}} \left[ c_{H u d} \right]_{j k} \delta_{\alpha \beta}, \\
    \left[\epsilon_S^{j k}\right]_{\alpha \beta} &\approx -\frac{1}{2 V_{j k}}\left( V_{i k} \left[c_{l e q u}^{(1)}\right]_{\beta \alpha i j}^* + \left[c_{l e d q}\right]_{\beta \alpha k j}^* \right), \\
    \left[\epsilon_P^{j k}\right]_{\alpha \beta} &\approx -\frac{1}{2 V_{j k}}\left( V_{i k} \left[c_{l e q u}^{(1)}\right]_{\beta \alpha i j}^* - \left[c_{l e d q}\right]_{\beta \alpha k j}^* \right), \\
    \left[\epsilon_T^{j k}\right]_{\alpha \beta} &\approx -\frac{2}{V_{j k}} V_{i k} \left[c_{l e q u}^{(3)}\right]_{\beta \alpha i j}^*.
\end{align}
For the NC interactions given in \cref{eq:WEFT_NC}, the matching conditions read
\begin{align}
    \left[\epsilon_L^{m,u}\right]_{\alpha\beta} &= \delta_{\alpha \beta} \left[\delta  g_L^{Z u} \right]_{11} 
    + \left( 1 - \frac{4}{3} s_{\mathrm{w}}^2 \right) \left[ \delta g_L^{Z \nu} \right]_{\alpha \beta} 
    - \frac{1}{2} \left( \left[ c_{l q}^{(3)} \right]_{\alpha \beta 1 1} + \left[ c_{l q} \right]_{\alpha \beta 1 1} \right), \\
    \left[\epsilon_R^{m,u}\right]_{\alpha\beta} &= -\frac{4}{3} s_{\mathrm{w}}^2 \left[ \delta g_L^{Z \nu} \right]_{\alpha \beta} 
    + \delta_{\alpha \beta} \left[ \delta g_R^{Z u} \right]_{11} 
    - \frac{1}{2} \left[ c_{l u} \right]_{\alpha \beta 1 1}, \\
   \left[\epsilon_L^{m,d}\right]_{\alpha\beta} &= \delta_{\alpha \beta} \left[ \delta g_L^{Z d} \right]_{11} 
    + \frac{1}{3} \left( -3 + 2 s_{\mathrm{w}}^2 \right) \left[ \delta g_L^{Z \nu} \right]_{\alpha \beta} 
    + \frac{1}{2} \left( \left[ c_{l q}^{(3)} \right]_{\alpha \beta 1 1} - \left[ c_{l q} \right]_{\alpha \beta 1 1} \right), \\
    \left[\epsilon_R^{m,d}\right]_{\alpha\beta} &= \frac{2}{3} s_{\mathrm{w}}^2 \left[ \delta g_L^{Z \nu} \right]_{\alpha \beta} 
    + \delta_{\alpha \beta} \left[ \delta g_R^{Z d} \right]_{11} 
    - \frac{1}{2} \left[ c_{l d} \right]_{\alpha \beta 1 1}, \\
    \left[\epsilon_L^{m,e}\right]_{\alpha\beta} &= \delta_{e \beta} \left[ \delta g_L^{W e} \right]_{\alpha 1} 
    + \delta_{\alpha \beta} \left[ \delta g_L^{Z e} \right]_{11} 
    + \left( -1 + 2 s_{\mathrm{w}}^2 \right) \left( \left[ \delta g_L^{W e} \right]_{\alpha \beta} + \left[ \delta g_L^{Z e} \right]_{\alpha \beta} \right) \nonumber \\
    &\quad - \frac{1}{2} \left[ c_{ll} \right]_{11 \alpha \beta} 
    + \delta_{e \alpha} \left( \left[ \delta g_L^{W e} \right]_{1 \beta} 
    + \frac{1}{2} \delta_{e \beta} \left( -2 \left( \left[ \delta g_L^{W e} \right]_{11} + \left[ \delta g_L^{W e} \right]_{22} \right) 
    + \left[ c_{ll} \right]_{1221} \right) \right) \nonumber \\
    &\quad - \frac{1}{2} \left[ c_{ll} \right]_{\alpha \beta 11}, \\
    \left[\epsilon_R^{m,e}\right]_{\alpha\beta} &= 2 s_{\mathrm{w}}^2 \left( \left[ \delta g_L^{W e} \right]_{\alpha \beta} 
    + \left[ \delta g_L^{Z e} \right]_{\alpha \beta} \right) 
    + \delta_{\alpha \beta} \left[ \delta g_R^{Z e} \right]_{11} 
    - \frac{1}{2} \left[ c_{le} \right]_{\alpha \beta 11}.
\end{align}
Finally, the vertex corrections are given by~\cite{Breso-Pla:2023tnz}
\begin{align}\label{eq:vertexCorMap}
    \delta g_L^{W e} &= c_{H e}^{(3)} + f\left(\frac{1}{2}, 0\right) - f\left(-\frac{1}{2}, -1\right), \\
    \delta g_L^{Z e} &= -\frac{1}{2} c_{H e}^{(3)} - \frac{1}{2} c_{H e} + f\left(-\frac{1}{2}, -1\right), \\
    \delta g_R^{Z e} &= -\frac{1}{2} c_{H e} + f(0, -1), \\
    \delta g_L^{Z u} &= \frac{1}{2} c_{H q}^{(3)} - \frac{1}{2} c_{H q}^{(1)} + f\left(\frac{1}{2}, \frac{2}{3}\right), \\
    \delta g_L^{Z d} &= -\frac{1}{2} c_{H q}^{(3)} - \frac{1}{2} c_{H q}^{(1)} + f\left(-\frac{1}{2}, -\frac{1}{3}\right), \\
    \delta g_R^{Z u} &= -\frac{1}{2} c_{H u} + f\left(0, \frac{2}{3}\right), \\
    \delta g_R^{Z d} &= -\frac{1}{2} c_{H d} + f\left(0, -\frac{1}{3}\right), \\
    \delta g_R^{W q} &= -\frac{1}{2} c_{H u d}.\\
    \delta g_L^{Z \nu} &=\delta g_L^{W e}+\delta g_L^{Z e},
\end{align}
where
\begin{align}
f\left(T^3, Q\right) & =-Q \frac{g_L g_Y}{g_L^2-g_Y^2} c_{H W B} \mathbb{1} \nonumber\\
& +\left(\frac{1}{4}\left[c_{l l}\right]_{e \mu \mu e}-\frac{1}{2}\left[c_{H l}^{(3)}\right]_{e e}-\frac{1}{2}\left[c_{H l}^{(3)}\right]_{\mu \mu}-\frac{1}{4} c_{H D}\right)\left(T^3+Q \frac{g_Y^2}{g_L^2-g_Y^2}\right) \mathbb{1}
\end{align}

\section{Renormalization Group Running}
\label{sec:running}

In this appendix, we provide the numerical results for the renormalization group evolution of the Wilson coefficients relevant to this work. Together with the matching conditions in \cref{sec:matching}, they determine operator mixing between $\mu = \SI{1}{TeV}$ and $\mu = \SI{2}{GeV}$, indicating which WEFT operators are generated by any given SMEFT operator. The results listed here are taken from ref.~\cite{Gonzalez-Alonso:2017iyc}; they include three-loop QCD running as well as one-loop electroweak running.

\subsection*{SMEFT Running from \SI{1}{TeV} to the Electroweak Scale}

The Wilson coefficients of the chirality-flipping SMEFT operators are evolved from the high scale $\mu = \SI{1}{TeV}$ down to the electroweak scale $\mu = m_Z$, following the RG equation in eq.~(2.5) of ref.~\cite{Gonzalez-Alonso:2017iyc}, which, when solved, gives
\begin{align}
\begin{pmatrix}
    c_{ledq} \\
    c_{lequ}^{(1)}\\
    c_{lequ}^{(3)} \\
\end{pmatrix}_{\!(\mu=m_Z)}
= 
\begin{pmatrix}
    1.19 & 0. & 0. \\
    0. & 1.20 & -0.185 \\
    0. & -0.00381 & 0.959 \\
\end{pmatrix} 
\begin{pmatrix}
    c_{ledq}\\
    c_{lequ}^{(1)} \\
    c_{lequ}^{(3)} \\
\end{pmatrix}_{\!(\mu=\SI{1}{TeV})},
\label{eq:RGEsmeft}
\end{align}
where flavour indices have been suppressed for clarity. Note the significant mixing of the tensor operator $c_{lequ}^{(3)}$ into $c_{lequ}^{(1)}$. The electroweak running of the chirality-conserving SMEFT operators is smaller and not expected to be relevant for our results, so we neglect it here.

\subsection*{WEFT Running from the EW Scale to 2 GeV}

After matching SMEFT onto WEFT at $\mu=m_Z$, the WEFT Wilson coefficients are run down to the low-energy scale $\mu = \SI{2}{GeV}$, which is characteristic for long-baseline neutrino experiments. The evolution of the CC WEFT Wilson coefficients is governed by the RG equation in eq.~(2.7) of ref.~\cite{Gonzalez-Alonso:2017iyc}, which, when solved, gives
\begin{align}
\begin{pmatrix}
    \epsilon_L^{jk} \\
    \epsilon_R^{jk} \\
    \epsilon_S^{jk} \\
    \epsilon_P^{jk} \\
    \epsilon_T^{jk}
\end{pmatrix}_{\!(\mu=\SI{2}{GeV})}
= 
\begin{pmatrix}
    1 & 0 &0&0&0\\
    0 & 1.0046&0&0&0\\
    0 & 0 & 1.72 & 2.46\times 10^{-6} & -0.0242 \\
    0 & 0 & 2.46\times 10^{-6} & 1.72 & -0.0242 \\
    0 & 0 & -2.17 \times 10^{-4} & -2.17 \times 10^{-4}& 0.825 \\
\end{pmatrix}
\begin{pmatrix}
    \epsilon_L^{jk} \\
    \epsilon_R^{jk} \\
    \epsilon_S^{jk} \\
    \epsilon_P^{jk} \\
    \epsilon_T^{jk}
\end{pmatrix}_{\!(\mu=m_Z)}.
\label{eq:RGEweft}
\end{align}
where lepton flavour indices have been suppressed for clarity. The NC WEFT Wilson coefficients are expected to have only very small electroweak running, and therefore we neglect it in this work.

\section{Complete SMEFT Sensitivities}
\label{sec:full_set_constraints}

\begin{figure}
  \centering
  \includegraphics[width=1\textwidth]{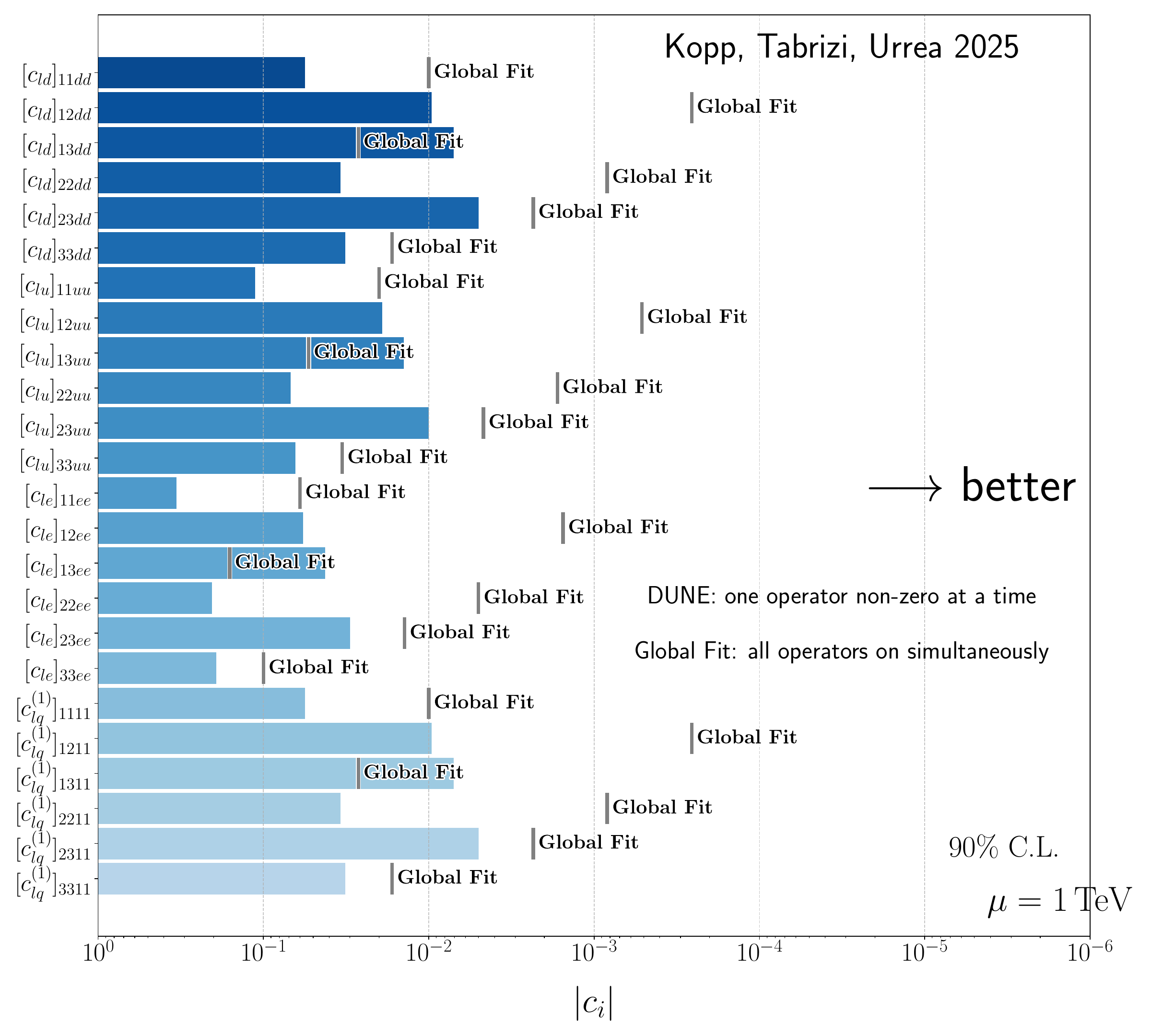}
  \caption{Expected DUNE sensitivity to the Wilson coefficients of the 4-fermion SMEFT operators $c_{ld}$, $c_{lu}$, $c_{le}$ and $c_{l q}^{(1)}$ within the SMEFT framework. We also show in black the current constraint taken from the global fit in ref.~\cite{Coloma:2023ixt}.  Note, however, that in this global fit, all NC Wilson coefficients are allowed to vary simultaneously, while in our analysis, we take only one of them non-zero at a time.}
  \label{fig:smeft1}
\end{figure}

\begin{figure}
  \centering
  \includegraphics[width=1\textwidth]{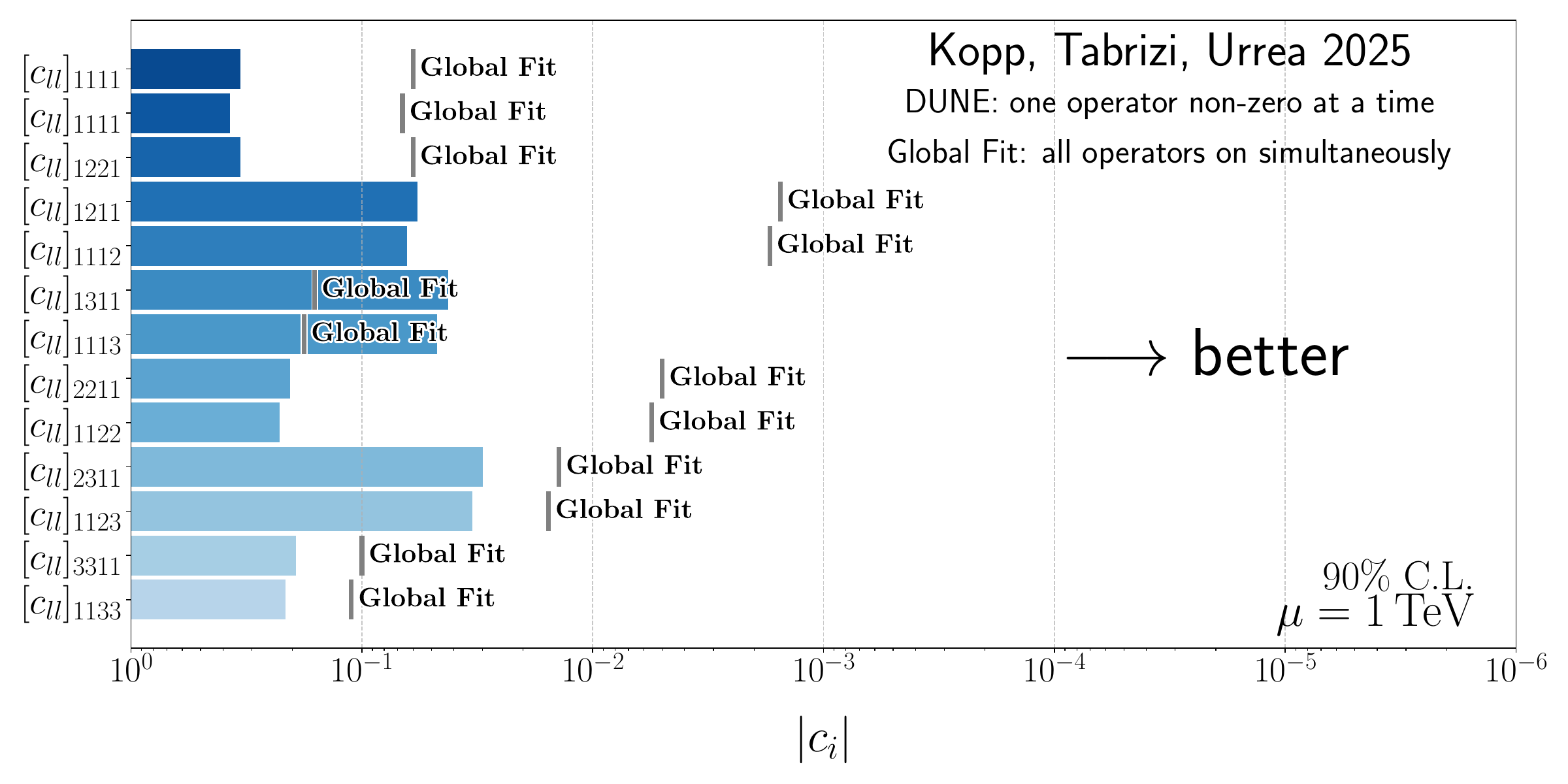}
  \caption{Expected DUNE sensitivity to the Wilson coefficients of the 4-fermion operator $c_{ll}$ within the SMEFT framework. We also show in black the current constraint taken from the global fit in ref.~\cite{Coloma:2023ixt}.  Note, however, that in this global fit, all NC Wilson coefficients are allowed to vary simultaneously, while in our analysis, we take only one of them non-zero at a time.}
  \label{fig:smeft2}
\end{figure}

\begin{figure}
  \centering
  \includegraphics[width=1\textwidth]{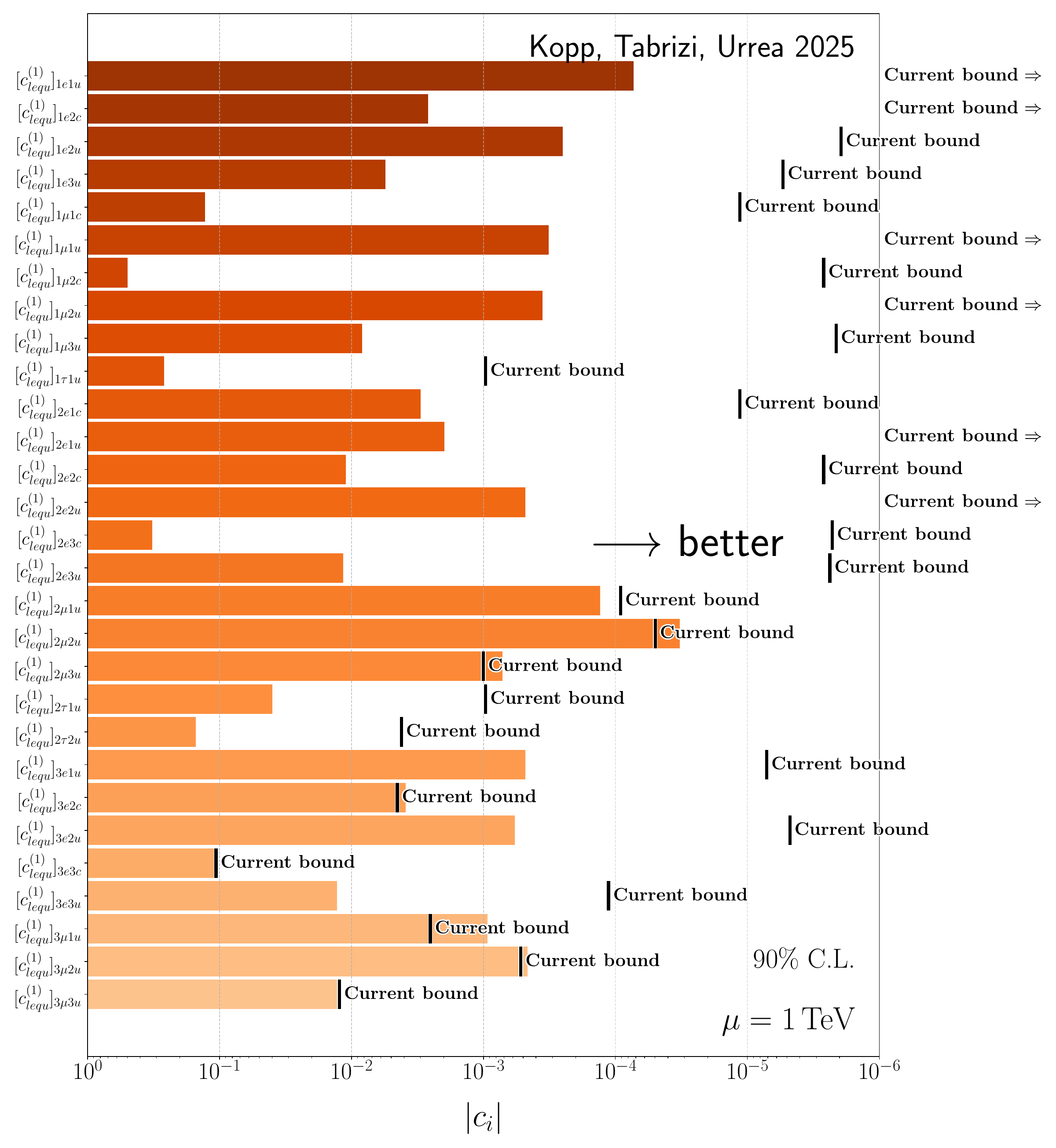}
  \caption{Expected DUNE sensitivity to the SMEFT Wilson coefficients $c_{lequ}^{(1)}$. Vertical black lines indicate current constraints, mostly taken from refs.~\cite{Falkowski:2021bkq, Coloma:2023ixt}.}
  \label{fig:smeft3}
\end{figure}

\begin{figure}
  \centering
  \includegraphics[width=1\textwidth]{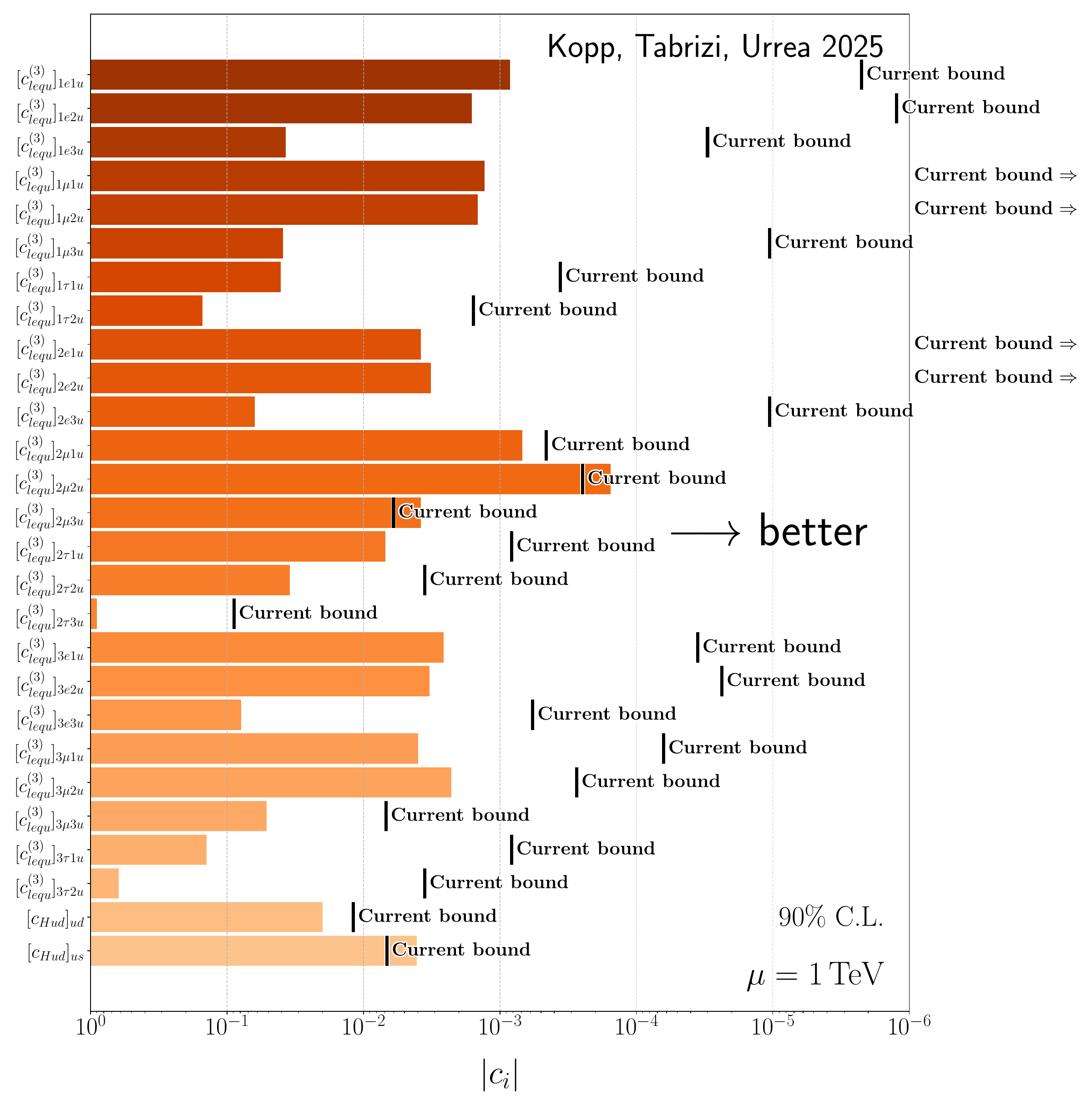}
  \caption{Expected DUNE sensitivity to the SMEFT Wilson coefficients $c_{lequ}^{(3)}$ and $c_{Hud}$. Vertical black lines indicate current constraints, mostly taken from refs.~\cite{Falkowski:2021bkq, Coloma:2023ixt}.}
  \label{fig:smeft4}
\end{figure}

\begin{figure}
  \centering
  \includegraphics[width=1\textwidth]{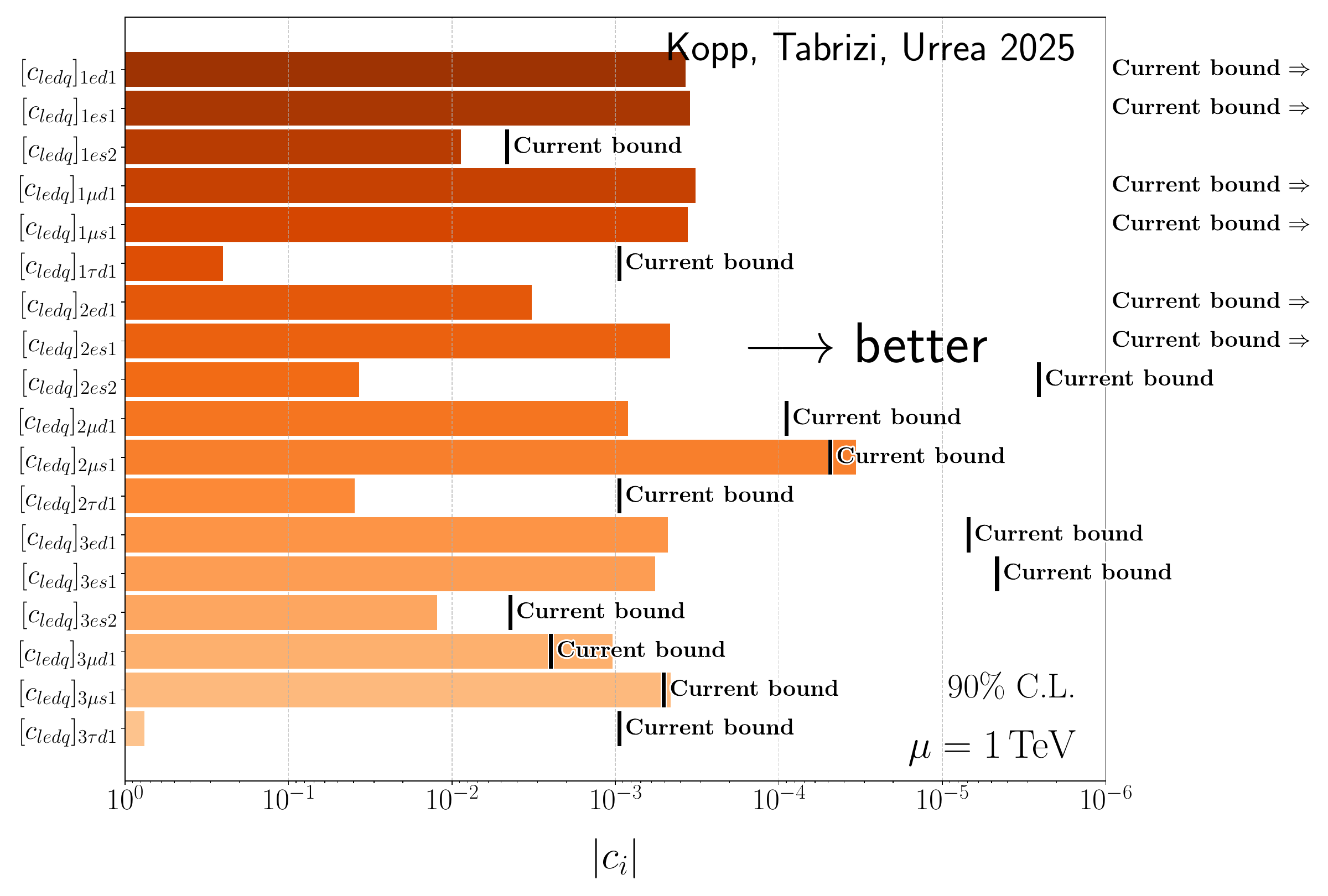}
  \caption{Expected DUNE sensitivity to the SMEFT Wilson coefficients $c_{ledq}$. Vertical black lines indicate current constraints, mostly taken from refs.~\cite{Falkowski:2021bkq, Coloma:2023ixt}.}
  \label{fig:smeft5}
\end{figure}

In this appendix, we give our full set of projected SMEFT constraints. \Cref{fig:smeft1} shows the sensitivities to the Wilson coefficients $c_{ld}$, $c_{lu}$, $c_{le}$ and $c_{l q}^{(1)}$, while \cref{fig:smeft2} displays corresponding results for $c_{ll}$. All these operators give rise only to NC interactions when matched onto WEFT. Similarly, \cref{fig:smeft3,fig:smeft4,fig:smeft5} display constraints on $c_{lequ}^{(1)}$, $c_{lequ}^{(3)}$, $c_{Hud}$, and $c_{ledq}$.

\clearpage
\section{GLoBES-EFT Documentation}
\label{sec:GLoBES-EFT}

The phenomenological results presented in this paper were obtained using a custom-built probability engine interfacing with the GLoBES library, which we call \texttt{GLoBES-EFT}. This plugin implements both WEFT and SMEFT in a consistent manner. It correctly models new physics effects in neutrino production, propagation, and detection, and handles the renormalisation group evolution (RGE) using results from ref.~\cite{Gonzalez-Alonso:2017iyc} and the matching between WEFT and SMEFT discussed in \cref{sec:matching}. In this appendix, we provide a technical manual for users who wish to employ \texttt{GLoBES-EFT} in their own analyses. 

New physics in the neutrino production and detection processes is implemented via production coefficients $p_{XY,\alpha}^{S,jk}$ and detection coefficients $d_{XY, \beta}^{D,\mathrm{rs}}$, as introduced in \cref{sec:Productions} and \cref{sec:Detection}, respectively. As $p_{XY,\alpha}^{S,jk}$ depends on the neutrino production process, it is typically necessary to split the total fluxes employed in standard GLoBES simulations into their individual components, as we have done in this paper. Regarding detection, we have considered QES as the sole detection process in this work. \texttt{GLoBES-EFT}, however, can handle corrections also to other detection processes if the user provides tabulated detection coefficients and SM cross sections. \texttt{GLoBES-EFT} reads the production and detection coefficients from external files, which we provide for the production and detection processes considered in this paper. The first column in each file specifies the neutrino energy (in units of \si{GeV} for the production coefficients and in $\log_{10}(E/\si{GeV})$ for the detection coefficients. The remaining columns contain the EFT coefficient values, arranged in a fixed and structured order:
\begin{itemize}
    \item The outermost loop is over the Lorentz structure pairs $XY \in \{\text{LL}, \text{LR}, \text{LS}, \text{LP}, \text{LT}, \text{RL}, \text{RR}, \dots\}$.
    
    \item For each Lorentz structure, there are six columns corresponding to the lepton flavour indices $\alpha = e, \mu, \tau$, first for neutrinos and then for antineutrinos.
\end{itemize}
To link these files to the fluxes and cross sections in a GLoBES experiment definition, we have introduced two new directives: \texttt{@eft\_coeff\_file} points to a coefficient file and \texttt{@quark\_flavors} specifies the quark flavor indices. These directives are used in the context of \texttt{nuflux} and \texttt{cross} environments as follows:

\begin{verbatim}
  nuflux(#DplusNuMode)<
     @flux_file="./FluxesDplusNuModeND.txt"
     @time = 3.5 /* years */
     @power = 11.0  
     @eft_coeff_file = "./plus_ProdALLDpDecay.txt"
     @norm = 4.415e8
    @quark_flavors = 1 : 0
  > 
\end{verbatim}

\begin{verbatim}
  cross(#CC)<
    @cross_file = "./xsec_cc_Oxygen_LQCD_LL_divided_by_16.txt"
    @eft_coeff_file = "./DetAllCCQEOxygenLQCDS1Correct2.txt"
    @quark_flavors = 0 : 0
  >
\end{verbatim}

\noindent
In the user's code, the \texttt{GLoBES-EFT} engine needs to be initialized via
\begin{verbatim}
  smeft_init_probability_engine_3();
\end{verbatim}
and then registered with GLoBES using
\begin{verbatim}
  glbRegisterProbabilityEngine(MAX_PARAMS, &smeft_probability_matrix,
                               &smeft_set_oscillation_parameters,
                               &smeft_get_oscillation_parameters, NULL);

\end{verbatim}
Here \texttt{MAX\_PARAMS} is the number of oscillation parameters. With \texttt{GLoBES-EFT} supporting both WEFT and SMEFT, this number is 2405. As a first check, to verify that coefficients have been successfully loaded, the following functions can be used:
\begin{verbatim}
  double glbEFTFluxCoeff(int exp, int flux_ident, int X, int Y,
                         int alpha, double E)
\end{verbatim}
\begin{verbatim}
  double glbEFTXSecCoeff(int exp, int xsec_ident, int X, int Y,
                         int alpha,int cp, double E)
\end{verbatim}
These functions return the values of $p_{XY,\alpha}(E)$ and $d_{XY,\alpha}(E)$, respectively. \texttt{cp} is $+1$ for neutrinos and $-1$ for antineutrinos, while the \texttt{flux\_ident} and \texttt{xsec\_ident} arguments specify the flux or cross section; they are indexed according to the order in which they appear in the GLoBES files.  

\texttt{GLoBES-EFT} supports both WEFT and SMEFT modes. The default is WEFT. To switch to SMEFT mode, use:
\begin{verbatim}
  glbSetOscParamByName(test_values, 1, "SMEFT_FLAG");
\end{verbatim}
In WEFT mode, the Wilson coefficients are assumed to be defined at a renormalization scale of \SI{2}{GeV} and no renormalization group evolution is performed. In SMEFT mode, the code assumes that the Wilson coefficients are given at $\mu = \SI{1}{TeV}$, and the code automatically takes care of the RGE evolution to the electroweak scale, the matching onto WEFT, and the continued running  to \SI{2}{GeV} (or to the $b$-quark mass, depending on the operator). 

To define model parameters, it is recommended to use \texttt{glbSetOscParamByName} to refer to model parameters using a human-readable name as shown in example \texttt{example-smeft}.
A list of appropriate names is provided in the global array \texttt{char smeft\_param\_strings[][64]}. The naming conventions are detailed in \cref{tab:weft_params_globes} for WEFT and \cref{tab:smeft_params_globes} for SMEFT.

\begin{table}
    \centering
    \begin{tabular}{lll}
        \toprule
        \textbf{WEFT Coefficient} & \textbf{GLoBES Name (Modulus)} & \textbf{GLoBES Name (Phase)} \\
        \midrule
        \multicolumn{3}{c}{\textit{Charged Current (CC) Operators}} \\
        $[\epsilon_L^{jk}]_{\alpha\beta}$ & \texttt{ABS\_EPS\_CC\_L\_$jk$\_$p r$} & \texttt{ARG\_EPS\_CC\_L\_$jk$\_$p r$} \\
        $[\epsilon_R^{jk}]_{\alpha\beta}$ & \texttt{ABS\_EPS\_CC\_R\_$jk$\_$p r$} & \texttt{ARG\_EPS\_CC\_R\_$jk$\_$p r$} \\
        $[\epsilon_S^{jk}]_{\alpha\beta}$ & \texttt{ABS\_EPS\_CC\_S\_$jk$\_$p r$} & \texttt{ARG\_EPS\_CC\_S\_$jk$\_$p r$} \\
        $[\epsilon_P^{jk}]_{\alpha\beta}$ & \texttt{ABS\_EPS\_CC\_P\_$jk$\_$p r$} & \texttt{ARG\_EPS\_CC\_P\_$jk$\_$p r$} \\
        $[\epsilon_T^{jk}]_{\alpha\beta}$ & \texttt{ABS\_EPS\_CC\_T\_$jk$\_$p r$} & \texttt{ARG\_EPS\_CC\_T\_$jk$\_$p r$} \\
        \midrule
        \multicolumn{3}{c}{\textit{Neutral Current (NC) Operators}} \\
        $[\epsilon_{L}^{m,f}]_{\alpha\beta}$ & \texttt{ABS\_EPS\_NC\_L\_$f f$\_$p r$} & \texttt{ARG\_EPS\_NC\_L\_$f f$\_$p r$} \\
        $[\epsilon_{R}^{m,f}]_{\alpha\beta}$ & \texttt{ABS\_EPS\_NC\_R\_$f f$\_$p r$} & \texttt{ARG\_EPS\_NC\_R\_$f f$\_$p r$} \\
        $[\epsilon_{L}^{m,f}]_{\alpha\alpha}$ & \texttt{EPS\_NC\_L\_$f f$\_$p p$} & --- \\
        $[\epsilon_{R}^{m,f}]_{\alpha\alpha}$ & \texttt{EPS\_NC\_R\_$f f$\_$p p$} & --- \\
        \bottomrule
    \end{tabular}
    \caption{WEFT parameter names available in \texttt{GLoBES-EFT}. Lepton flavor indices $\alpha$, $\beta$ correspond to GLoBES strings \texttt{E, MU, TAU, S1, ...}. Quark indices $j,k$ correspond to \texttt{u, c} and \texttt{d, s, b} respectively, forming pairs like \texttt{ud} or \texttt{us}. Matter fermion fields $f$ are \texttt{E, u, d}.}
    \label{tab:weft_params_globes}
\end{table}

\begin{table}
    \centering 
    \begin{tabular}{lll}
        \toprule
        \textbf{SMEFT Coefficient} & \textbf{GLoBES Name (Modulus)} & \textbf{GLoBES Name (Phase)} \\
        \midrule
        $[c_{HWB}]$ & \texttt{ABS\_C\_HWB} & \texttt{ARG\_C\_HWB} \\
        $[c_{HD}]$ & \texttt{ABS\_C\_HD} & \texttt{ARG\_C\_HD} \\
        $[c_{H l}^{(3)}]_{pr}$ & \texttt{ABS\_C3\_PHI\_L\_$pr$} & \texttt{ARG\_C3\_PHI\_L\_$pr$} \\
        $[c_{H l}^{(1)}]_{pr}$ & \texttt{ABS\_C\_PHI\_L\_$pr$} & \texttt{ARG\_C\_PHI\_L\_$pr$} \\
        $[c_{H e}]_{pr}$ & \texttt{ABS\_C\_PHI\_E\_$pr$} & \texttt{ARG\_C\_PHI\_E\_$pr$} \\
        $[c_{H q}^{(3)}]_{j_q k_q}$ & \texttt{ABS\_C3\_PHI\_Q\_$j_q k_q$} & \texttt{ARG\_C3\_PHI\_Q\_$j_q k_q$} \\
        $[c_{H q}^{(1)}]_{j_q k_q}$ & \texttt{ABS\_C\_PHI\_Q\_$j_q k_q$} & \texttt{ARG\_C\_PHI\_Q\_$j_q k_q$} \\
        $[c_{H u}]_{j_u k_u}$ & \texttt{ABS\_C\_PHI\_U\_$j_u k_u$} & \texttt{ARG\_C\_PHI\_U\_$j_u k_u$} \\
        $[c_{H d}]_{j_d k_d}$ & \texttt{ABS\_C\_PHI\_D\_$j_d k_d$} & \texttt{ARG\_C\_PHI\_D\_$j_d k_d$} \\
        $[c_{H ud}]_{j_u k_d}$ & \texttt{ABS\_C\_PHI\_UD\_$j_u k_d$} & \texttt{ARG\_C\_PHI\_UD\_$j_u k_d$} \\
        $[c_{ll}]_{prst}$ & \texttt{ABS\_C\_LL\_$pr$\_$st$} & \texttt{ARG\_C\_LL\_$pr$\_$st$} \\
        $[c_{le}]_{prst}$ & \texttt{ABS\_C\_LE\_$pr$\_$st$} & \texttt{ARG\_C\_LE\_$pr$\_$st$} \\
        $[c_{lq}^{(3)}]_{pr, j_q k_q}$ & \texttt{ABS\_C3\_LQ\_$pr$\_$j_q k_q$} & \texttt{ARG\_C3\_LQ\_$pr$\_$j_q k_q$} \\
        $[c_{lq}^{(1)}]_{pr, j_q k_q}$ & \texttt{ABS\_C\_LQ\_$pr$\_$j_q k_q$} & \texttt{ARG\_C\_LQ\_$pr$\_$j_q k_q$} \\
        $[c_{lu}]_{pr, j_u k_u}$ & \texttt{ABS\_C\_LU\_$pr$\_$j_u k_u$} & \texttt{ARG\_C\_LU\_$pr$\_$j_u k_u$} \\
        $[c_{ld}]_{pr, j_d k_d}$ & \texttt{ABS\_C\_LD\_$pr$\_$j_d k_d$} & \texttt{ARG\_C\_LD\_$pr$\_$j_d k_d$} \\
        $[c_{ledq}]_{pr, j_d k_q}$ & \texttt{ABS\_C\_LEDQ\_$pr$\_$j_d k_q$} & \texttt{ARG\_C\_LEDQ\_$pr$\_$j_d k_q$} \\
        $[c_{lequ}^{(1)}]_{pr, j_q k_u}$ & \texttt{ABS\_C\_LEQU\_$pr$\_$j_q k_u$} & \texttt{ARG\_C\_LEQU\_$pr$\_$j_q k_u$} \\
        $[c_{lequ}^{(3)}]_{pr, j_q k_u}$ & \texttt{ABS\_C3\_LEQU\_$pr$\_$j_q k_u$} & \texttt{ARG\_C3\_LEQU\_$pr$\_$j_q k_u$} \\
        \bottomrule
    \end{tabular}
    \caption{SMEFT parameter names available in \texttt{GLoBES-EFT}. Lepton indices $p$, $r$, $s$, $t$ correspond to \texttt{E, MU, TAU} in \texttt{GLoBES-EFT}. Up-type quark indices $j_u, k_u$ can take the values \texttt{u, c, t}, down-type indices $j_d, k_d$ the values \texttt{d, s, b}; quark doublet families $j_q, k_q$ are denoted \texttt{Q1, Q2, Q3}.}
    \label{tab:smeft_params_globes}
\end{table}

To directly access the oscillation (pseudo-)probabilities $\tilde{P}_{\alpha\beta}^S(E_\nu, L)$ defined in \cref{eq:tildeP}, \texttt{GLoBES-EFT} provides the functions
\begin{verbatim}
  double smeft_glbVacuumProbability(int initial_flavour, int final_flavour,
            int cp_sign, double E, double L,int flux_id, int cross_id);
  double smeft_glbConstantDensityProbability(int initial_flavour, int final_flavour,
            int cp_sign, double E, double L, double rho,int flux_id, int cross_id);
  double smeft_glbProfileProbability(int exp,int initial_flavour, int final_flavour,
            int panti, double energy,int flux_id, int cross_id);
  double smeft_glbFilteredConstantDensityProbability(int exp,int initial_flavour,
            int final_flavour, int panti, double energy,int flux_id, int cross_id);
\end{verbatim}
These functions behave like their native counterparts, but require two extra arguments \texttt{flux\_id} and \texttt{cross\_id} to specify the flux and cross section by order of definition in the input files. These extra arguments are necessary to identify the correct set of production and detection coefficients for the computation of $\tilde{P}_{\alpha\beta}^S(E_\nu, L)$. 

After using the \texttt{GLoBES-EFT} engine, it is good practice to release the small amount of memory it allocates by calling
\begin{verbatim}
  smeft_free_probability_engine();
\end{verbatim}

\clearpage
\bibliographystyle{JHEP} 
\bibliography{ref}

\end{document}